\DocumentMetadata{}
\PassOptionsToPackage{table,xcdraw}{xcolor}
\documentclass[sigconf]{acmart}

\usepackage{multirow}
\usepackage{makecell}
\usepackage{longtable}
\usepackage{enumitem}
\usepackage{soul}
\usepackage{listings}
\usepackage{makecell}

\usepackage{textcomp}
\usepackage{tabularx} 
\usepackage[table]{xcolor}

\usepackage{fancyhdr}

\setlistdepth{8}
\setlist[itemize,1]{label=\textbullet}
\setlist[itemize,2]{label=\textopenbullet}
\setlist[itemize,3]{label=$\blacktriangleright$}
\setlist[itemize,4]{label=$\smalltriangleright$}
\setlist[itemize,5]{label=$\sqbullet$}
\setlist[itemize,6]{label=$\square$=}
\setlist[itemize,7]{label=$\blackdiamond$}
\setlist[itemize,8]{label=$\diamond$}
\renewlist{itemize}{itemize}{8}

\definecolor{blockcolor}{HTML}{555555}

\newenvironment{block}%
  {\list{}{\leftmargin=0.2in\rightmargin=0in}\item[]\color{blockcolor}}%
  {\endlist}

\newcommand{\sysname}{\textsc{Proxona}}
\newcommand{\baseline}{\textsc{Baseline}}
\newcommand{\docs}{\textsc{Docs}}

\newcommand{\dimension}{\texttt{dimension}}
\newcommand{\dimensions}{\texttt{dimensions}}
\newcommand{\dimval}{\texttt{value}}
\newcommand{\dimvals}{\texttt{values}}

\definecolor{Dimension}{HTML}{86469C} 
\definecolor{Value}{HTML}{DFD7FC} 
\definecolor{Proxona}{HTML}{FFD485} 

\newcommand{\dimcol}{\textcolor{Dimension}}
\newcommand{\valcol}{\colorbox{Value}}

\AtBeginDocument{%
  \providecommand\BibTeX{{%
    \normalfont B\kern-0.5em{\scshape i\kern-0.25em b}\kern-0.8em\TeX}}}

\copyrightyear{2025}
\acmYear{2025}
\setcopyright{cc}
\setcctype{by}
\acmConference[CHI '25]{CHI Conference on Human Factors in Computing Systems}{April 26-May 1, 2025}{Yokohama, Japan}
\acmBooktitle{CHI Conference on Human Factors in Computing Systems (CHI '25), April 26-May 1, 2025, Yokohama, Japan}
\acmDOI{10.1145/3706598.3714034}
\acmISBN{979-8-4007-1394-1/25/04}

\begin{document}


\title{Proxona: Supporting Creators' Sensemaking and Ideation with LLM-Powered Audience Personas}


\author{Yoonseo Choi}
\email{yoonseo.choi@kaist.ac.kr}
\affiliation{%
    \institution{School of Computing, KAIST}
    \country{Republic of Korea}
}

\author{Eun Jeong Kang}
\email{ek646@cornell.edu}
\affiliation{%
    \institution{Information Science, Cornell University}
    \country{Ithaca, NY, United States}
}

\author{Seulgi Choi}
\email{seulgi@kaist.ac.kr}
\affiliation{%
    \institution{School of Computing, KAIST}
    \country{Republic of Korea}
}

\author{Min Kyung Lee}
\email{minkyung.lee@austin.utexas.edu}
\affiliation{%
    \institution{The University of Texas at Austin}
    \country{Austin, TX, United States}
}

\author{Juho Kim}
\email{juhokim@kaist.ac.kr}
\affiliation{%
    \institution{School of Computing, KAIST}
    \country{Republic of Korea}
}

\renewcommand{\shortauthors}{Yoonseo Choi et al.}

\begin{teaserfigure}
  \centering
\includegraphics[trim=0cm 1cm 0cm 0cm, clip=true, width=\textwidth]{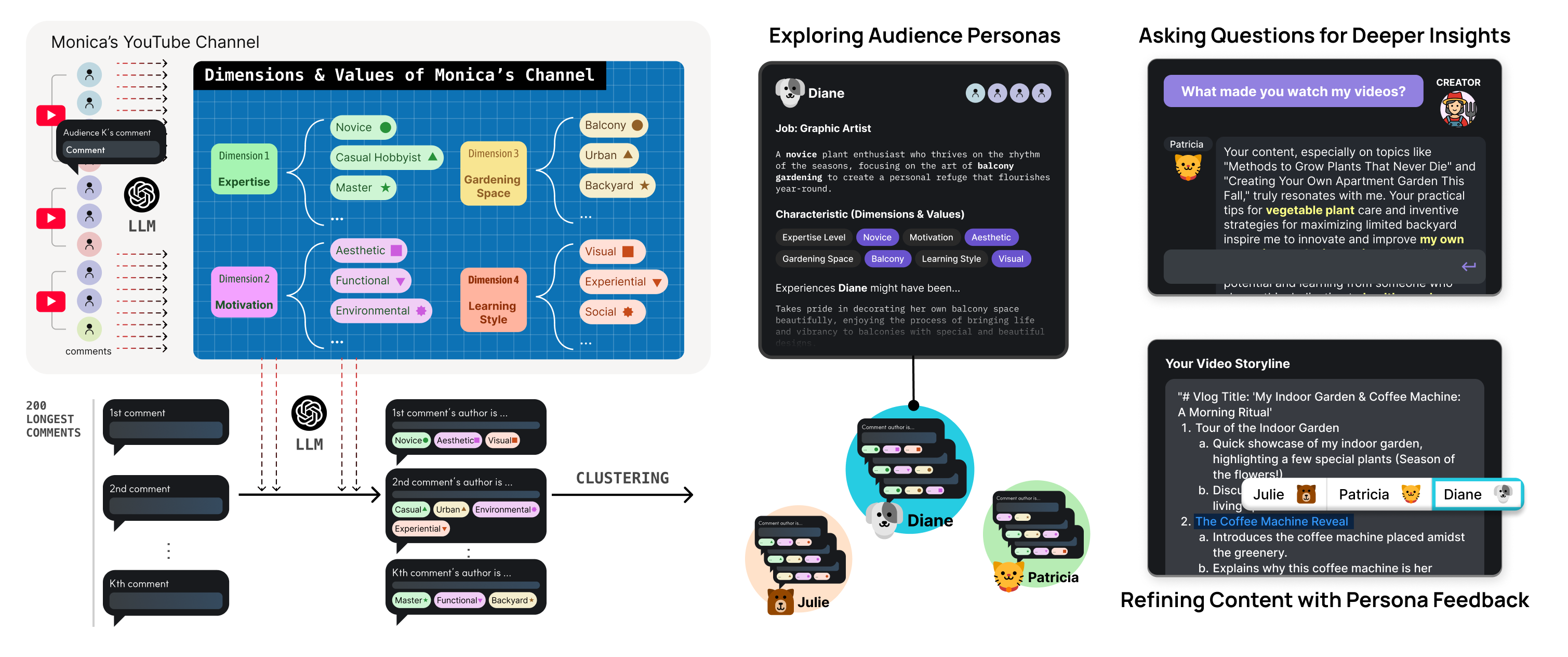}
  \caption{\sysname{}, an LLM-powered system that transforms audience comments into interactive, multi-dimensional personas, enabling creators to gain deeper insights, gather simulated feedback, and refine their early-stage creative work.}
  \label{fig:teaser}
  \Description{Teaser image. \sysname{}, an LLM-powered system that transforms audience comments into interactive, multi-dimensional personas, enabling creators to gain deeper insights, gather simulated feedback, and refine their early-stage creative work.}
\end{teaserfigure}

\begin{abstract}

A content creator's success depends on understanding their audience, but existing tools fail to provide in-depth insights and actionable feedback necessary for effectively targeting their audience. 
We present \sysname{}, an LLM-powered system that transforms static audience comments into interactive, multi-dimensional personas, allowing creators to engage with them to gain insights, gather simulated feedback, and refine content.
\sysname{} distills audience traits from comments, into \dimensions{} (categories) and \dimvals{} (attributes), then clusters them into interactive personas representing audience segments.
Technical evaluations show that \sysname{} generates diverse \dimensions{} and \dimvals{}, enabling the creation of personas that sufficiently reflect the audience and support data-grounded conversation.
User evaluation with 11 creators confirmed that \sysname{} helped creators discover hidden audiences, gain persona-informed insights on early-stage content, and allowed them to confidently employ strategies when iteratively creating storylines. 
\sysname{} introduces a novel creator-audience interaction framework and fosters a persona-driven, co-creative process. 




\end{abstract}




\maketitle


\section{Introduction}

\begin{quote}
    ``\textit{A show without an audience is nothing, after all. In the response of the audience, that is where the power of performance lives.}'' (Erin Morgenstern, 2012 ~\cite{Morgenstern_2012}) 
\end{quote}
With the rise of the creator economy, competition for audience attention on digital platforms has grown~\cite{rossi2021measuring}. As viewer engagement directly influences creators' popularity, revenue, and content strategy, creators must capture and retain the audience's interest~\cite{hodl2023content, ormen2023towards, biel2011vlogsense}. Platform data analytics tools like YouTube Studio~\footnote{https://studio.youtube.com/} offer an overview of audience behaviors, such as view counts, watch time, and demographic data to support content decisions. 

However, these tools focus on quantitative data, which often do not capture the deeper, contextual aspects of viewer behavior, such as motivations or preferences. Creators are left with valuable yet abstract data, which requires significant effort to interpret and act upon. In contrast, comments provide direct, qualitative insights into viewers' sentiments and reactions~\cite{luo2020emotional}. Yet, the sheer volume of comments can make it difficult for creators to extract insights that aid ideations considering their specific audiences. Our formative study (N = $13$) confirmed that creators face challenges in understanding the motivations behind audience behavior ---e.g., why viewers engaged with or disengaged from their content---, seek more direct audience feedback beyond basic reactions such as likes or comments and often feel uncertain about how to adapt their content to meet audience expectations better. These challenges are particularly pronounced during the early stages of content creation, when creators explore ideas and make initial decisions about how to target and engage their audiences.

To address these challenges, we introduce the concept of \emph{audience personas}, built on the concept of personas from user-centered design~\cite{salminen2018personas}. Audience personas synthesize large-scale and unstructured audience data--- comments ---into compact and actionable representations of distinct audience segments. Instead of requiring manual analysis of a large number of comments or relying on general metrics, these personas provide a structured way to discover audience insights. \sysname{} enables conversational interactions, allowing creators to explore audience personas dynamically while engaging with simulated feedback that reflects plausible audience preferences and opinions. 

We instantiate the concept of audience personas in our system, \sysname{}, which generates interactive audience personas grounded on real audience data, bridging the gap between abstract analytics and relatable audience insights. With \sysname{}, creators can chat with audience personas to better understand their audiences' characteristics and receive targeted support for content creation. To create consistent and interpretable personas, we propose a framework that organizes audience data into \dimensions{} (broad characteristic categories) and \dimvals{} (specific attributes within each dimension). This structured organization uncovers patterns in audience traits that might otherwise be overlooked, enabling systematic analysis and clustering of audience comments into distinct, meaningful audience segments. Then, \sysname{} uses a Large Language Model (LLM) to simulate personas and enable interactive feedback based on these structured segments. This structured approach allows creators to gain insights into overlooked or niche audience traits, which might be hard to discern through traditional data analytics. 

Technical evaluations show that \sysname{} produced multidimensional personas that provided unexpected and diverse perspectives of audiences when compared to personas generated by a baseline. Furthermore, persona-generated responses had a low likelihood of hallucinations (below 5\%), which we define as inaccuracies in referencing specific video or channel content. This demonstrates the system’s reliability in providing grounded, evidence-based insights.

To evaluate the practical impact of \sysname{} on creators' early-stage creative practices, we conducted a user evaluation with 11 YouTube creators, asking them to create a video storyline on a given topic (e.g., a promotion video for consumer products or digital tools), comparing \sysname{} with their current practices as a baseline (\docs{}). The creators reported that \sysname{} helped them explore diverse and in-depth audience characteristics and gain confidence in tailoring content ideas to plausible audience preferences. They valued that \sysname{} supported the identification of audience segments they might have previously overlooked. By interacting with audience-driven personas, creators gathered potential audience opinions, refined content logic through iterative feedback, and made informed decisions throughout their creative process. Overall, audience personas of \sysname{} empowered exploratory audience engagement, enabling creators to envision their audience more effectively during content ideation, making them feel as if they were genuinely collaborating with their virtual audience.

Our contributions are as follows: 
\begin{itemize}
    \item Insights from a formative study that highlights design opportunities to help creators better understand and target their audience. 
    \item \sysname{}, an LLM-powered system that supports creators in exploring plausible audience traits and patterns by interacting with data-driven personas and making informed decisions in content creation.
    \item A technical pipeline that effectively generates relevant, distinct, and audience-centric personas with our persona construction framework (\dimensions{} \& \dimvals{}).
    \item Empirical findings from a user study showing how \sysname{} enhanced creators' sensemaking of their audience and helped them make audience-driven decisions in their creative practices.
\end{itemize}


\section{Background and Related Work}
We review prior work on creator context, person-based user testing methods, and the use of LLMs for simulating agents. We begin by investigating the unique context of creators, who have distinct characteristics and backgrounds. Then, as we propose system-empowered LLM and persona methods, we discuss how each method is utilized in our context.

\subsection{Involving Audience to Catalyze Creator's Creativity}
How creativity emerges and operates in society remains an ongoing area of research~\cite{csikszentmihalyi1997flow}, and Henriksen et al.~\cite{henriksen2016systems} suggest that platforms like YouTube redefine `creativity' by connecting creators with their audience, offering more opportunities for self-expression~\cite{henriksen2016systems}. This highlights the complex interplay between creators and their audiences~\cite{mcroberts2016viewers, li2021drives, Wohn2018}, requiring creators to put extra effort into comprehending their audience when creating content, such as through analyzing audience engagement metrics~\cite{mallari2021understanding} or perusing comments~\cite{jhaver2022designing}. 

The HCI community has introduced several creative support tools to aid content creators in their creative endeavors, from gathering feedback to refining their work~\cite{frich2019mapping, chung2022talebrush, choi2023creativeconnect, kim2017mosaic}. However, from a practical perspective, these tools are most beneficial for experienced creators open to experimenting to connect with their target audience effectively. 
Given that creators produce diverse content for varying audiences~\cite{ma2023multi}, it is essential for them to develop the skills to filter feedback from a wide range of reactions, drawing upon their own experiences~\cite{choi2023creator}. 
Also, with platforms increasingly using algorithms for recommendation, strategic content planning becomes essential for gaining audience exposure~\cite{bishop2020algorithmic}. 
We designed \sysname{} that helps creators process and interpret audience feedback more effectively through LLM-driven personas, offering clearer insights into audience preferences. Highlighting the significance and challenges of audience comprehension in content creation, we propose an interactive system that allows creators to explore their audience personas and inform their creative work.

\subsection{Creating and Testing with Personas}

Creating and testing with an `imagined user' is often practiced by product managers and user experience (UX) designers who need to tailor products to meet user needs~\cite{salminen2018personas}. 
The `imagined user', or `persona' is not solely created by designers' imagination; it integrates users' actual characteristics with the designers' ideal product goals. 
To understand the users' actual characteristics, designers conduct user research meticulously, such as user surveys or focus group interviews~\cite{cooper1999inmates}. 
Based on the evidence, they detail attributes of a `persona,' such as lifestyle and preferred brands, effectively communicating actual users' needs.
However, the simplification process involved in creating these representations may overlook marginalized perspectives in individuals~\cite{turner2011stereotyping, marsden2016stereotypes}.

In recent work, researchers also suggest data-driven methods to generate `persona', which utilizes users' behavioral data via algorithms~\cite{mcginn2008chi, salminen2020template, zhu2019creating, salminen2020literature, salminen2020persona}. 
These methods create personas close to presented users efficiently, based on data.
On the other hand, recent advances in LLMs have introduced new methods for persona generation, enabling a more profound analysis of qualitative data, offering richer insights into user behavior and preferences compared to traditional quantitative approaches~\cite{Shin2024dis}.

Similarly, content creators form an `imagined audience'—a mental representation of their viewers based on both online and offline contextual clues~\cite{duffy2017platform, litt2012knock}. 
They can access audience data, such as comments or demographic information, which shapes creators' decisions regarding activities in platforms and creatives~\cite{ma2023multi}.
However, their audience groups are diverse and hard to reach, challenging creators to integrate their perspectives in content creation.
In \sysname{}, we leverage this potential by combining LLM-driven qualitative analysis with quantitative methods --- blending the scalability of algorithmic approaches with the depth of qualitative data. 

\subsection{Simulating Agents with LLMs}
Recent studies suggest that LLMs can simulate aspects of human behavior, offering potential utility in situations where engaging human resources might be costly or inaccessible~\cite{park2022social, park2023generative, benharrak2023writer}. In addition to surveys~\cite{hamalainen2023evaluating} and annotation tasks~\cite{salminen2023can} that demand significant human effort, LLMs can be used to comprehend human perspectives and cognition as an agent~\cite{park2023generative}, for example, in user interviews~\footnote{https://www.syntheticusers.com/} and feedback sessions~\cite{benharrak2023writer}.

Advancing beyond human simulation, incorporating personality traits into agents is thought to boost engagement by reflecting specific character viewpoints in LLM simulations~\cite{deshpande2023anthropomorphization, jiangPersonaLLMInvestigatingAbility2024a, shao2023character, wei2023leveraging}. 
This can open up ways for users with limited access to human resources to engage in their workflow~\cite{benharrak2023writer, markel2023gpteach, jin2024teach}. 
For instance, `GPTeach'~\cite{markel2023gpteach} helps junior teaching assistants prepare for real-life situations, by simulating students' questions and perspectives. 
Similarly, Benharrak~\cite{benharrak2023writer} presents a tool for writers to refine their work using feedback from persona-based agents. 
Still, without a grounded contextual background, such agents risk miscommunication, misleading users with incorrect information, and potentially distorting users' views~\cite{chengCoMPosTCharacterizingEvaluating2023a, deshpande2023anthropomorphization}. 
When designing feedback and interaction processes for LLM-based personas, it is crucial to consider how stereotyping specific audience groups can influence these designs, due to potential biases inherent in foundation models~\cite{cheng2023marked} and the automated generation process~\cite{salminen2019detecting}.

While previous work has focused on simulating virtual feedback in specific scenarios, such as education~\cite{markel2023gpteach} or writing assistance~\cite{benharrak2023writer}, \sysname{} differentiates itself by grounding persona simulations in real-world comments collected directly from the creator’s channel. This approach provides creators with a more contextually relevant and personalized experience, tailored to their unique audience. 
Persona generation methods vary in their trade-offs between depth, scalability, and flexibility. Park et al.~\cite{park2024generative} leverages extensive interviews to create highly accurate personas suited for predictive insights, while Shin et al.~\cite{Shin2024dis} combines human-driven clustering with LLM summarization to balance representativeness and empathy. In contrast, \sysname{} synthesizes audience comments into actionable personas, offering a lightweight and scalable solution tailored for early-stage creative workflows. Unlike generic feedback simulations that may overlook subtle cultural or contextual nuances in audience interactions, \sysname{} leverages the richness of real audience data to enhance the relevance and authenticity of its generated personas.

While grounding persona simulations in real audience data enhances their relevance, it also introduces challenges, such as handling unstructured or noisy data and addressing biases inherent in user-generated content. 
To mitigate these issues, \sysname{} employs a dimension-value framework that systematically organizes audience traits into interpretable categories, thereby improving the interpretability and usability of persona simulations. By anchoring persona generation in actual audience data while incorporating structured organization, \sysname{} aims to reduce the risk of miscommunication or bias, offering creators a complementary tool for exploring audience traits and motivations.


\section{Formative Study}

To explore the challenges creators face in understanding their audiences, we conducted semi-structured interviews with 13 YouTube creators (N = 13) who have been actively managing their channels for over a year (Appendix~\ref{app:formative-study}-Table~\ref{tab:formative_study}). We recruited them through social media and direct cold emails. Each interviewee participated in a 50-minute Zoom session and was compensated with KRW 50,000 (approx. USD 38). The interviews focused on understanding their current practices for defining and analyzing their target audience during content creation. Participants shared their experiences using existing tools (e.g., YouTube Studio), the challenges they faced, and how these limitations shaped their creative process. We also explored the types of feedback and data they rely on, as well as the additional insights they desired to better align their content with audience preferences. For the data analysis, two researchers conducted a thematic analysis~\cite{thematic2006} by reviewing transcripts, identifying key themes, and consolidating them through iterative discussions to highlight creators' main challenges and practices.

\subsection{Findings} 
All interviewees (I1 - I13) unanimously agreed on the importance of understanding their audience for their work. Predominantly, they used YouTube Studio and comments to gauge audience engagement and demographics, such as tracking which types of videos received the most likes or analyzing patterns in audience retention rates. However, they found these tools provided only surface-level data, leaving them uncertain about the deeper motivations, such as why certain content resonated with their audience or what specific topics they found engaging. This limitation hindered their ability to create targeted and engaging videos based on audience insights.

\subsubsection{Difficulty in Gaining In-depth Audience Insights}
Only a few interviewees were able to go beyond surface-level analysis to gain a deeper insights into their audience. For example, I4 (M, car reviews) firmly defined his target audience demographic as `white-collar males in the U.S., aged between 40-60, nearing retirement, predominantly white'. This is mostly based on heuristic analysis of viewer comments and demographic data from YouTube Studio. However, most interviewees found that the such available tools primarily provide quantitative data—like views and demographics—that does not fully capture motivations or preferences behind audience behaviors. These tools highlight `what' the audience is doing, but fail to answer `why' they engage with specific content or disengage from others. Most interviewees desired more detailed audience knowledge to contextualize and tailor their content.

\subsubsection{Hard to Expect Meaningful Feedback from the Real Audience}
While one interviewee, I2 (nail arts), successfully gleaned insights about viewer preferences from reading comments, such as requests for more detailed techniques or specific video editing styles, accessing this level of useful feedback was not a universal experience. Most interviewees felt that comments often focus on surface-level aspects, such as emotional reactions, the video's topics, or editing quality, rather than providing deeper insights into viewers' motivations and needs. So, interviewees expressed a desire for more focused feedback on individual content pieces (I3) or the strength of their channel against others (I5), but they recognized the difficulty in obtaining such feedback directly from their audience.

\subsubsection{Translating Insights into Actionable Plans Is Challenging}
Interviewees often attempted to adjust their video elements and content based on their understanding of the audience, such as resizing transcript fonts for older viewers (I7), replicating popular video formats (I1), or following trends among younger audiences (I3, I11). While performance metrics and demographic information provided a useful snapshot of current engagement, they lacked the depth needed to guide future content strategies. For instance, I3 wanted to expand his channel's audience to women and people in their 30s but was unsure how to proceed. He wanted insights into `\emph{which content should I make for widening the spectrum of my audience?}'.

\subsection{Design Goals}
To enhance creators' understanding of their audience and support informed decision-making for creative practices, we formulate three design goals to address existing challenges.

\subsubsection{DG 1: Provide In-depth Insights of Their Audience in Easily Digestible Formats}
Comments are used by viewers to express their opinions and feelings publicly, helping creators gauge audience reactions to their channels~\cite{wu2019agent}. Even though creators want to understand their audience, it typically requires hands-on experience with data analysis to extract meaningful insights from the vast amount of comments. The system must efficiently process large-scale comment data and deliver insights relevant to individual creators in an easily digestible format (e.g., personas), minimizing the time and effort needed to understand their audience.

\subsubsection{DG 2: Facilitate Creation through Communication with Simulated Audience}
Content creators have different challenges and creative stages in producing content, and they have different levels of understanding on audiences that they can refer to in content production. However, reaching a real audience during the iterative creative process is challenging for creators seeking sufficient feedback or reactions. These interactions should be flexible and interactive, allowing creators to ask diverse questions tailored to their creative challenges and audience understanding. The system must provide responses that help creators adapt their content at hand, based on the audience's perspective.

\subsubsection{DG 3: Foster On-demand, Applicable Feedback to Creative Process}
To support creators in building actionable plans, the system must provide feedback that is timely, easy to understand, and directly applicable to content creation. Providing on-demand feedback enables creators to iterate and refine their content while maintaining their creative flow, minimizing disruptions and allowing them to make informed adjustments based on audience insights. For instance, the system could suggest adjustments to better match audience preferences or explore new content themes. This ensures that creators receive tailored, actionable suggestions to align their content with their goals efficiently.


\section{\sysname{}}

Based on these design goals, we present \sysname{} (Figure~\ref{fig:interface1} \& ~\ref{fig:interface2}), an interactive system that helps creators explore and engage with multiple audience personas generated from comments and video metadata in the early-stage of creative decision-making. 
While creators typically plan and develop content based on prior assumptions about audiences, \sysname{} enhances this process by structuring audience insights into a more actionable format. This allows creators to iteratively refine their video storylines through conversations with and feedback from audience personas.
Powered by Large Language Models (LLMs), the system generate audience personas that reflect unique characteristics and preferences, structuring them around \dimensions{} and \dimvals{} specific to the channel’s audience (DG 1). This enables creators to easily digest complex audience data, receiving targeted suggestions (DG 2) that help them make informed content decisions (DG 3). \sysname{} was implemented as a web application using React~\footnote{https://react.dev/} and includes a text editor for storyline creation built with the Lexical framework~\footnote{https://lexical.dev/}.

\subsection{What is the Audience Persona?}

The concept of audience personas in \sysname{} draws inspiration from `personas'~\cite{salminen2020persona}--- which are fictional yet data-driven representations of different user groups. 
In \sysname{}, audience personas serve as exploratory tools that highlight plausible audience traits and motivations, rather than attempting to replicate real-world viewers. This distinction ensures that personas guide creators in sense-making and early-stage ideation without attempting to predict individual audience behaviors.
Unlike traditional persona creation methods that rely on structured and explicit observations gained from user surveys or interviews, audience personas are generated by transforming unstructured, implicit audience comments into digestible and relatable formats. Comments provide direct insights from viewers, capturing their opinions, preferences, and reactions that are often missed in conventional data analytics~\cite{sun2024influence, hanusch2019comments}. 
By analyzing and augmenting large-scale audience comments, \sysname{} provides a lens to explore potential audience characteristics and motivations, offering plausible interpretations to guide the creative exploration process.

The use of LLMs could be particularly effective for constructing audience personas, as they can extract latent characteristics and discern subtle patterns from large volumes of unstructured text data (e.g., comments) that traditional methods (e.g., sentiment analysis, topic modeling) might overlook~\cite{Shin2024dis, ma-etal-2023-insightpilot, saikia2023unveiling}. 
To construct audience personas, we developed a \dimension{}-\dimval{} framework, which systematically organizes audience data into structured categories.
Dimensions represent broad characteristic themes (e.g., hobbies, expertise levels), while \dimvals{} are specific attributes within those dimensions (e.g., 'novice' or 'expert' for the dimension of 'expertise levels'). This framework helps capture the detailed traits of different audience segments by organizing unstructured audience data into meaningful categories. For example, for a cooking channel, `Culinary Expertise' (\dimension{}) may include `Amateur Cook' or `Professional Chef' \dimvals{}), enabling creators to acknowledge and target distinct audience groups with specific needs or preferences.

Introducing such an intermediate layer enhances LLM performance by providing structured representations~\cite{wei2022chain, lee2024aligning}, reducing ambiguity and inconsistency in persona generation. This approach aligns with sensemaking practices, where categorizing information into interpretable units facilitates the derivation of targeted and actionable insights.
Employing a \dimension{}-\dimval{} framework facilitates more consistent persona generation by mitigating the ambiguity and inconsistency that might arise from directly transforming comments into personas. It also provides creators with explanations of how audience traits are derived from data, thereby improving transparency of process and interpretability of results.

With audience personas, creators can engage with them to make informed content decisions. By simulating audience personas, the system allows creators to tailor their content strategy to the specific needs and preferences of their audience. Through interactions with LLM-generated personas, creators can explore audience preferences with natural language and gather potential audience perspectives and insights on their work-in-progress content.



\subsection{Interface}
\begin{figure*}[h]
\centering
\includegraphics[width=\textwidth]{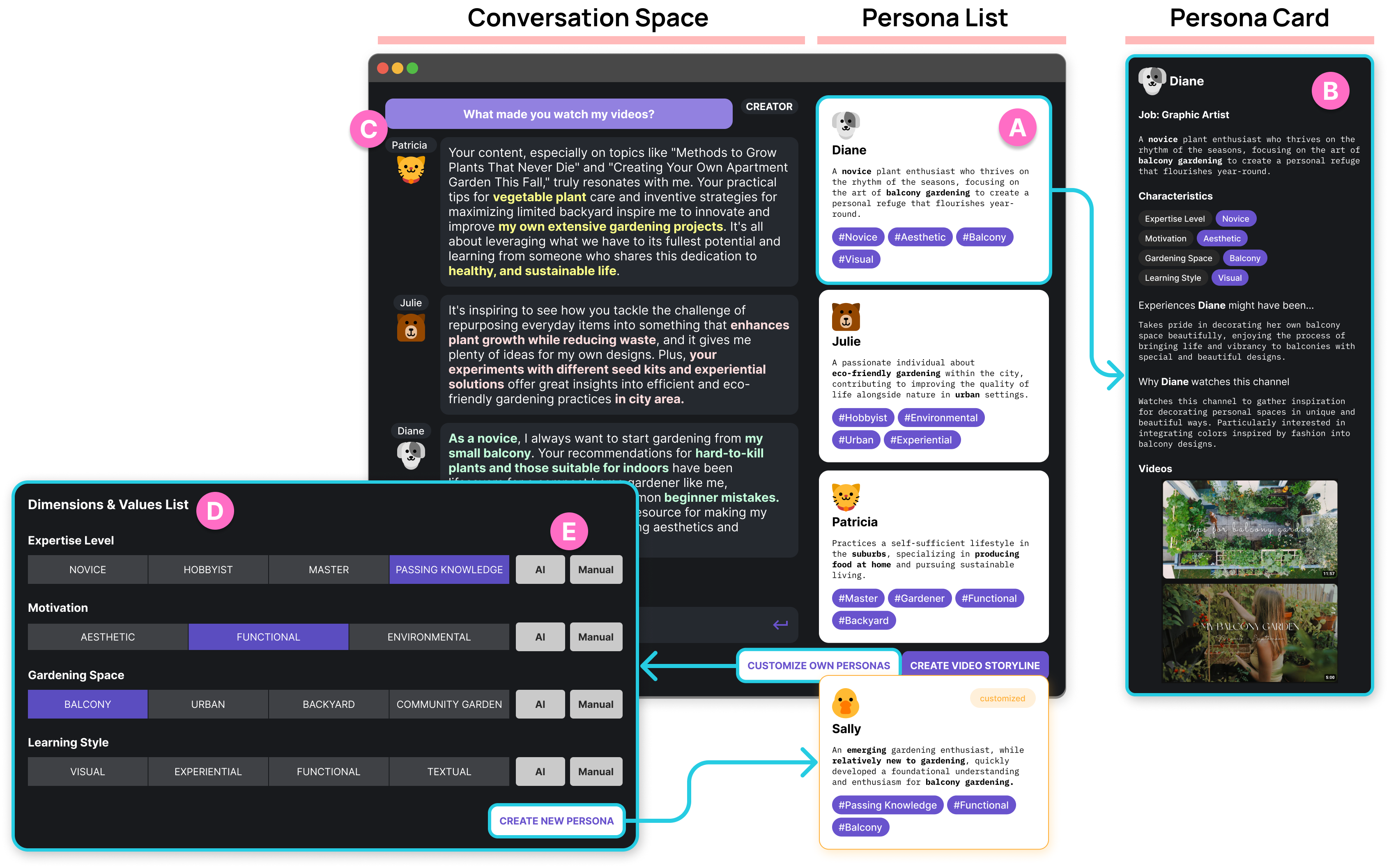}
\caption{Exploration page where creators can explore and interact with audience personas. (A) Persona list showing multiple audience personas. (B) Persona cards with details of persona profile. (C) Conversation space to freely chat with personas. Creators can ask questions to personas and request their opinions. (D) Dimensions and values list showing diverse audience attributes. (E) Customization options for creating new personas by selecting or generating different dimension-value combinations.}
\label{fig:interface1}
\Description{Exploration page where creators can explore and interact with audience personas. (A) Persona list displaying the generated audience personas with a summary of key dimensions (e.g., motivation and expertise). Creators can click on each persona to gain more detailed insights. (B) Upon selecting a persona, a detailed persona card opens, providing a fixed set of information about the persona's job, motivations, experiences, and relevant video preferences. (C) Conversation space where creators can engage in real-time dialogue with personas by asking open-ended questions, allowing the personas to respond based on audience data. (D) Dimensions and values list presenting the various customizable audience attributes (e.g., expertise level, motivation). (E) Customization options for creating new personas by selecting different dimensions and values manually or using AI-generated suggestions, enabling creators to explore new audience personas beyond the pre-generated ones.}
\end{figure*}

\begin{figure*}[h]
\centering
\includegraphics[width=\textwidth]{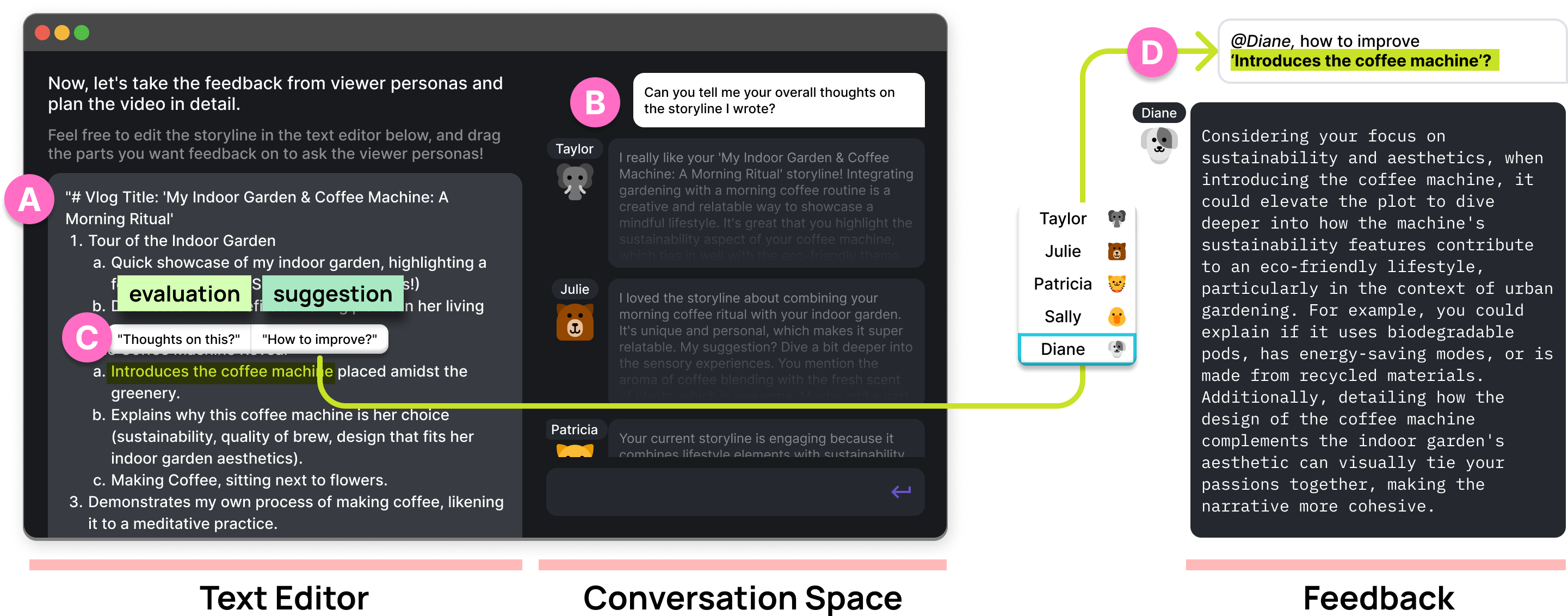}
\caption{Creation page where creators write a video storyline, proceed with conversations about their plot and request feedback on their written content. (A) Text editor where creators can draft their video storylines. (B) Conversation space, where creators ask personas for thoughts on the storyline. (C) Feedback feature allows creators to get [evaluation] or [suggestions] on particular sections of the text from a particular persona. (D) Example feedback provided by a chosen persona (Diane).}
\label{fig:interface2}
\Description{Creation page where creators write a video storyline, proceed with conversations about their plot, and request feedback on their written content. (A) Text editor where creators can draft their video storylines. (B) Conversation space, where creators ask personas for a holistic review of the storyline. (C) Feedback feature allows creators to get [evaluation] or [suggestions] on particular sections of the text from a particular persona. This inline feature provides more granular control for creators to receive targeted feedback on specific parts of the storyline. (D) Example feedback provided by a chosen persona (Diane). Feedback is provided in a conversation format, along with the directly revised text in the text editor.}
\end{figure*}

The \sysname{} interface consists of two parts: \textbf{1) Exploration Page} for exploring the audience personas by interacting with them (Figure~\ref{fig:interface1}), and \textbf{2) Creation Page} for drafting a storyline for the video and refining it with persona feedback (Figure~\ref{fig:interface2}). To demonstrate how \sysname{} works, we present a user scenario featuring Monica, a YouTube creator who has run a gardening channel [\textit{Welcome to Monica's Garden}] for over two years. Despite her experience, she remains uncertain about how to attract and engage her audience more effectively.

\subsubsection{Configuring Dimension-Value of Own Audience}
The creator first provides data about their channel, such as audience comments and video metadata. These data sources allow the system to generate a channel-specific \dimension{}-\dimval{} set that reflects the unique characteristics and preferences of the audience as shown in Figure~\ref{fig:interface1} - D. The personalized overview of \dimension{}-\dimval{} sets provides a clear, visual representation of the audience's unique characteristics, making it easier for creators to grasp key insights at a glance.

\hfill
\begin{block}
    Four dimensions specify Monica's channel audience: 
    \begin{itemize}
        \item \texttt{\dimcol{Expertise Level}} 
            \begin{itemize}
                \item \valcol{Novice} Audience new to gardening, seeking basic tips and easy-to-grow plants.
                \item ...
            \end{itemize}
        \item \texttt{\dimcol{Motivation}}
            \begin{itemize}
            \item \valcol{Aesthetic} Audience attracted by the visual transformation gardening provides.
            \item \valcol{Functional} Audience values the practical benefits, like homegrown food or herbal remedies.
            \item \valcol{Environmental} Audience motivated by the positive impact on the ecosystem, such as attracting pollinators or improving soil health.
            \end{itemize}
        \item \texttt{\dimcol{Gardening Space}} 
            \begin{itemize}
                \item \valcol{Balcony} Audience has some outdoor space for container gardening or small planter boxes.
                \item ...
            \end{itemize}
        \item \texttt{\dimcol{Learning Style}} 
            \begin{itemize}
                \item \valcol{Visual} Audience prefers content with many images, diagrams, and video walkthroughs.
                \item ...
            \end{itemize}
    \end{itemize}
\end{block}

\subsubsection{Exploring Audience Personas.}
After reviewing \dimension{}-\dimval{} set, the creator moves on to the \emph{Exploration Page} (Figure ~\ref{fig:interface1}), which provides multiple {Persona Cards} generated by \sysname{}. Personas are represented with the distinct combinations of \dimensions{} and \dimvals{}. Each persona card provides a snapshot of the persona's name, one-line introduction, and top-5 relevant \dimvals{} (Figure ~\ref{fig:interface1} - A). By clicking on a persona card, creators can view detailed information, including the job, recent experiences, and motivations behind enjoying this channel. In addition, based on the creators' video list, videos frequently watched by the original comment owners are presented together, to help the creator understand which video topics are preferred by the audience (Figure ~\ref{fig:interface1} - B). 

\hfill
\begin{block}
    Monica discovers three personas --- \textsc{Diane}, \textit{the balcony beautifier} (\valcol{Novice}, \valcol{Aesthetic}, \valcol{Balcony}, \valcol{Visual}), \textsc{Julie}, \textit{the urban eco-gardener} (\valcol{Casual Hobbyist}, \valcol{Environmental}, \valcol{Urban},\\ \valcol{Experiential}), \textsc{Patricia}, \textit{the suburban homesteader} (\valcol{Master}, \valcol{Functional}, \valcol{Backyard}). Monica explores their distinct motivations and experiences by clicking on each persona, learning which videos each persona prefers and why. 
\end{block}

\subsubsection{Asking Questions to Personas}
Alongside the Persona Card list, \sysname{} provides the \emph{Conversation Space}, where creators can initiate a natural language chat with a set of audience personas to ask any questions they want to ask their audiences. Creators can discuss specific concerns or ideas about their content. To initiate the chat easily, example questions are provided--- `\textit{Why do you watch my videos?}', `\textit{What videos do you like on my channel?}', `\textit{What's your daily routine?}' (Figure~\ref{fig:interface1} - C). 

\hfill
\begin{block}
    Since Monica wants to know more about her audience, she starts to ask multiple questions to understand her audience's engagement (``\textit{Why did you skip my pruning tutorial video?}'') and collect audiences' preferences (``\textit{Instead, would you prefer to see my daily Vlog?}''). For the latter question, one of her audience personas, \textsc{Patricia},\emph{the suburban homesteader}, responds with ``\textit{Your Vlog video sounds very interesting! Why don't you show how you prepare meals with homegrown vegetables?}'' among others. Through this conversation, Monica discovers that her audiences are interested in more personalized, lightweight, and everyday content. 
\end{block}

\subsubsection{Customizing Persona with Dimension-Value Configurations.}
Beyond the data-driven personas generated by our pipeline, creators can customize, test, and interact with desirable personas with different \dimension{}-\dimval{} configurations. By clicking a button \emph{Discover More Persona}, creators can easily create a custom persona by choosing a combination of existing \dimvals{} (Figure~\ref{fig:interface1} - D). Using this feature, creators can envision audiences that share \dimvals{} relevant to their channel.

Understanding various audience groups is essential for creators who want to reach a broader audience or identify niche segments that could be targeted with tailored content. Inspired by Luminate~\cite{suh2023structured}, we also enable creators to extend the set of \dimvals{} in each dimension to customize new personas. Creators can extend the \dimvals{} under specific dimensions in two ways: (a) manual addition and (b)system-generated suggestions, where our pipeline intelligently suggests additional \dimvals{} and highlights those distinct from the current set when creators find it difficult to define new ones (Figure~\ref{fig:interface1} - E). This feature encourages creators to explore untapped audience segments, broadening the scope of their content creation.  After creating a custom persona, the creator can interact with it in the Conversation Space, just like other existing personas.

\hfill
\begin{block}
    To gain a broader range of understanding of her audience, Monica clicks on the ``Discover More Persona'' button and manually enters a \dimval{} under the Dimension \texttt{\dimcol{Motivation}}. She also wants to extend the \texttt{\dimcol{Expertise Level}} but is unsure about what other options could be relevant. She decides to get recommendations from the system by clicking the button; the new \dimval{} of \valcol{Passing Knowledge} is appended to the list of the \texttt{\dimcol{Expertise}} \texttt{\dimcol{Level}} dimension. 
    Based on the dimensions and \dimvals{} manually added and recommended through the interface, Monica forms a new persona by clicking `Passing Knowledge (Expertise Level),' `Functional (Motivation),' and `Balcony (Gardening Space).' Finally, she obtains a new persona \textsc{Sally}, who is \textit{a practical urban gardener}. 
\end{block}

\subsubsection{Creating and Revising Content with Audience Personas.}
Finally, \sysname{} supports audience-centered content creation for creators ---planning and drafting a video storyline with the perspectives of personas. Entering \emph{Creation Page}, creators can quickly draft a video storyline for a given topic on their own, similar to their common practices (Figure ~\ref{fig:interface2} - A). Once the creator has finished writing a rough storyline, the system initiates a conversation between the existing personas to provide overall feedback on the creator's draft (Figure ~\ref{fig:interface2} - B). 

Throughout the process, creators can directly ask personas for feedback on specific sections of their draft (Figure ~\ref{fig:interface2} - C). When the creator selects a specific part of their draft, a floating menu appears with two options: (a) What are your thoughts on this part?' and (b) How can I revise or improve this section?'. This targeted feedback allows creators to refine specific sections of their content, ensuring that each part aligns with audience expectations (Figure ~\ref{fig:interface2} - D).

\hfill
\begin{block}
    One day, Monica gets a request from an advertising agency to make a promotion video with \emph{``Nespresso Virtuo Pop coffee machine''}. She starts to write a short storyline considering her audience whose main interests are gardening but varied in their \texttt{\dimcol{Motivation}} and \texttt{\dimcol{Learning Style}}.  As she reflects on her personas, Monica realizes that her audience persona \textsc{Diane}, who is highly interested in aesthetics, would likely appreciate the visual appeal of the coffee machine. So, Monica decides to include more detailed explanations about the aesthetic features of the coffee machine and searches for stylish promotional clips to incorporate into her video.

    After completing a rough draft for the storyline, her audience personas start providing holistic feedback on her draft without explicit request. While editing and improving the video storyline, she becomes uncertain about focusing too heavily on the technical specifications of the coffee machine. Monica assumed her audience would have low technical interests, but she had no confidence. She requests a suggestion on the part of the current storyline about technical specification from \textsc{Julie}. \textsc{Julie} gave a suggestion --- ``Even though I drink a lot of coffee during my work day, viewers like me might not be interested in the technical abilities of the coffee machine. How about adding your own experiences using it at home?'' Monica finds \textsc{Julie}'s suggestion both reasonable and engaging, so she adds more personal anecdotes about her daily life usages, making the content more relatable and appealing to her audience.
\end{block}


\subsection{Technical Pipeline}

\begin{figure*}[h]
\centering
\includegraphics[width=\textwidth]{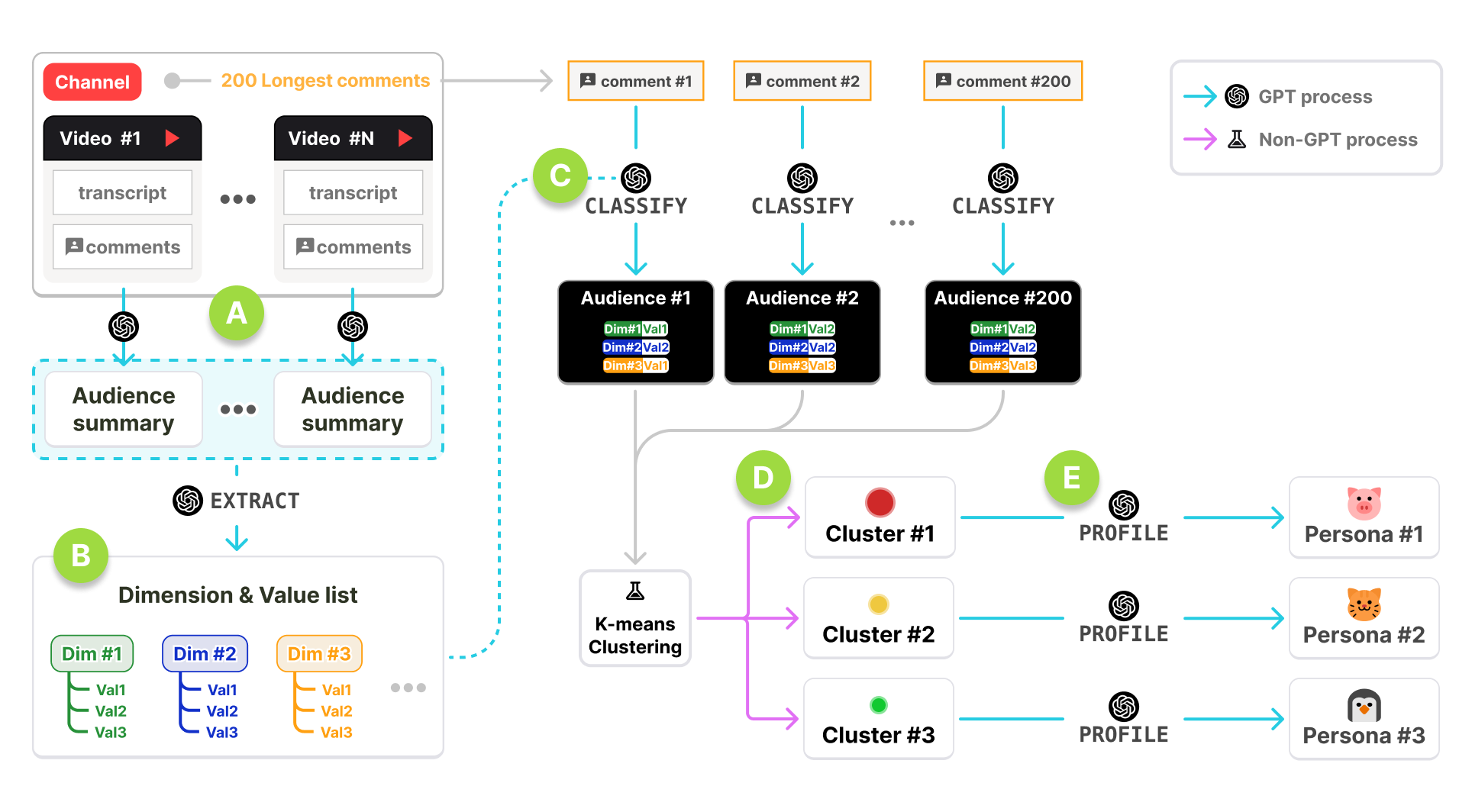}
\caption{Our pipeline generates audience personas with GPT-4 and k-means clustering. Our pipeline first builds audience summaries and transcript summaries and constructs a dimension \& values list with GPT-4. With comments, the pipeline predicts the audience characteristics of each comment based on the dimension \& values list. Using pre-trained BERT embeddings and k-means clustering, the 200 comments are clustered in k groups of predicted audiences. In the end, our pipeline generates a persona profile that consists of a job, a short persona description, and recent experiences of generated personas with GPT-4.}
\label{fig:pipeline1}
\Description{Our proposed pipeline for generating audience personas. Our pipeline first builds audience summaries and constructs dimension \& values list with GPT-4. With the 200 longest comments, the pipeline predicts the audience characteristics of each comment based on the dimension \& values list. Using BERT embedding and k-means clustering, the 200 comments are clustered in k-groups of the predicted audience. In the end, our pipeline generates a persona profile (job, short description, recent experiences) for each cluster of audiences with GPT-4.}
\end{figure*}

\begin{figure}
\centering
\includegraphics[width=0.5\textwidth]{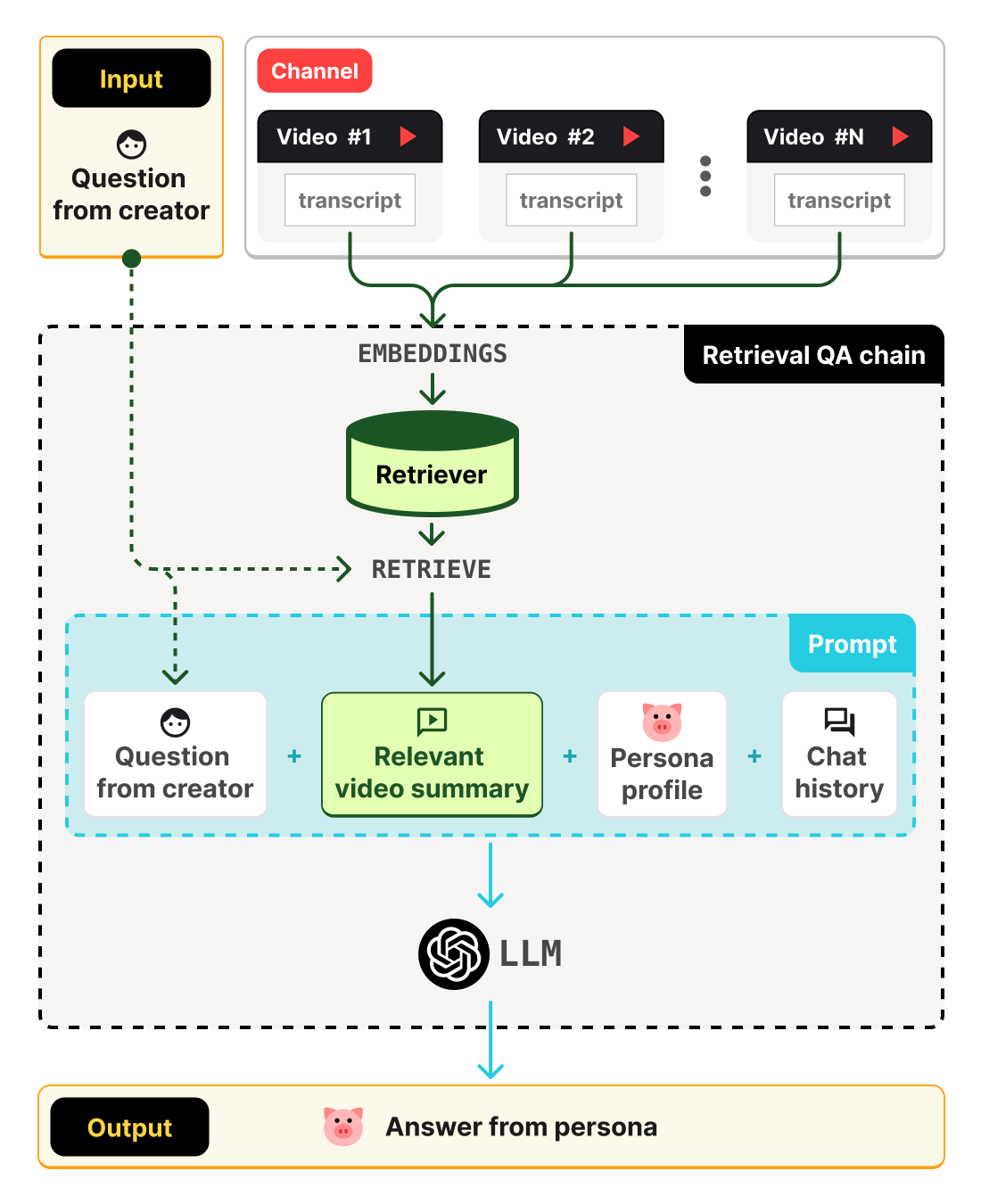}
\caption{For interacting with personas, our pipeline retrieves relevant video transcripts and context data using LangChain and RetrievalQA, enabling the generation of context-aware responses. 
It also provides evaluative critiques or actionable suggestions, ensuring that the feedback is relevant and grounded in real insights.}
\label{fig:pipeline2}
\Description{For interaction with personas, our pipeline retrieves relevant video transcripts and context data using LangChain and RetrievalQA, enabling the generation of context-aware responses. Creators can engage with these personas by either asking questions or highlighting specific content for feedback. The system then provides evaluations or actionable suggestions based on the audience data, ensuring that the responses are relevant and grounded in real insights.}
\end{figure}

In this section, we present our technical pipeline, detailing the process for persona generation (Section~\ref{sec:pipeline1-a}, ~\ref{sec:pipeline1-b}, ~\ref{sec:pipeline1-cde}; Figure~\ref{fig:pipeline1}) and persona interaction, including message and feedback simulation (Section~\ref{sec:context_retention}; Figure~\ref{fig:pipeline2}). The technical pipeline was developed using OpenAI's \texttt{gpt-4-1106-preview} model. The prompts used for the pipeline are described in the Supplementary Materials.

\subsubsection{Data Collection}
\label{sec:pipeline1-a}

Our pipeline first crawls data from users' YouTube channels. Using the channel-specific handle ID, the pipeline collects public metadata of the channel (e.g., channel name, description, categories, number of subscribers, total view counts), all of the videos and their information (e.g., video ID, title, description, related comments), and comments for each video (e.g., content, writer ID, date created). For large-scale YouTube data collection, we used the yt-dlp library~\footnote{https://github.com/yt-dlp/} and stored it in a Django SQLite~\footnote{https://docs.djangoproject.com/en/5.1/ref/databases/\#sqlite-notes}.

\subsubsection{Inferring Audience Characteristics from Comments} 
\label{sec:pipeline1-b}
Due to the high volume of comments that exceed the context window limit of GPT-4 (currently capped at 8,192 tokens), our pipeline applies a summarization technique with GPT-4 to compress extensive data and extract key insights necessary for inferring audience characteristics from comments with our pipeline.

Our pipeline first generates a \textit{transcript summary} of each video to obtain contextual information in a compressed version. Then, from the title, \textit{transcript summary}, and the raw comments from the video, the pipeline generates an \textit{audience observation summary} for each video using a summarization technique similar to those used for a transcript summarization (Figure~\ref{fig:pipeline1} - A). 
The purpose of the audience observation summary is to infer the characteristics and preferences of the video's audiences. 

Our pipeline then extracts key \dimensions{} and \dimvals{} from the audience observation summaries to represent the possible audience characteristics for each creator’s channel (Figure~\ref{fig:pipeline1} - B). Inspired by the principles of constrained generation and chain-of-thought prompting~\cite{yang2022doc, yang2022re3, yin2017syntactic, gandhi2023strategic}, the pipeline decomposes summaries into \dimensions{} and \dimvals{} as intermediate representations. This structured approach aims that the final personas are both based on the audience data and aligned with the creator's channel characteristics.
Within prompts, we asked to infer audience characteristics by providing the definition of the \dimensions{} and \dimvals{} as below, with few-shot examples: 
\begin{itemize}
    \item \dimensions{} : Categories that describe the audience's characteristics. 
    \item \dimvals{} : Distinct attributes that predict the audience's specific traits or preferences for each dimension.
\end{itemize}

To implement this, the pipeline prompts GPT-4 to generate \dimensions{} and \dimvals{} that are relevant to the channel audience, and mutually exclusive and unique to each other. By leveraging a structured intermediate step of dimension-value pairs (\dimensions{} and \dimvals{}), the system ensures coherence and relevance in persona generation.

\subsubsection{Generating Audience Personas}
\label{sec:pipeline1-cde}

To create prototypical personas that capture diverse but distinct audience traits, our pipeline clusters comments based on similar combinations of inferred characteristics before creating detailed personas. The clustering process organizes unstructured audience data into distinct groups, enabling the creation of personas that highlight traits and preferences within each cluster.
To achieve this, our pipeline chooses the longest 200 comments across all the videos, as longer comments tend to contain richer, more detailed insights about the audience~\cite{zhang2021comments}. To ensure a balanced representation of comments from different videos, we also include the three longest comments from each video not covered by the initial sampling (Figure~\ref{fig:pipeline1} - A). Once the comments are selected, our pipeline utilizes GPT-4 to infer the audience's characteristics for each comment. 
As each comment is treated as a proxy for an individual audience member's perspective, we represent each comment as a combination of \dimvals{} across several audience \dimensions{}. If specific \dimensions{} are challenging to infer from the comment, GPT-4 classifies them as 'None' to avoid forcing inferences from insufficient data (Figure~\ref{fig:pipeline1} - B, C).

To meaningfully organize the inferred traits of diverse comment-driven audience groups into personas, we chose to use clustering as a supporting mechanism within our pipeline. While the primary contribution of \sysname{} lies in its \dimension{}-\dimval{} framework, the clustering step enhances usability by grouping audience traits into actionable and comprehensible forms that support creating detailed personas.
Our pipeline focuses on considering `audience similarity' when clustering, rather than solely using traditional criteria like `semantic similarity' or `syntactic similarity'. Audience similarity refers to the similarity in the traits of the audience that we infer from the comments. The rationale for this approach is that relying only on semantic similarity often leads to clustering based on superficial language patterns (e.g., tone, keyword), which may not accurately reflect the implicit characteristics of the audience. Thus, our pipeline concatenates each comment after \dimensions{} and \dimval{} to ensure that both the semantic similarity and the inferred audience traits are considered together. 

Our pipeline embeds this combined input using a Transformer-based text embedding model (\texttt{all-MiniLM-L6-v2}~\footnote{https://huggingface.co/sentence-transformers/all-MiniLM-L6-v2}) and applies k-means clustering to organize audience traits into distinct clusters (Figure~\ref{fig:pipeline1} - D). The k-means clustering algorithm was chosen to complement the \dimension{}-\dimval{}  framework by grouping audience comments into actionable personas based on inferred traits. While the LLM performs clustering based on varying reasoning and produces different results, k-means produces stable and consistent cluster results. We decided on the optimal k by calculating an inertia value.

Finally, our pipeline generates detailed audience personas for each cluster using GPT-4. The pipeline creates persona profiles that include information such as the persona's name, job, a short biography, reasons for watching the videos and channel, and personal experiences (Figure~\ref{fig:pipeline1} - E). To create these profiles based on actual audience data, we provide prompts with \dimension{} and \dimval{} information, actual comments from the cluster, and definitions of \dimension{} and \dimval{} sets. Additionally, we incorporate \dimension{} and \dimval{} information from other personas to highlight unique aspects or specific \dimvals{}.

\subsubsection{Retaining Context for Persona Conversations and Contextual Feedback} 
\label{sec:context_retention}

Our pipeline provides natural language chat with the personas, allowing creators to engage with them in real-time and receive contextual feedback on their work-in-progress content (Figure~\ref{fig:interface1} - C \& Figure~\ref{fig:interface2} - B). 
To ensure the generated content retains relevant and coherent storylines in long-form formats, previous research suggested prompting techniques~\cite{yang2022doc, yang2022re3}, such as constructing detailed outlines and aligning story drafts with the outline. 
In our study, as content creators often plan the content based on previous videos that contain relevant topics or audience reactions, we implement context retention by embedding transcript summaries using OpenAI Embedding~\footnote{https://python.langchain.com/v0.2/docs/integrations/text\_embedding/openai/}, to ensure the conversations remain coherent and relevant to the actual videos. These embeddings are stored in an FAISS~\footnote{https://python.langchain.com/v0.2/docs/integrations/vectorstores/faiss/} database, allowing the system to quickly retrieve relevant content. 
We then employed a chain-of-thought method~\cite{gandhi2023strategic}, utilizing RetrievalQA chain from LangChain~\footnote{https://python.langchain.com/v0.2/docs/versions/migrating\_chains/retrieval\_qa/} (Figure~\ref{fig:pipeline2}) in generating responses to the creator’s questions. 
Based on the retrieved data relevant to the questions, answers are generated via reasoning with the persona profiles and chat history.
By this approach, we aimed to reduce the likelihood of hallucinations such as generating content that does not exist in actual user data—a method increasingly recognized for its reliability~\cite{ayala2024reducing}. 
 

\section{Technical Evaluation}

To assess the effectiveness of our LLM-powered pipeline, we conducted a comprehensive technical evaluation focusing on four key aspects: (1) the quality of generated dimensions \& values (Section~\ref{sec:techeval_1}), (2) the validity of audience comment clustering methods (Section~\ref{sec:techeval_2}), (3) the quality of generated personas with our pipeline (Section~\ref{sec:techeval_3}), and  (4) the presence of hallucinations in persona responses (Section~\ref{sec:techeval_4}). We define high-quality personas as those that are highly relevant, reflect real audiences, and are distinct from each other. 
Through these evaluations, we aim to address our core research question:
\begin{itemize}
    \item \textbf{RQ 1: Can \sysname{} effectively generate relevant, distinct, and audience-reflecting personas that provide evidence-based responses?}
\end{itemize}

\subsection{Evaluating Dimension-Value Generation Pipeline}
\label{sec:techeval_1}

We evaluated the relevance and mutual exclusiveness of the dimensions and values generated by our pipeline with external raters. We did not perform a comparative evaluation against a baseline, as existing tools like YouTube Studio primarily focus on quantitative metrics (e.g., views, demographics) and do not support persona generation or a dimension-value framework or analyze audience comments through a structured dimension-value framework. As a result, there is no direct comparison for the qualitative insights our system offers. We measured relevance to determine if the pipeline accurately captured audience traits unique to the channel, which is crucial for creating \emph{channel-relevant} personas. Mutual exclusiveness was evaluated to ensure that the dimensions and values were sufficiently distinct, supporting the construction of \emph{diverse} audience personas.

As evaluators, we recruited three YouTube creators with experience in understanding their audience by reviewing comments. To reduce bias, evaluators analyzed other creators' channels for a more neutral perspective, avoiding preconceptions they might have about their own audience. 

We sampled six channels (Channel A-F) across diverse channel topics for generalizability. On average, our pipeline produced five dimensions ($min$ = $4$, $max$ = $6$) and 17.5 values ($min$ = $15$, $max$ = $24$) for each channel (Appendix~\ref{app:techeval1}-Table~\ref{tab:techeval1}).  
Evaluators watched at least five videos per channel and reviewed their comment sections to develop a basic understanding of the audience. They were then asked to rate:
\begin{itemize}
    \item Relevance: How well does each dimension/value relate to the channel's viewers? (Rated on a 5-point Likert scale, 1 = `Highly Irrelevant', 5 = `Highly Relevant')
    \item Mutual exclusiveness: Are the dimensions distinct, or do they overlap with other dimensions? Do the values overlap or are they too similar across different dimensions? If so, which ones? (Binary scale with rationale provided)
\end{itemize}
Ratings were aggregated using average scores for relevance and by counting the number of overlapping pairs noted by the majority of evaluators for mutual exclusiveness.

\subsubsection{Result: Quality of Generated Dimension \& Values}

Results indicated that our pipeline produced dimensions and values generally relevant to the channel audience while being distinct from each other. The average relevance score was $3.68$/$5$ (SD = $0.35$, min = $3.07$, max = $4$; Appendix~\ref{app:techeval1} - Figure~\ref{fig:techeval1_dimension}) for dimensions and $3.6$/$5$ (SD = $0.23$, min = $3.30$, max = $3.82$; Appendix~\ref{app:techeval1} - Figure~\ref{fig:techeval1_value}) for values, indicating that our pipeline effectively generated artifacts describing channel audiences.

Our evaluation of mutual exclusiveness showed that there were low overlaps in dimension and values: $6.67$\% for all dimensions and $7.62$\% for all values were marked as overlaps by at least two evaluators, indicating a high level of distinctiveness. Only one pair of dimensions was investigated as overlapped in Channel A, and overlapping values were noted in Channels A (Flavor innovator \& Experimental Gastronome, Culinary adventurer \& Recipe seeker), E (Show seeker \& Festival Fanatics), and F (Creative explorations vs. DIY home projects).  While these overlaps suggest some redundancy, they often reflected real-world similarities in audience behaviors. For instance, in Channel E, the overlap between \textit{Show Seeker} and \textit{Festival Fanatic} was due to similar interests expressed by viewers in attending events, rather than an error in the pipeline. Detailed results are reported in Appendix~\ref{app:techeval1} - Table~\ref{tab:techeval1}. 



\subsubsection{Further Validation: User Study Ratings}

We also asked user study participants (N=$11$) to rate the overall quality of dimensions and values generated for their own channels with a 5-point Likert scale (Appendix~\ref{app:question_dimension_values}). They rated them as highly relevant to their audience ($M_{dimension}$ = $4.55$, $M_{value}$ = $4.45$) and helpful for understanding their audience ($M_{dimension}$ = $4.36$, $M_{value}$ = $4.18$), with new perspectives provided ($M_{dimension}$ = $3.91$, $M_{value}$ = $4.36$). Notably, user study results showed a moderate score of mutual exclusiveness, $3.36$ ($M_{dimension}$, SD = $1.21$) and $3.64$ ($M_{value}$, SD = $1.12$). Channel A’s owner, P5, rated mutual exclusiveness as low (1/5), explaining that some dimensions and values, while semantically distinct, were contextually intertwined. For example, \textit{Culinary Curiosity} and \textit{Local Culinary Scene} described similar audience behaviors that overlapped in practice, even though they represented different interests on the surface. This suggests that some content domains, particularly niche topics, may naturally produce overlapping audience traits.

\subsection{Evaluating Comment Clustering Pipeline}
\label{sec:techeval_2}

To address RQ1, we evaluated whether our clustering pipeline effectively captured `audience resemblance', a key factor in generating \emph{audience-reflecting} personas. We define `audience similarity' as the extent to which comments within a cluster reflect similar audience characteristics. We compared our clustering method (\sysname{} (right) in Appendix~\ref{app:techeval2} - Figure~\ref{fig:techeval2-procedure}) to a baseline approach that applied k-means clustering solely based on the semantic similarity of comments, without incorporating the \dimensions{}-\dimvals{} (dim-val) information that represents audience characteristics (\baseline{} (left) in Appendix~\ref{app:techeval2} - Figure~\ref{fig:techeval2-procedure}).
Both methods used the same SentenceTransformer model (\texttt{all-MiniLM-L6-v2}) for embedding, followed by k-means clustering. The key difference was that \sysname{} included audience-related information (dim-val sets), while the baseline focused only on comment semantics. Since evaluating clustering quality is challenging due to the absence of ground truth data, and because assessing audience similarity between comments can be unfamiliar for evaluators, we also included `linguistic similarity' as a comparative measure to help distinguish it from audience similarity. 
Linguistic similarity refers to how closely the language and wording of the comments resemble each other. By contrasting linguistic similarity with audience similarity, we aimed to highlight the distinctiveness of our pipeline’s focus on capturing implicit audience traits rather than just surface-level comment semantics.

We tested our pipeline across five different channels for various channel topics (Appendix~\ref{app:techeval2} - Table~\ref{tab:techeval2-result}). For each channel, we selected 10 reference comments and retrieved the four closest comments to the cluster centroid from both methods. 
As evaluators, we recruited three YouTube creators who have experience understanding their audience by reviewing comments. For each reference comment, evaluators were presented with two comment groups, one from each clustering method. They were asked to assess:
\begin{itemize}
    \item Linguistic similarity: Which group shows greater similarity in language and wording?  
    \item Audience similarity: Which group reflects more closely aligned audience traits or behaviors?
\end{itemize}
Evaluators were not allowed to tie groups, and ratings were aggregated using majority voting. The evaluation setup and procedure are illustrated in Appendix~\ref{app:techeval2} - Figure~\ref{fig:techeval2-procedure}. 

\subsubsection{Result: Quality of Clustering Methods}
Our evaluation showed that the clusters generated with our pipeline were generally perceived as more homogeneous in terms of audience similarity compared to the baseline method. On average, evaluators marked the \sysname{} clusters as more aligned with audience characteristics than baseline (M = $6.4$/$10$). All results are shown in Appendix~\ref{app:techeval2} - Table~\ref{tab:techeval2-result}. 

For Channel D, only 5 out of 10 clusters created by \sysname{} were perceived as having greater audience similarity than those generated by the baseline method. This result may be influenced by the fact that, in some cases, the baseline method clustered comments with similar keywords or content, unintentionally reflecting audience traits. For example, one baseline cluster for Channel D grouped comments that shared the value ``Brand Analyst,'' despite the method focusing on semantics rather than audience characteristics. 
This finding suggests that, while semantic clustering can sometimes result in seemingly homogeneous groups, \sysname{} consistently provided clusters that better reflected implicit audience traits. Example clusters of comments of Channel D is shown in Appendix~\ref{app:techeval2} - Table~\ref{tab:techeval2_ex}.


\subsection{Evaluating Generated Persona}
\label{sec:techeval_3}

To compare the quality of personas generated by \sysname{} with those produced by a pure LLM approach (\baseline{}), we conducted a comparative evaluation. Both methods used the same number of comments as input. However, unlike \sysname{}, which employed the dimension-value framework and clustering techniques, the \baseline{} method directly generated personas from comments, as in the LLM-Auto condition~\cite{Shin2024dis}. Both approaches generated the same number of personas (N) based on optimal clustering determined by \sysname{}. To ensure a fair comparison, the format of persona profiles was controlled across both methods, containing name, job, introduction, characteristics, past experiences, and reasons for engaging with the channel. Both methods used OpenAI's \texttt{gpt-4-1106-preview} model for persona generation.

We randomly selected 7 YouTube channels previously used in other technical evaluations and recruited 6 YouTube creators with prior experience understanding their audience through comments as evaluators. Evaluators were randomly assigned to complete evaluations on 5 or 6 channels. The assessment was conducted through two types of surveys: (1) a comparative survey where evaluators chose which persona set---\sysname{} or \baseline{}--- better fulfilled certain criteria and (2) a standalone survey where evaluators rated each persona set on a 5-point Likert scale (1 = `Strongly Disagree', 5 = `Strongly Agree'). To provide enough context for third-party assessments, evaluators reviewed the 200 comments used for persona generation for at least 10 minutes before their evaluation. 

Personas were evaluated on completeness, realism, clarity, and empathy (adapted from Persona Perception Scale~\cite{salminen2020persona}); We also added criteria specific to our goals, including diversity, novelty, serendipity, multi-dimensionality, and generalizability. These criteria aimed to assess the overall utility, informativeness, and strength of the personas compared to those generated by a baseline LLM approach.
For the comparative survey, we used majority voting to determine which method was preferred on each channel for each criterion. For the standalone survey, we conducted Wilcoxon signed-rank tests to compare the ratings of each question across both methods. A summary of the results is presented in Appendix~\ref{app:techeval3} - Table~\ref{tab:techeval3}. 

\subsubsection{Result: Quality of Personas}

The comparative survey results indicate that \sysname{} consistently outperformed the \baseline{} method across most evaluation criteria except for realism (Q2) and comprehensiveness (Q9) (Appendix~\ref{app:techeval3} - Table~\ref{tab:techeval3}). Evaluators rated \sysname{} higher in the ability to provide sufficiently complete profiles (Q1), with \sysname{} preferred in 7 out of 7 channels compared to 0 for \baseline{}. In terms of organizing persona information (Q3), \sysname{} was seen as more structured and clearer, allowing evaluators to better understand the personas (Q4). \sysname{} also did better in promoting diversity (Q5) and serendipity (Q6), and marked lower scores in generalizability (Q10) and predictability (Q7). Evaluators felt that \sysname{} personas felt more multi-dimensional (Q8), even though the ability to easily compare personas (Q9) was rated slightly lower. Still, in terms of making the personas feel like real individuals (Q2), the \baseline{} method outperformed \sysname{}, with evaluators feeling that \baseline{} personas seemed more like real viewers. 

To confirm these findings, the standalone survey revealed that \sysname{} significantly outperformed \baseline{} in several key areas (Appendix~\ref{app:techeval3} -Table~\ref{tab:techeval3-descriptive}). Wilcoxon signed-rank tests showed that \sysname{} was rated higher in terms of persona completeness (Q1, \emph{p} < .01), the diversity of personas (Q5, \emph{p} < .01), and the multi-dimensionality of personas (Q8, \emph{p} < .01). Additionally, evaluators were more satisfied with the novel insights provided by \sysname{} personas (Q6, \emph{p} < .01), which corresponds to the result that \baseline{} is more significantly predictable than \sysname{} (Q7, \emph{p} < .01). 

These standalone results not only confirm the findings from the comparative survey but also highlight how \sysname{} provides more detailed, diverse, and novel personas that support better decision-making and audience understanding. While \baseline{} personas were perceived as more ``real'' in some aspects, \sysname{} offered the depth and structure necessary for creators to make informed, strategic decisions about their content. 

\subsection{Evaluating Hallucinations in Persona Chat Responses}
\label{sec:techeval_4}

LLM-based chat generation is prone to hallucinations, which can hinder the process of understanding the audience, decreasing the creator's trust in our system. To measure the performance of the chat generation pipeline (Figure~\ref{fig:pipeline2}), we evaluated the \emph{groundedness} of responses generated by our personas. In this context, hallucinations refer to responses that inaccurately mention resources or content, particularly when referencing specific video or channel content. To mitigate the hallucination, we used RetrievalQA which has been effective in reducing hallucination by referring to existing data~\cite{ayala2024reducing}.

To assess the groundedness of the responses, we conducted a post-hoc evaluation involving two external evaluators who reviewed chat responses for both direct and indirect references to video or channel content. The goal was to determine whether the mentioned resources in the responses could be accurately found in the referenced videos or channels. This evaluation included responses from five randomly selected channels used in our user study, 203 responses in total. A response was classified as a hallucination if (1) the referred video title was incorrect, (2) the referred video content did not match the actual YouTube video content, or (3) the relevant video could not be found despite being indirectly mentioned. Evaluators are asked to review responses and find the original source as evidence. We computed inter-rater reliability (IRR; Cohen's Kappa), and aggregated the labels by counting a response as a hallucination if at least one evaluator considered it to be so.

\subsubsection{Result: Hallucinations of Persona Responses}

Results showed that $4.93$\% ($10$ out of $203$) of the responses were classified as hallucinations by at least one evaluator (IRR = $0.804$). This indicates a relatively low occurrence of hallucinations in the generated responses, highlighting the effectiveness of our methods in minimizing inaccuracies in the personas' chat outputs.



\section{User Evaluation}
\label{sec:user_study}
We conducted a user study with 11 YouTube creators to evaluate how \sysname{} facilitates creators in exploring audience traits and integrates persona-informed insights into the content ideation process. Participants first used their existing methods to understand their audience and plan new content, then repeated the task with \sysname{}. The task for both conditions was identical: ``Begin by exploring your audience using [your existing methods / \sysname{}], and then develop a video storyline for the provided topic, tailored to your audience.'' This evaluation aimed to answer two research questions:
\begin{itemize}
    \item \textbf{RQ2: How effectively can \sysname{} support creators in exploring audience traits for sensemaking and ideation?}
    \item \textbf{RQ3: With \sysname{}, how do creators integrate persona-informed insights into their creative practices?}
\end{itemize}

The study aimed to evaluate how \sysname{} enhances exploratory audience understanding and to identify its potential impact on creators' workflows.

\subsection{Recruitment}
We recruited YouTube creators who: (1) have maintained an active channel~\footnote{We defined an active channel with at least one video upload per month within the last year.} for at least a year, (2) run an informational video channel with a specific topic, (3) use subtitles or audio narration, and (4) have received more than 400 comments on their videos. The third and fourth criteria were to utilize our technical pipeline to secure enough quantity of data for clustering comments. We recruited participants through social media, university boards, and cold emails to creators, following a pre-survey to select eligible participants satisfying all four criteria. Participants received KRW 150,000 (approx. USD 112) for up to two hours of participation. All user studies took place in Korean, including the content provided in the studies.

\subsection{Participants}
We recruited 11 YouTube creators (5 females, 6 males). All participants were running their channels in Korean, were aged between their 20s and 40s. Their content covered various informational domains such as electronics and studying abroad. Four participants were full-time creators, and seven were part-time. Channel activity varied from 1 year to over 6 years (See Table~\ref{tab:demo_study}).

\begin{table*}[h]
\centering
\scalebox{0.8}{
\begin{tabular}{cccccc}
\toprule
\textbf{\makecell[c]{Participant ID \\ (Gender, Age)}} & \textbf{\makecell[c]{Channel \\ category}} & \textbf{\makecell[c]{Start \\ date}} & \textbf{\makecell[c]{Total number of\\subscribers}} & \textbf{\makecell[c]{Total number of\\comments}} & \textbf{\makecell[c]{Level of\\commitment}} \\
\midrule
P1 (F, 20s) & Studying abroad & 2022. 03 & 8.2k & 586 & Part-time \\
\midrule
P2 (M, 30s) & Music production tutorial & 2021. 10 & 12.7k & 1371 & Full-time \\
\midrule
P3 (M, 30s) & Pop music review & 2018. 02 & 16.4k & 3267 & Full-time \\
\midrule
P4 (F, 30s) & Baking tutorial & 2020. 10 & 20.7k & 1657 & Full-time \\
\midrule
P5 (F, 30s) & Home interior & 2018. 10 & 38.9k & 4354 & Part-time \\
\midrule
P6 (M, 30s) & Electronics & 2022. 09 & 1.1k & 1050 & Full-time \\
\midrule
P7 (M, 40s) & Music industry \& K-pop & 2023. 03 & 6.7k & 2350 & Part-time \\
\midrule
P8 (F, 20s) & Single-person households \& Economics & 2023. 01 & 9.9k & 1721 & Part-time \\
\midrule
P9 (M, 30s) & Skin care \& Men fashion & 2023. 05 & 2.2k & 610 & Part-time \\
\midrule
P10 (M, 20s) & Electronics & 2020. 05 & 56.2k & 9089 & Part-time \\
\midrule
P11 (F, 20s) & Travel \& Life tips & 2021. 05 & 2.1k & 1313 & Part-time \\
\bottomrule
\end{tabular}
}
\vspace*{2mm}
\caption{
Participants’ demographics, channel information, and their level of commitment at the time of user studies.
}
\label{tab:demo_study}
\Description{This table provides an overview of the demographics, channel information, and commitment levels of 11 participants involved in the user studies. Each row represents a participant's ID, gender, age group, channel category, channel start date, total number of subscribers, total number of comments, and whether they are part-time or full-time creators. Notably, P10 has the highest number of subscribers (56.2k) and comments (9089), while P6 has the fewest subscribers (1.1k).}
\end{table*}

\subsection{Study Protocol}

\begin{figure*}[h]
\centering
\includegraphics[width=\textwidth]{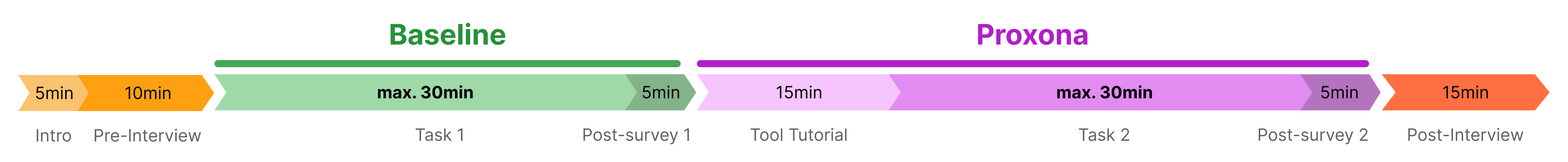}
\caption{User study procedure including tasks using the \docs{} and \sysname{}, surveys, and post-interview. The study lasted for approximately 2 hours.}
\label{fig:procedure}
\Description{User study procedure. The study was divided into two main phases: the \docs{} phase and the \sysname{} phase. Participants first completed a brief introduction (5 min) followed by a pre-interview (10 min). In the \docs{} phase, participants performed Task 1 (up to 30 min) using Google Docs, followed by a post-survey (5 min). After a brief Tool Tutorial for \sysname{} (15 min), participants proceeded with Task 2 (up to 30 min) using \sysname{}. Finally, participants completed a second post-survey (5 min) and a post-interview (15 min) to conclude the study.}

\end{figure*}

Figure~\ref{fig:procedure} illustrates the study procedure. Participants performed video storyline creation tasks in two different settings: \sysname{} and \docs{} (baseline of user evaluation). In the \docs{} condition, participants could use any methods they currently use, including data analytics and channel pages when planning for their new content. The \docs{} condition was designed to provide a standardized baseline that reflects the text-based ideation processes commonly used by creators, while also allowing flexibility for participants to incorporate their preferred tools where applicable. 
Then, they completed the same task with \sysname{}. We avoided counterbalancing conditions to prevent learning effects from \sysname{} influencing \docs{} performance. The task was provided as ``Please first explore your audience [as current practices / with our system], then create a video storyline with provided topic, targeting your audience.'' To minimize prior knowledge effects, each participant was assigned topics unrelated to their channel (See Appendix~\ref{app:userstudy} - Table~\ref{tab:user_study_topics}). An example task is shown in Appendix~\ref{app:userstudy} - Figure~\ref{fig:userstudy-task}. 

For the \docs{} condition, participants used Google Docs to document their audience understanding and storyline creation. For the \sysname{} condition, participants received a 15-minute tutorial to familiarize themselves with the system. To ensure a consistent mental model among participants during the study, we explicitly described the audience personas as virtual constructs generated from channel data, rather than representations of real audience members. This clarification emphasized the exploratory purpose of \sysname{} and guided participants to engage with the system's feedback with the appropriate expectations.

After each round, participants completed a 5-minute post-task survey. Upon completing both rounds, a semi-structured interview was conducted to discuss their experience with the two conditions and how each impacted their ideation process.

The entire study was conducted remotely over Zoom, with screen and audio recordings collected for analysis. The study was approved by the Institutional Review Board (IRB) at our institution.

\subsection{Measures and Analyses}

In both conditions, participants completed a questionnaire assessing the system's usability in audience exploration, sensemaking, and early-stage ideation support (7-point Likert scale; Appendix~\ref{app:question_usability} - Table~\ref{tab:question_usability}). We also used NASA-TLX~\cite{hart1988development} to measure the cognitive load (Appendix~\ref{app:question_usability} - Table~\ref{tab:question_nasatlx}). Finally, they scored the completeness of the content they created. 
Under the \sysname{} condition, we further inquired about the perceived quality of dimensions and values (adjusted from ~\cite{suh2023structured}; Appendix~\ref{app:question_dimension_values} - Table~\ref{tab:question_dimension} \& ~\ref{tab:question_value}), and persona chat and their feedback (Appendix~\ref{app:question_persona_chat} - Table~\ref{tab:question_persona_chat} \& ~\ref{tab:question_persona_feedback}). Also, participants evaluated the effectiveness of the human-AI collaboration and core user-centered challenges in human-AI systems (adjusted from ~\cite{wu2022ai}; Appendix~\ref{app:ai_chains} - Table~\ref{tab:ai_chains}). 

To supplement the quantitative data, we conducted post-interviews about the overall experience with the system, their perceptions of audience personas, dimensions, and values, along with any difficulties they faced while using the system. We also inquired about the potential impact of the system on their content creation process.

To analyze the responses to the survey questions including usability, quality of the attributes (personas, \dimensions{}, \dimvals{}), and NASA-TLX, Wilcoxon signed-rank test~\cite{wilcoxon1970critical}, which is a non-parametric statistics test that compares paired data, was used. For interview analysis, we transcribed audio recordings using Clova Note and structured the findings on a Miro board. Then, two authors consolidated themes based on the specific research questions through the thematic analysis process~\cite{thematic2006}.


\subsection{\textbf{How Effectively Can \sysname{} Support Creators in Exploring Audience Traits for Sensemaking and Ideation? (RQ2)}}

\label{rq2}
User study results demonstrated that \sysname{} significantly enhanced participants' ability to explore plausible audience segments and traits and gain actionable insights through persona-based interactions. This section presents how \sysname{} enabled creators to uncover audience diversity, identify hidden audience segments, and gain detailed insights into audience preferences and motivations through persona-based interactions. 

In the baseline (\docs{}) condition, participants used Google Docs and other familiar tools (e.g., YouTube Studio) typically used in their real practices. However, participants struggled to explore and make sense of audience information within the limited timeframe (max. 30 minutes). The overwhelming amount of unstructured data, combined with the lack of integrated analysis tools, hindered their ability to derive actionable insights. While some participants mentioned that existing tools provided them with sufficient general knowledge about their audience, they also noted difficulties in refining this knowledge further. As a result, participants often relied on their prior knowledge and assumptions about their audience to complete the task. For instance, participants shared general attributes like ``20-30 year-old women interested in interior design'' or ``music enthusiasts who are fans of indie bands and festivals.''

In contrast, with \sysname{}, participants felt that personas provided a tangible and relatable way to engage with their audience. P7 mentioned, \textit{``Seeing my audience represented as personas was incredibly moving. Normally, I just see usernames and their comments, but with personas, I felt like I could finally picture who my viewers were. It gave me a sense of comfort, almost like having someone real to talk to when creating content.''} Other participants echoed similar sentiments, emphasizing that the presence of personas emphasizing that personas facilitated a deeper exploration of their audience.

In the quantitative data, participants engaged with \sysname{} to explore potential audience characteristics, interacting with the personas through $6.73$ turns of chat on average (SD = $2.8$, min = $3$, med = $6$, max = $11$). 
The majority of participants agreed that \sysname{} effectively helped them explore the aspects of the audience through personas, which also boosted their confidence when targeting their audience group. 
From the survey responses, participants felt that they were able to explore their audience better with \sysname{} compared to their current methods (\docs{}), as reflected in Q1 scores ($M_{proxona}$ = $6.09$, $M_{docs}$ = $4.73$, \emph{p} < .05). Participants mentioned that they could gain specific data about their audience, which was previously only vaguely sensed through comments and metrics (P6, P10). They also stated that they could learn about potential audience groups that they had never thought of (P3, P11).



\begin{figure*}[h]
\centering
  \includegraphics[width=0.8\textwidth]{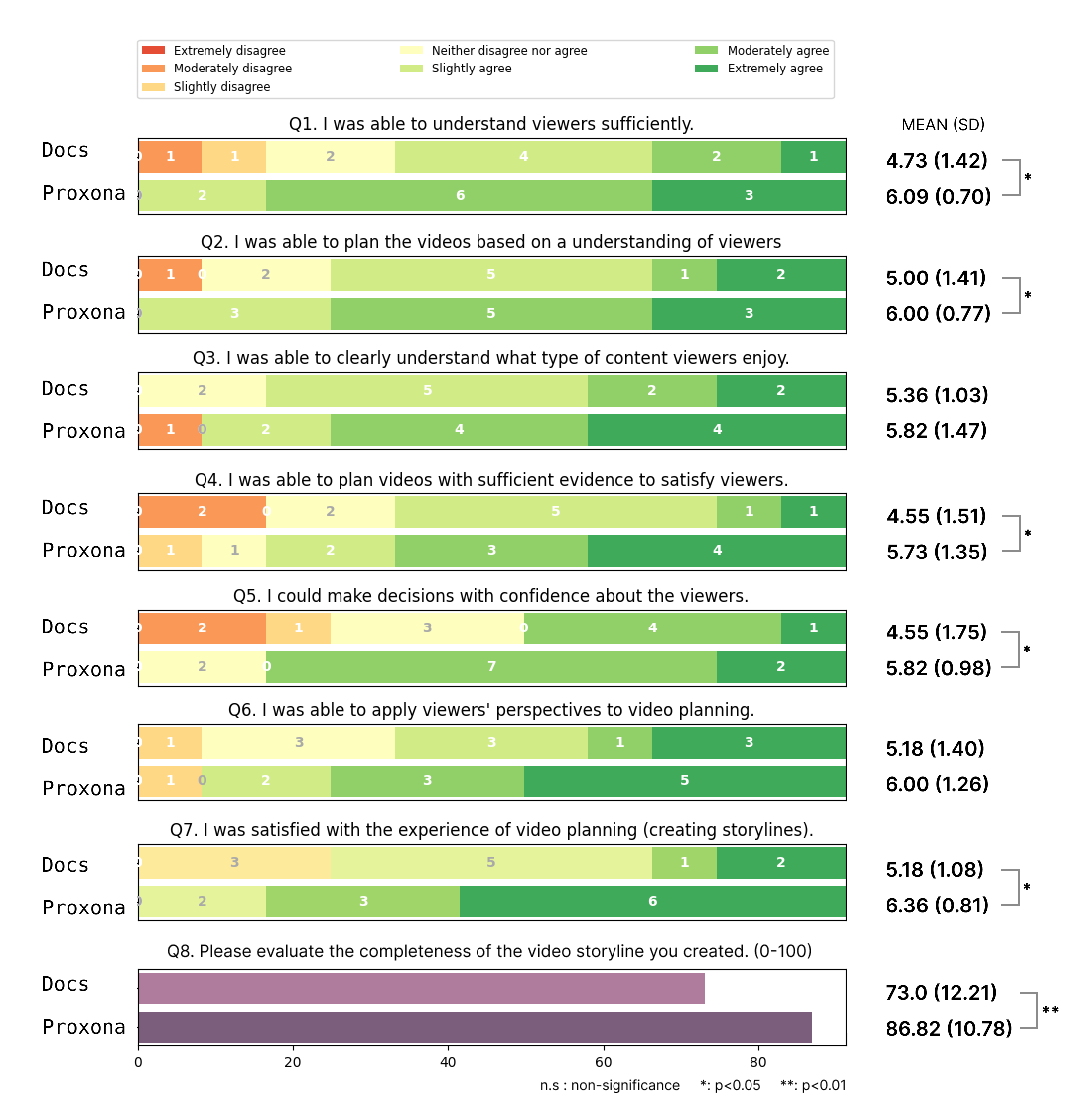}
  \caption{Overall usability survey questionnaires and results with each significance. 
  Note that Q8 is visualized with mean values calculated by participants' scores (0 - 100).}
  \label{fig:usability}
  \Description{Overall usability survey questionnaires and results with each significance. We present the participants' distribution with a colored bar graph. Note that Q8 is visualized with mean values calculated by participants' scores (0 - 100).}
\end{figure*}

\subsubsection{Gaining Deeper Insights into Audience Preferences and Motivations}

Through interacting with detailed personas, participants reported gaining richer insights into their audience's preferences and motivations. Unlike the traditional methods that rely on aggregated metrics or unstructured comment analysis, which often only provide surface-level and result-oriented insights, \sysname{} enabled participants to delve into the underlying reasons behind audience behaviors. 
Participants highlighted the significant role of our dimensions and values in clarifying their personas. They perceived dimensions and values were useful to figure out their audience (Q10: $M_{dimension}$ = $4.36$, Q16: $M_{value}$ = $4.18$), specific (Q19: $M_{value}$ = $4.64$), and diversely composed (Q15: $M_{dimension}$ = $4.36$, Q21: $M_{value}$ = $4.73$). One participant (P6) noted, \textit{``It felt like things that were vaguely in my mind got sorted out. You can’t gain this kind of deep insight from short comments.''} Similarly, P10 mentioned that \sysname{} helped refine and solidify their impressions, turning floating ideas into clear, actionable insights: \textit{``It made me think more clearly about my audience.''}
This deeper level of detail enabled participants to better align their content strategies, ensuring their content resonated more effectively with their target personas.

\subsubsection{Unveiling Diversity within a Seemingly Homogeneous Audience} 
Participants found that \sysname{} helped them recognize the heterogeneity in their audience. Previously, they perceived their audience as relatively homogeneous, but the analysis through audience personas revealed diverse and distinct segments (P1, P4, P6, P8). This provided crucial insights for content planning and targeting, such as targeting different personas to tailor content that addresses their viewers' specific needs or curiosities. For instance, P4 used \sysname{} to identify which audience segments were most interested in new baking methods versus those who preferred traditional and familiar topics. By discovering these distinctions, P4 could roughly realize why many viewers might enjoy content on subjects they are already familiar with, asking questions to personas to balance between introducing new ideas and revisiting popular ones as a potential strategy. 

These findings align with the overall survey results, where participants reported higher satisfaction in exploring audience preferences when using \sysname{} compared to the baseline method (Q2, $M_{proxona}$ = $5.45$, $M_{docs}$ = $4.36$, \emph{p} < .05). 
However, some participants (P9, P10) found interacting with multiple personas overwhelming, noting that the diverse opinions presented by different personas made it difficult to process and synthesize feedback. P9 and P10 mentioned feeling exhausted since they attempted to satisfy as many audience personas as possible, which led to difficulties applying the insights they gained. 

\subsubsection{Discovering Hidden or Unseen Audience Segments}
\sysname{} led participants to discover audience segments that they were previously unaware of, expanding their understanding beyond expected personas. By interacting with personas, participants identified potential audience traits and motivations that were not evident from traditional methods. For example, P6 (Electronics) encountered one persona that provided perspectives they had not considered (`eco-friendly lifestyle blogger who prefers products that minimize environmental impact'). By proceeding with the conversation, the persona provided compelling arguments and persuaded P6 to emphasize sustainable aspects in his future video. Similarly, P8 (Single-person households) was surprised by a persona's request to see `lessons learned from their daily mistakes', which she had previously edited out of her videos. P8 imagined that she would intentionally include more ``real'' moments in future videos to appeal to her personas. 

Survey results further support this qualitative finding. Participants rated their ability to discover new audience preferences significantly higher with \sysname{} compared to \docs{} (Q3, $M_{proxona}$ = $5.45$, $M_{docs}$ = $4.18$, \emph{p} < .05).
These unexpected discoveries enabled participants to broaden their content strategies, targeting new audience segments they had previously overlooked.


\subsection{\textbf{With \sysname{}, How Do Creators Integrate Persona-informed Insights into Their Creative Practices? (RQ3)}}

This section explores how participants utilized \sysname{} to enhance their creative practices, focusing on the system's impact while completing tasks e.g., video storyline planning. Participants spent significantly more time on the whole task in the \sysname{} condition, reaching the full 30-minute limit, whereas the baseline (\docs{}) condition averaged 14.64 minutes ($SD$ = 6.50). Although participants reported similar mental load levels between \sysname{} and \docs{} (NASA-TLX-Q1: $M_{proxona}$ = $2.73$,  $M_{docs}$ = $3.09$, \emph{p} = .66), survey results indicated that participants felt they accomplished their tasks more successfully with \sysname{} than \docs{} (NASA-TLX-Q4: $M_{proxona}$ = $5.73$,  $M_{docs}$ = $4.91$, \emph{p} < .05). 

Participants noted a distinct advantage of \sysname{} over \docs{} in facilitating dynamic and interactive exploration of audience traits. In the \docs{} condition, several participants described challenges in refining their storylines due to the lack of interactivity or integrated feedback, despite relying on their pre-existing understanding of their audience. 

In contrast, participants were highly satisfied with using \sysname{} when planning a video storyline (Q7: $M_{proxona}$ = 6.36, $M_{docs}$ = 5.18, \emph{p} < .05). Notably, participants reported high confidence in making decisions about planning a video storyline (Q5: $M_{proxona}$ = 5.82, $M_{docs}$ = 4.55, \emph{p} < .05). Participants felt that \sysname{} helped them plan a video more effectively compared to the \docs{} based on their understanding of viewers (Q2: $M_{proxona}$ = 6.0, $M_{docs}$ = 5.00, \emph{p} < .05). They also felt that a storyline created using \sysname{} was more complete (Q8: $M_{proxona}$ = 86.82, $M_{docs}$ = 73.00 on a 100-point scale, \emph{p} < .01) and able to plan videos with enough evidence (Q4: $M_{proxona}$ = 5.73, $M_{docs}$ = 4.55, \emph{p} < .05). As P9 noted, \sysname{} helped improve the quality of his storyline by highlighting overlooked aspects during his planning process.

\subsubsection{Using Persona Conversations to Enhance Creative Decisions.}
\hfill
\label{sec:6.6.1}

Participants freely utilized conversation features to inform their creative decisions, regardless of the stage of understanding and creation. While understanding their audience personas, participants also asked some questions to better plan their content. Compared to active chat usage in \emph{Exploration} stage ($M$ = $6.73$, $SD$ = $2.8$), participants rarely used the conversation feature at \emph{Creation} stage ($M$ = $0.91$, $SD$ = $1.2$, $min$ = $0$, $med$ = $1$, $max$ = $4$) as shown in Table~\ref{tab:user_study}. 

Some participants used the chat feature not only to gather opinions on their ideas but also to collaborate on achieving specific channel goals and refining overall channel strategies. Participants evaluated the quality of persona conversations as highly consistent (Q22: $M$ = $6.55$, $SD$ = $0.69$), indicating that personas maintained a reliable and coherent perspective throughout the conversations. Furthermore, participants found that personas reflected specific audience perspectives clearly (Q23: $M$ = $6.36$, $SD$ = $0.81$). Through chat interactions, participants felt they could directly engage with audience perspectives, enabling audience-focused and targeted content during the creative process.

\textbf{Gathering opinions on topic selection.}
Utilizing their own original ideas or dealing with unfamiliar topics of product placement, participants gauged audience personas' interests in potential topics, such as \textit{``Are you interested in buying an electric massage chair? (P2)''}, and \textit{``How do you learn English these days? (P8)''}. Through these questions, they sought to align their content with viewers' desires, making it more engaging and relevant.

\textbf{Collaborating to improve content for certain goals.}
Participants naturally involved their audience personas in writing content to align with their certain goals, as if they were doing `collaboration'. This includes requesting a title for the written content to attract specific audiences, along with tips for revision to enhance visibility (P2, P3, P9). By incorporating personas' ideas into their content to achieve the goal, participants believed they could make better video content and provide greater satisfaction to viewers. For instance, P9 asked ``What should I create as a video to increase the viewing duration from viewers?'' 

\textbf{Consulting on overall channel strategy.}
Participants not only questioned `what content to include in videos', but also considered the elements necessary for channel management, such as editing style of video (P7, P9), transition of channel topics (P4, P8), or even how to convey information effectively considering a channel's brand identity (P10). They asked these types of questions, considering the audience personas as the representatives of people expected to have viewership in their channel. For example, P4 asked that \textit{``Considering videos on well-known baking topics get higher views, do you [viewers] prefer familiar subjects over new ones?''} 

\textbf{Assessing performance from the audience's perspective.} 
\label{sec:predictive_case}
While \sysname{} was designed primarily as a sensemaking tool, the flexibility of natural language interactions led some participants to explore predictive use cases that were beyond their intended scope. For example, P9 asked \textit{``Will you watch my Vlog?''} seeking direct predictions about future audience behavior. Similarly, participants attempted to interpret content performance metrics (e.g., click-through rate, view counts) by gathering feedback from personas, seeking to understand the reasons behind them (P4, P5). For instance, P5 asked \textit{``How much does a thumbnail influence your decision to click on my video?''} to all audience personas; she expected whether her thumbnails were actually appealing to different personas.

\subsubsection{Ways to Utilize Persona Feedback in Storyline Creation}
\label{sec:6.6.2}
\hfill

In total, 11 participants used the feedback feature 28 times while drafting their video storyline ($M$ = $2.55$, $SD$ = $2.5$, $min$ = $0$, $med$ = $2$, $max$ = $6$), with notable variations in usage patterns. For instance, P5 requested feedback six times but did not use the chat feature at all; P3 asked four questions, with only one feedback request. Two participants (P8, P9) did not use the feedback feature at all. This discrepancy suggests that participants employed the feedback feature more selectively during the content creation stage, focusing on receiving evaluative input rather than exploratory dialogue. We present how participants utilized the features.

\textbf{Strengthening content logic with multiple persona perspectives.}
Participants sought multiple perspectives to strengthen their content's logic and secure their coverage of the audience. P11, for example, received evaluative feedback on various aspects of planning a trip to Nha Trang, including the choice of destination and tips for a harmonious family travel experience. By integrating these insights, P11 was able to refine the content's logical flow. P4 sought [suggestions] from more than one persona on pairing bread with Nespresso coffee, emphasizing the value of diverse viewpoints in enhancing content quality.

\textbf{Enriching content through iterative persona feedback.}
Iterative feedback is one of the benefits of co-creation, where participants refine specific content elements based on feedback from the audience personas. Based on a persona's feedback, participants made their own improvements. P5 and P11 exemplified iterative feedback by repeatedly adjusting and seeking [evaluation] on the same block of content, ensuring the revision is closely aligned with the personas’ intentions. 
During the content planning iteration, they aimed to engage deeply with a limited number of personas. For instance, P10 concentrated on a single persona to gather both evaluations and suggestions. This method ensures that content not only meets the initial quality standards but also evolves based on constructive feedback.


\textbf{Confirming choices through persona feedback.}
Some participants decided against pursuing a content idea based on persona's feedback. In P2's case, the negative reaction from personas led to abandoning the idea altogether, despite initial attempts to persuade the personas of its relevance. This highlights the perceived role of personas not only in guiding content creation but also in evaluating the potential success or failure of content ideas.

\subsubsection{Perceiving \sysname{} as Human-AI Co-creation Support}
\label{sec:6.6.3}

Among the AI Chains measures~\cite{wu2022ai} (Appendix~\ref{app:ai_chains}), the highest score was related to the collaboration, where participants perceived that the process was a collaborative process with the system (Q36: $M$ = $6.18$, $SD$ = $1.17$). This finding reinforces the idea that \sysname{} functioned not merely as a tool, but as a co-creator in the content development process. P4 mentioned, \textit{``Through interacting with them (audience personas), the process of targeting and understanding the audience becomes more concrete---it feels like we're collaborating, creating a sense of `teamwork'.''}
On the other hand, the lowest score was from the controllability measure (Q35: $M$ = $5.18$, $SD$ = $1.66$), where participants felt they did not have enough control in using the system. It can be connected to personas' consistency as P7 mentioned: \textit{``Even when talking to people now, there are those who steer the conversation only towards their area of interest. I found it sometimes disappointing that, regardless of direction, each persona attempted to lead the discussion solely towards the topic they wanted to talk about.''}

\section{Discussion} 
Based on the technical evaluation and user studies, we discuss the implications of using data-driven personas in content creation, focusing on how dimensions and values help characterize audiences, the challenges and strengths of using LLM-generated personas, desirable and undesirable use cases for \sysname{}, and the generalizability of our system to other creative practices. Lastly, we briefly discuss the limitations and future work.

\subsection{Constructing Audience Personas with Dimensions-Values Framework}

With recent advancements in LLMs, researchers have explored synthetic users across diverse fields, such as education~\cite{markel2023gpteach}, community design~\cite{park2022social}, writing~\cite{benharrak2023writer}, and presentation~\cite{park2023audilens}. Traditional persona creation methods often involve manual configuration~\cite{benharrak2023writer} or are generated in response to specific tasks~\cite{park2022social}, requiring creators to assume who their audience should be. This assumption-based approach can lead to abstract personas that may not describe real audience data, which is hard to apply to our creator economy context. In contrast, our approach generates personas directly from real audience data, such as comments, removing the need for initial assumptions. This allows for more authentic insights into the audience, ensuring that personas are grounded in actual audience traits, not hypothetical profiles.

The \dimensions{} and \dimvals{} framework plays a pivotal role in characterizing audience personas. By structuring audience traits into distinct categories, the framework helps creators identify nuanced characteristics that are often missed with traditional data analytics. The study participants appreciated this approach, noting how it revealed hidden audience segments, such as `novice vs. expert' or `entertainment-seeking vs. information-seeking.' These traits, which would have been difficult to identify through manual analysis of comments or typical analytics tools, empowered creators to diversify their content and appeal to a wider range of viewers.

To further structure audience traits into actionable personas, our system organizes these attributes using clustering techniques. While k-means clustering currently serves as a practical mechanism for grouping audience traits, it may not be strictly necessary for the functionality of the dimension-value framework. Future iterations of \sysname{} could explore alternatives, such as LLM-based clustering or other dynamic grouping methods, to streamline the process while maintaining usability and interpretability. Such refinements could reduce system complexity while preserving its effectiveness for end users.

By enhancing creators' sensemaking process, \sysname{} opens up new possibilities for discovering and engaging with diverse audience segments. Beyond identifying existing traits, the system could dynamically adapt personas based on real-time audience data, ensuring alignment with shifting preferences and behaviors. Future work could explore integrating real-time audience data or allowing for more iterative persona development, ensuring alignment with shifting preferences and behaviors.

\subsection{Navigating LLM-Generated Personas: Trust, Creativity, and Reliance}

Our research highlights the potential of integrating LLM-generated personas into the creator economy, offering creators a novel way to sensemake and ideate with their potential audience. Creators valued data-driven personas in their creative practices, noting that the personas offered insights that were often easier to grasp compared to raw comments or data analytics. In the user study (Section~\ref{rq2}; Figure~\ref{fig:usability}), creators expressed satisfaction in obtaining informative content about their audiences from the audience personas, as they confirmed that personas were grounded in real audience data. This trust allowed creators to perceive the personas as reliable reflections of their audience, enhancing their confidence in the persona's feedback and suggestions. 
  
While trust is essential for creators to embrace personas as valuable tools, it also introduces potential risks -- such as providing misleading responses. For example, creators might lean too heavily on persona feedback that reinforces their preconceptions, leading to confirmation bias. As noted in the previous research, LLMs often provide overly agreeable or positive responses~\cite{chen-etal-2023-say}. 
The inherent social value alignment and moderation mechanisms of the LLM introduce a potential bias when creating personas~\cite{zhang2023heterogeneous, bauer2020virtuous}. Even though the dimension-value framework effectively structures audience traits for persona generation with LLM, it might not capture all relevant attributes, particularly those that are obscure or sensitive. For example, certain nuanced audience characteristics might be filtered out during LLM processing due to internal moderation. And this under-represents critical, controversial, or dissenting opinions—traits that may be pivotal in creative practices. 
This may cause creators to overlook critical or diverse opinions, leading to skewed content strategies. Additionally, in our study, some participants occasionally distorted the use of audience personas such as asking actual audience view counts. Continued use for such a purpose may lead to the formation of incorrect beliefs and misconceptions about real people~\cite{american1997source}.

Our findings also reveal a mixed reaction to audience personas. While many creators expressed high trust in the personas' utility, some remained cautious. Vigilant participants were aware that LLM-generated personas, though informative, might not fully capture the diversity or dynamics of their real audience. Still, this skepticism was not entirely negative—creators emphasized that final content decisions ultimately rested with them, highlighting the agency they maintained over the persona's influence. This reflects the balance creators must strike between utilizing AI-generated insights and relying on their own judgment. Importantly, during the study, participants are often critical in understanding the exploratory nature of \sysname{}. They used persona-driven feedback to leverage their ideas rather than fully relying on it. This may be because they have held an understanding of their channels and agency in creating their content. As they had knowledge to judge AI-generated insights, they were not merely passive recipients of the feedback but actively evaluated their relevance and applicability to their content strategies. By approaching \sysname{} as a sensemaking tool, participants leveraged its potential to enrich their creative process while maintaining control over the final decisions.

This implies that, while LLM-generated personas hold significant potential for supporting creative practices, their utility largely depends on how well creators understand their capabilities and limitations. To effectively integrate LLM-based personas as tools rather than definitive authorities, it is important to enhance transparency, promote balanced reliance, and preserve creator agency. These steps remain key questions for better integration.
While \sysname{} is intended to support the early-stage creative process, we expect that the improved sensemaking and exploratory processes will positively impact execution performance. To expedite the process, it is essential to support creators sensibly in each creative process, helping them connect deeply with their target audiences beyond persona-driven insights.

To mitigate the risk of over-reliance and ensure balanced decision-making, we propose that LLM-based personas should provide transparent explanations about their construction process, capabilities, and limitations. For instance, audience personas could include the markers~\cite{burrus2024unmasking} indicating when responses are AI-generated, alongside metadata such as the source of audience insights (e.g., source comments) or the degree of confidence in certain predictions. By offering more explicit signals, these personas can help creators better gauge the relevance and reliability of the feedback, supporting a more balanced integration of AI into their creative processes.

\subsection{Desirable and Undesirable Use Cases of \sysname{}}

While \sysname{} is designed for audience exploration, sensemaking and ideation, we found that participants utilized the natural language chat for purposes beyond the initial goals due to its flexibility (Section ~\ref{sec:6.6.1}). Based on the observations, we discuss desirable use cases that our system can support the most, and undesirable cases.

\sysname{} can be used for \textit{exploring potential audience to learn about them}. Creators can use it to uncover audience traits and preferences, gaining insights into diverse audience segments that may have been overlooked. For example, through questions like ``Would viewers prefer a video focused solely on baking techniques or one that includes stories about the origins of bread? (P4)'', creators can gain insights into relevant audience information. Also, it can be used to \textit{generate creative ideas informed by audience perspectives}. For instance, a creator (P3) who received an offer to advertise a branded shoes might ask, ``How can I integrate this ad seamlessly into our music content?'' \sysname{} can provide plausible ideas that inspire creative approaches to such challenges. Finally, \sysname{} enables creators to \textit{iteratively refine their content} with simulated feedback from audience personas. Personas might offer suggestions for improving a draft storyline, such as addressing specific audience interests (``@Kendall, how can I efficiently include the necessary information in my vlog?'' (P9)). This can help creators make informed adjustments to their content.

Despite its strengths, creators can \textit{attempt to predict audience behavior}, such as asking ``How many people will click on this video?''. They can also \textit{seek guaranteed content outcomes}, such as asking ``Would it be okay to add this ad to increase view count?''. \sysname{} is not designed for predictive use cases or guaranteeing specific outcomes, but some creators might expect to earn revenues through view counts, not considering content types. While \sysname{} can offer inspiration and qualitative feedback, it does not provide assurances about measurable results like viewership or engagement metrics.

To avoid these undesirable cases, future research could explore the way how creators can interact with the personas. For example, creators could use a set of template for questions. In addition, integrating data-driven predictive analytics into the current system could be considered. This direction could enable the system to meet the needs of creators seeking more insights for making critical decisions, while maintaining its core strengths in exploratory ideation and sensemaking.

\subsection{Generalizing \sysname{}: Applications in Diverse Creative Practices}

We believe our approach can be generalized to a broader range of creative practices, by including different data sources and content types for persona generation, and engaging creators of diverse expertise levels. 

\emph{Beyond Comments.} 
While \sysname{} utilizes comments as the primary source of persona construction, integrating additional platform analytics data could enhance the completeness of the generated personas. For example, behavioral data such as view duration, timestamps, and watch patterns would allow \sysname{} to create even more detailed and realistic personas by combining both. Additionally, integrating data from social media platforms like X, Instagram, or forums like Reddit---where users often discuss a creator's content more informally---could further enrich the personas, capturing a wider range of audience opinions and interactions.

\emph{Beyond Transcripts and Videos.} 
Our \sysname{} pipeline, which currently processes transcripts to generate audience persona's feedback and conversations, can be easily applicable to support other forms of text-based content beyond video storylines. This opens up the possibility of applying \sysname{} to other media types such as blogs, podcasts, and articles, where audience engagement primarily occurs through comments or discussions. In addition, as videos are multimedia in nature, there is potential to incorporate visual and auditory modalities into the system. For instance, if personas could analyze multimodal content, creators might receive feedback on visual aesthetics, audio quality, or editing styles, thus supporting decision-making across the entire video production process. One potential expansion lies in applying \sysname{} to live-streaming platforms, where real-time audience interaction is a core feature. \sysname{} could be adapted to provide synchronous feedback during live broadcasts, allowing creators to receive real-time insights from audience personas based on live chat interactions.

\emph{Beyond Experienced Creators.} 
While primarily designed for channels with a large volume of audience comments, \sysname{} can be adapted for creators with smaller viewership or early-stage channels. For novice creators, \sysname{} could be used to generate personas from aspirational channels and gain opportunities to explore the characteristics and motivations of the audience they hope to attract. Alternatively, \sysname{} could simulate audience comments based on similar channels or content, like beta viewers~\cite{choi2023creator}, providing valuable audience personas even under a cold start setting. These adaptations for newer creators would allow them to benefit from persona-driven insights, helping them better understand potential audiences and refine their content strategies as they grow.

\emph{Beyond Audience Personas.}
The \dimension{}-\dimval{} framework used in \sysname{} could be extended beyond audience persona generation. Its structured approach, which organizes unstructured data into dimensions and values, makes it adaptable to various contexts where large amounts of qualitative data need to be synthesized into meaningful insights. For example, this structure could be widely applied to generate student personas from class forums or consumer personas from product reviews. The framework's flexibility makes it suitable for structuring insights in diverse contexts, opening opportunities to create useful, context-specific personas.

\subsection{Limitations \& Future Work}
While \sysname{} introduces a novel approach to integrating LLM-driven audience personas into creative practices, we acknowledge several limitations. 

\sysname{} relies on audience comments to generate personas, which potentially introduces biases. Comments represent only a subset of audience behaviors and opinions, excluding non-commenting viewers who may have different preferences or perspectives. Future research could explore integrating additional data sources, such as passive audience behaviors (e.g., view duration, likes, and shares) or external audience surveys, to ensure a more holistic representation of the entire audience. Plus, the initial step of our pipeline involves filtering comments based on length--- which might omit valuable insights from shorter comments, potentially skewing the personas. Future research should explore more comprehensive comment filtering techniques--- e.g., judging the helpfulness or constructiveness of the comments. 

The LLM performance can affect the user experience when handling non-English languages. While participants engaged with \sysname{} personas in Korean, and the comments were also primarily in Korean, it is important to acknowledge that large language models often perform better in English than in other languages. This disparity may impact the quality of personas generated, particularly for nuanced or idiomatic expressions in Korean. Future work will explore strategies to enhance multilingual support, such as fine-tuning models on additional language-specific data or leveraging multilingual datasets.

While the \sysname{} generates data-driven personas, they cannot be mapped to real-world audience populations. Our goal is not to mirror the audience perfectly but to enable the exploration of prototypical audiences to inform content creation, which can be solved by adjusting the mental model and clarifying the role of the system. 

When there are few or no comments in creators' channels, the creators may not benefit from \sysname{}. This cold-start problem limits the applicability of the system across a wide range of creators, underscoring the necessity of developing additional strategies to support those creators. Future work could utilize the data from existing channels with similar content or audience demographics to simulate personas.


\section{Conclusion}

We introduce \sysname{}, a novel system that employs LLMs to generate data-driven audience personas, enhancing creators' exploration and sensemaking into their audience and supporting the development of more informed, audience-centered content strategies. Through technical evaluations and a user study with YouTube creators, we demonstrated how \sysname{} generates high-quality, multi-dimensional that provides more unusual and diverse perspectives of audiences. The user study highlighted \sysname{}'s potential to bridge the gap between creators and their audiences, promoting a collaborative approach where creators engage with audience personas in a dynamic exchange of ideas and feedback.



\begin{acks}

This work was supported by the National Research Foundation of Korea (NRF) grant funded by the Korea government (MSIT) (No. RS-2024-00406715) and by Institute of Information \& communications Technology Planning \& Evaluation (IITP) grant funded by the Korea government (MSIT) (No. RS-2024-00443251, \textit{Accurate and Safe Multimodal, Multilingual Personalized AI Tutors}). We also acknowledge support from the Office of Naval Research (ONR: N00014-24-1-2290). Additionally, this work was supported by the National Science Foundation (USA) (DGE-2125858) and Good Systems, a UT Austin Grand Challenge for developing responsible AI technologies~\footnote{https://goodsystems.utexas.edu}.


We all thank the members of KIXLAB (KAIST Interaction Lab) and CSTL (Collaborative Social Technologies Lab) at KAIST for their encouragement, support, invaluable feedback, and infinite cups of coffee. We are deeply grateful to all of our interviewees and study participants for their time and insights.
The first author sincerely thanks all team members for their unwavering support and dedication, especially during the multi-timezone collaboration leading up to the paper submission. Yoonseo also expresses her deepest appreciation to her wonderful research friends--- Hyuntak, Jeongeon, Haesoo, Evey, Mathias, Hoon, Nyoungwoo, KJ, Jisoo, and Renkai--- for their friendship, inspiration, and constant encouragement.

\end{acks}

\bibliographystyle{ACM-Reference-Format}
\bibliography{chi2025paper-proxona}


\begin{thebibliography}{71}


\ifx \showCODEN    \undefined \def \showCODEN     #1{\unskip}     \fi
\ifx \showDOI      \undefined \def \showDOI       #1{#1}\fi
\ifx \showISBNx    \undefined \def \showISBNx     #1{\unskip}     \fi
\ifx \showISBNxiii \undefined \def \showISBNxiii  #1{\unskip}     \fi
\ifx \showISSN     \undefined \def \showISSN      #1{\unskip}     \fi
\ifx \showLCCN     \undefined \def \showLCCN      #1{\unskip}     \fi
\ifx \shownote     \undefined \def \shownote      #1{#1}          \fi
\ifx \showarticletitle \undefined \def \showarticletitle #1{#1}   \fi
\ifx \showURL      \undefined \def \showURL       {\relax}        \fi
\providecommand\bibfield[2]{#2}
\providecommand\bibinfo[2]{#2}
\providecommand\natexlab[1]{#1}
\providecommand\showeprint[2][]{arXiv:#2}

\bibitem[Ayala and Bechard(2024)]%
        {ayala2024reducing}
\bibfield{author}{\bibinfo{person}{Orlando Ayala} {and} \bibinfo{person}{Patrice Bechard}.} \bibinfo{year}{2024}\natexlab{}.
\newblock \showarticletitle{Reducing hallucination in structured outputs via Retrieval-Augmented Generation}. In \bibinfo{booktitle}{\emph{Proceedings of the 2024 Conference of the North American Chapter of the Association for Computational Linguistics: Human Language Technologies (Volume 6: Industry Track)}}. \bibinfo{pages}{228--238}.
\newblock


\bibitem[Bauer(2020)]%
        {bauer2020virtuous}
\bibfield{author}{\bibinfo{person}{William~A Bauer}.} \bibinfo{year}{2020}\natexlab{}.
\newblock \showarticletitle{Virtuous vs. utilitarian artificial moral agents}.
\newblock \bibinfo{journal}{\emph{AI \& SOCIETY}} \bibinfo{volume}{35}, \bibinfo{number}{1} (\bibinfo{year}{2020}), \bibinfo{pages}{263--271}.
\newblock


\bibitem[Benharrak et~al\mbox{.}(2024)]%
        {benharrak2023writer}
\bibfield{author}{\bibinfo{person}{Karim Benharrak}, \bibinfo{person}{Tim Zindulka}, \bibinfo{person}{Florian Lehmann}, \bibinfo{person}{Hendrik Heuer}, {and} \bibinfo{person}{Daniel Buschek}.} \bibinfo{year}{2024}\natexlab{}.
\newblock \showarticletitle{Writer-Defined AI Personas for On-Demand Feedback Generation}. In \bibinfo{booktitle}{\emph{Proceedings of the CHI Conference on Human Factors in Computing Systems}} (Honolulu, HI, USA) \emph{(\bibinfo{series}{CHI '24})}. \bibinfo{publisher}{Association for Computing Machinery}, \bibinfo{address}{New York, NY, USA}, Article \bibinfo{articleno}{1049}, \bibinfo{numpages}{18}~pages.
\newblock
\showISBNx{9798400703300}
\urldef\tempurl%
\url{https://doi.org/10.1145/3613904.3642406}
\showDOI{\tempurl}


\bibitem[Biel and Gatica-Perez(2011)]%
        {biel2011vlogsense}
\bibfield{author}{\bibinfo{person}{Joan-Isaac Biel} {and} \bibinfo{person}{Daniel Gatica-Perez}.} \bibinfo{year}{2011}\natexlab{}.
\newblock \showarticletitle{VlogSense: Conversational behavior and social attention in YouTube}.
\newblock \bibinfo{journal}{\emph{ACM Transactions on Multimedia Computing, Communications, and Applications (TOMM)}} \bibinfo{volume}{7}, \bibinfo{number}{1} (\bibinfo{year}{2011}), \bibinfo{pages}{1--21}.
\newblock


\bibitem[Bishop(2020)]%
        {bishop2020algorithmic}
\bibfield{author}{\bibinfo{person}{Sophie Bishop}.} \bibinfo{year}{2020}\natexlab{}.
\newblock \showarticletitle{Algorithmic experts: Selling algorithmic lore on YouTube}.
\newblock \bibinfo{journal}{\emph{Social Media+ Society}} \bibinfo{volume}{6}, \bibinfo{number}{1} (\bibinfo{year}{2020}), \bibinfo{pages}{2056305119897323}.
\newblock


\bibitem[Braun and Clarke(2006)]%
        {thematic2006}
\bibfield{author}{\bibinfo{person}{Virginia Braun} {and} \bibinfo{person}{Victoria Clarke}.} \bibinfo{year}{2006}\natexlab{}.
\newblock \showarticletitle{Using thematic analysis in psychology}.
\newblock \bibinfo{journal}{\emph{Qualitative Research in Psychology}} \bibinfo{volume}{3}, \bibinfo{number}{2} (\bibinfo{year}{2006}), \bibinfo{pages}{77--101}.
\newblock
\urldef\tempurl%
\url{https://doi.org/10.1191/1478088706qp063oa}
\showDOI{\tempurl}


\bibitem[Burrus et~al\mbox{.}(2024)]%
        {burrus2024unmasking}
\bibfield{author}{\bibinfo{person}{Olivia Burrus}, \bibinfo{person}{Amanda Curtis}, {and} \bibinfo{person}{Laura Herman}.} \bibinfo{year}{2024}\natexlab{}.
\newblock \showarticletitle{Unmasking AI: Informing Authenticity Decisions by Labeling AI-Generated Content}.
\newblock \bibinfo{journal}{\emph{Interactions}} \bibinfo{volume}{31}, \bibinfo{number}{4} (\bibinfo{year}{2024}), \bibinfo{pages}{38--42}.
\newblock


\bibitem[Chen et~al\mbox{.}(2023)]%
        {chen-etal-2023-say}
\bibfield{author}{\bibinfo{person}{Jiangjie Chen}, \bibinfo{person}{Wei Shi}, \bibinfo{person}{Ziquan Fu}, \bibinfo{person}{Sijie Cheng}, \bibinfo{person}{Lei Li}, {and} \bibinfo{person}{Yanghua Xiao}.} \bibinfo{year}{2023}\natexlab{}.
\newblock \showarticletitle{Say What You Mean! Large Language Models Speak Too Positively about Negative Commonsense Knowledge}. In \bibinfo{booktitle}{\emph{Proceedings of the 61st Annual Meeting of the Association for Computational Linguistics (Volume 1: Long Papers)}}, \bibfield{editor}{\bibinfo{person}{Anna Rogers}, \bibinfo{person}{Jordan Boyd-Graber}, {and} \bibinfo{person}{Naoaki Okazaki}} (Eds.). \bibinfo{publisher}{Association for Computational Linguistics}, \bibinfo{address}{Toronto, Canada}, \bibinfo{pages}{9890--9908}.
\newblock
\urldef\tempurl%
\url{https://doi.org/10.18653/v1/2023.acl-long.550}
\showDOI{\tempurl}


\bibitem[Cheng et~al\mbox{.}(2023)]%
        {cheng2023marked}
\bibfield{author}{\bibinfo{person}{Myra Cheng}, \bibinfo{person}{Esin Durmus}, {and} \bibinfo{person}{Dan Jurafsky}.} \bibinfo{year}{2023}\natexlab{}.
\newblock \showarticletitle{Marked personas: Using natural language prompts to measure stereotypes in language models}.
\newblock \bibinfo{journal}{\emph{arXiv preprint arXiv:2305.18189}} (\bibinfo{year}{2023}).
\newblock


\bibitem[Cheng et~al\mbox{.}({[n.\,d.]})]%
        {chengCoMPosTCharacterizingEvaluating2023a}
\bibfield{author}{\bibinfo{person}{Myra Cheng}, \bibinfo{person}{Tiziano Piccardi}, {and} \bibinfo{person}{Diyi Yang}.} \bibinfo{year}{[n.\,d.]}\natexlab{}.
\newblock \showarticletitle{{CoMPosT}: Characterizing and Evaluating Caricature in {LLM} Simulations}. In \bibinfo{booktitle}{\emph{Proceedings of the 2023 Conference on Empirical Methods in Natural Language Processing}} (Singapore, 2023-12), \bibfield{editor}{\bibinfo{person}{Houda Bouamor}, \bibinfo{person}{Juan Pino}, {and} \bibinfo{person}{Kalika Bali}} (Eds.). \bibinfo{publisher}{Association for Computational Linguistics}, \bibinfo{pages}{10853--10875}.
\newblock
\urldef\tempurl%
\url{https://doi.org/10.18653/v1/2023.emnlp-main.669}
\showDOI{\tempurl}


\bibitem[Choi et~al\mbox{.}(2023a)]%
        {choi2023creativeconnect}
\bibfield{author}{\bibinfo{person}{DaEun Choi}, \bibinfo{person}{Sumin Hong}, \bibinfo{person}{Jeongeon Park}, \bibinfo{person}{John Joon~Young Chung}, {and} \bibinfo{person}{Juho Kim}.} \bibinfo{year}{2023}\natexlab{a}.
\newblock \showarticletitle{CreativeConnect: Supporting Reference Recombination for Graphic Design Ideation with Generative AI}.
\newblock \bibinfo{journal}{\emph{arXiv preprint arXiv:2312.11949}} (\bibinfo{year}{2023}).
\newblock


\bibitem[Choi et~al\mbox{.}(2023b)]%
        {choi2023creator}
\bibfield{author}{\bibinfo{person}{Yoonseo Choi}, \bibinfo{person}{Eun~Jeong Kang}, \bibinfo{person}{Min~Kyung Lee}, {and} \bibinfo{person}{Juho Kim}.} \bibinfo{year}{2023}\natexlab{b}.
\newblock \showarticletitle{Creator-friendly Algorithms: Behaviors, Challenges, and Design Opportunities in Algorithmic Platforms}. In \bibinfo{booktitle}{\emph{Proceedings of the 2023 CHI Conference on Human Factors in Computing Systems}}. \bibinfo{pages}{1--22}.
\newblock


\bibitem[Chung et~al\mbox{.}(2022)]%
        {chung2022talebrush}
\bibfield{author}{\bibinfo{person}{John Joon~Young Chung}, \bibinfo{person}{Wooseok Kim}, \bibinfo{person}{Kang~Min Yoo}, \bibinfo{person}{Hwaran Lee}, \bibinfo{person}{Eytan Adar}, {and} \bibinfo{person}{Minsuk Chang}.} \bibinfo{year}{2022}\natexlab{}.
\newblock \showarticletitle{TaleBrush: visual sketching of story generation with pretrained language models}. In \bibinfo{booktitle}{\emph{CHI Conference on Human Factors in Computing Systems Extended Abstracts}}. \bibinfo{pages}{1--4}.
\newblock


\bibitem[Cooper(1999)]%
        {cooper1999inmates}
\bibfield{author}{\bibinfo{person}{Alan Cooper}.} \bibinfo{year}{1999}\natexlab{}.
\newblock \bibinfo{booktitle}{\emph{The inmates are running the asylum}}.
\newblock \bibinfo{publisher}{Springer}.
\newblock


\bibitem[Csikszentmihalyi(1997)]%
        {csikszentmihalyi1997flow}
\bibfield{author}{\bibinfo{person}{Mihaly Csikszentmihalyi}.} \bibinfo{year}{1997}\natexlab{}.
\newblock \showarticletitle{Flow and the psychology of discovery and invention}.
\newblock \bibinfo{journal}{\emph{HarperPerennial, New York}}  \bibinfo{volume}{39} (\bibinfo{year}{1997}), \bibinfo{pages}{1--16}.
\newblock


\bibitem[Deshpande et~al\mbox{.}(2023)]%
        {deshpande2023anthropomorphization}
\bibfield{author}{\bibinfo{person}{Ameet Deshpande}, \bibinfo{person}{Tanmay Rajpurohit}, \bibinfo{person}{Karthik Narasimhan}, {and} \bibinfo{person}{Ashwin Kalyan}.} \bibinfo{year}{2023}\natexlab{}.
\newblock \showarticletitle{Anthropomorphization of AI: opportunities and risks}.
\newblock \bibinfo{journal}{\emph{arXiv preprint arXiv:2305.14784}} (\bibinfo{year}{2023}).
\newblock


\bibitem[Duffy et~al\mbox{.}(2017)]%
        {duffy2017platform}
\bibfield{author}{\bibinfo{person}{Brooke~Erin Duffy}, \bibinfo{person}{Urszula Pruchniewska}, {and} \bibinfo{person}{Leah Scolere}.} \bibinfo{year}{2017}\natexlab{}.
\newblock \showarticletitle{Platform-specific self-branding: Imagined affordances of the social media ecology}. In \bibinfo{booktitle}{\emph{Proceedings of the 8th international conference on social media \& society}}. \bibinfo{pages}{1--9}.
\newblock


\bibitem[for Research~into Nervous et~al\mbox{.}(1997)]%
        {american1997source}
\bibfield{author}{\bibinfo{person}{American~Association for Research~into Nervous}, \bibinfo{person}{Mental Diseases}, {and} \bibinfo{person}{Marcia~K Johnson}.} \bibinfo{year}{1997}\natexlab{}.
\newblock \showarticletitle{Source monitoring and memory distortion}.
\newblock \bibinfo{journal}{\emph{Philosophical Transactions of the Royal Society of London. Series B: Biological Sciences}} \bibinfo{volume}{352}, \bibinfo{number}{1362} (\bibinfo{year}{1997}), \bibinfo{pages}{1733--1745}.
\newblock


\bibitem[Frich et~al\mbox{.}(2019)]%
        {frich2019mapping}
\bibfield{author}{\bibinfo{person}{Jonas Frich}, \bibinfo{person}{Lindsay MacDonald~Vermeulen}, \bibinfo{person}{Christian Remy}, \bibinfo{person}{Michael~Mose Biskjaer}, {and} \bibinfo{person}{Peter Dalsgaard}.} \bibinfo{year}{2019}\natexlab{}.
\newblock \showarticletitle{Mapping the landscape of creativity support tools in HCI}. In \bibinfo{booktitle}{\emph{Proceedings of the 2019 CHI Conference on Human Factors in Computing Systems}}. \bibinfo{pages}{1--18}.
\newblock


\bibitem[Gandhi et~al\mbox{.}(2023)]%
        {gandhi2023strategic}
\bibfield{author}{\bibinfo{person}{Kanishk Gandhi}, \bibinfo{person}{Dorsa Sadigh}, {and} \bibinfo{person}{Noah~D Goodman}.} \bibinfo{year}{2023}\natexlab{}.
\newblock \showarticletitle{Strategic reasoning with language models}.
\newblock \bibinfo{journal}{\emph{arXiv preprint arXiv:2305.19165}} (\bibinfo{year}{2023}).
\newblock


\bibitem[H{\"a}m{\"a}l{\"a}inen et~al\mbox{.}(2023)]%
        {hamalainen2023evaluating}
\bibfield{author}{\bibinfo{person}{Perttu H{\"a}m{\"a}l{\"a}inen}, \bibinfo{person}{Mikke Tavast}, {and} \bibinfo{person}{Anton Kunnari}.} \bibinfo{year}{2023}\natexlab{}.
\newblock \showarticletitle{Evaluating large language models in generating synthetic hci research data: a case study}. In \bibinfo{booktitle}{\emph{Proceedings of the 2023 CHI Conference on Human Factors in Computing Systems}}. \bibinfo{pages}{1--19}.
\newblock


\bibitem[Hanusch and Tandoc~Jr(2019)]%
        {hanusch2019comments}
\bibfield{author}{\bibinfo{person}{Folker Hanusch} {and} \bibinfo{person}{Edson~C Tandoc~Jr}.} \bibinfo{year}{2019}\natexlab{}.
\newblock \showarticletitle{Comments, analytics, and social media: The impact of audience feedback on journalists’ market orientation}.
\newblock \bibinfo{journal}{\emph{Journalism}} \bibinfo{volume}{20}, \bibinfo{number}{6} (\bibinfo{year}{2019}), \bibinfo{pages}{695--713}.
\newblock


\bibitem[Hart and Staveland(1988)]%
        {hart1988development}
\bibfield{author}{\bibinfo{person}{Sandra~G Hart} {and} \bibinfo{person}{Lowell~E Staveland}.} \bibinfo{year}{1988}\natexlab{}.
\newblock \showarticletitle{Development of NASA-TLX (Task Load Index): Results of empirical and theoretical research}.
\newblock In \bibinfo{booktitle}{\emph{Advances in psychology}}. Vol.~\bibinfo{volume}{52}. \bibinfo{publisher}{Elsevier}, \bibinfo{pages}{139--183}.
\newblock


\bibitem[Henriksen et~al\mbox{.}(2016)]%
        {henriksen2016systems}
\bibfield{author}{\bibinfo{person}{Danah Henriksen}, \bibinfo{person}{Megan Hoelting}, {and} \bibinfo{person}{Deep-Play~Research Group}.} \bibinfo{year}{2016}\natexlab{}.
\newblock \showarticletitle{A systems view of creativity in a YouTube world}.
\newblock \bibinfo{journal}{\emph{TechTrends}}  \bibinfo{volume}{60} (\bibinfo{year}{2016}), \bibinfo{pages}{102--106}.
\newblock


\bibitem[H{\"o}dl and Myrach(2023)]%
        {hodl2023content}
\bibfield{author}{\bibinfo{person}{Tatjana H{\"o}dl} {and} \bibinfo{person}{Thomas Myrach}.} \bibinfo{year}{2023}\natexlab{}.
\newblock \showarticletitle{Content Creators Between Platform Control and User Autonomy: The Role of Algorithms and Revenue Sharing}.
\newblock \bibinfo{journal}{\emph{Business \& Information Systems Engineering}} (\bibinfo{year}{2023}), \bibinfo{pages}{1--23}.
\newblock


\bibitem[Jhaver et~al\mbox{.}(2022)]%
        {jhaver2022designing}
\bibfield{author}{\bibinfo{person}{Shagun Jhaver}, \bibinfo{person}{Quan~Ze Chen}, \bibinfo{person}{Detlef Knauss}, {and} \bibinfo{person}{Amy~X Zhang}.} \bibinfo{year}{2022}\natexlab{}.
\newblock \showarticletitle{Designing word filter tools for creator-led comment moderation}. In \bibinfo{booktitle}{\emph{Proceedings of the 2022 CHI conference on human factors in computing systems}}. \bibinfo{pages}{1--21}.
\newblock


\bibitem[Jiang et~al\mbox{.}({[n.\,d.]})]%
        {jiangPersonaLLMInvestigatingAbility2024a}
\bibfield{author}{\bibinfo{person}{Hang Jiang}, \bibinfo{person}{Xiajie Zhang}, \bibinfo{person}{Xubo Cao}, \bibinfo{person}{Cynthia Breazeal}, \bibinfo{person}{Jad Kabbara}, {and} \bibinfo{person}{Deb Roy}.} \bibinfo{year}{[n.\,d.]}\natexlab{}.
\newblock \bibinfo{title}{{PersonaLLM}: Investigating the Ability of Large Language Models to Express Personality Traits}.
\newblock
\newblock
\showeprint[arxiv]{2305.02547 [cs]}
\urldef\tempurl%
\url{http://arxiv.org/abs/2305.02547}
\showURL{%
\tempurl}


\bibitem[Jin et~al\mbox{.}(2024)]%
        {jin2024teach}
\bibfield{author}{\bibinfo{person}{Hyoungwook Jin}, \bibinfo{person}{Seonghee Lee}, \bibinfo{person}{Hyungyu Shin}, {and} \bibinfo{person}{Juho Kim}.} \bibinfo{year}{2024}\natexlab{}.
\newblock \showarticletitle{Teach AI How to Code: Using Large Language Models as Teachable Agents for Programming Education}. In \bibinfo{booktitle}{\emph{Proceedings of the CHI Conference on Human Factors in Computing Systems}} (Honolulu, HI, USA) \emph{(\bibinfo{series}{CHI '24})}. \bibinfo{publisher}{Association for Computing Machinery}, \bibinfo{address}{New York, NY, USA}, Article \bibinfo{articleno}{652}, \bibinfo{numpages}{28}~pages.
\newblock
\showISBNx{9798400703300}
\urldef\tempurl%
\url{https://doi.org/10.1145/3613904.3642349}
\showDOI{\tempurl}


\bibitem[Kim et~al\mbox{.}(2017)]%
        {kim2017mosaic}
\bibfield{author}{\bibinfo{person}{Joy Kim}, \bibinfo{person}{Maneesh Agrawala}, {and} \bibinfo{person}{Michael~S Bernstein}.} \bibinfo{year}{2017}\natexlab{}.
\newblock \showarticletitle{Mosaic: designing online creative communities for sharing works-in-progress}. In \bibinfo{booktitle}{\emph{Proceedings of the 2017 ACM conference on computer supported cooperative work and social computing}}. \bibinfo{pages}{246--258}.
\newblock


\bibitem[Lee et~al\mbox{.}(2024)]%
        {lee2024aligning}
\bibfield{author}{\bibinfo{person}{Seongyun Lee}, \bibinfo{person}{Sue~Hyun Park}, \bibinfo{person}{Seungone Kim}, {and} \bibinfo{person}{Minjoon Seo}.} \bibinfo{year}{2024}\natexlab{}.
\newblock \showarticletitle{Aligning to thousands of preferences via system message generalization}.
\newblock \bibinfo{journal}{\emph{arXiv preprint arXiv:2405.17977}} (\bibinfo{year}{2024}).
\newblock


\bibitem[Li and Peng(2021)]%
        {li2021drives}
\bibfield{author}{\bibinfo{person}{Yi Li} {and} \bibinfo{person}{Yi Peng}.} \bibinfo{year}{2021}\natexlab{}.
\newblock \showarticletitle{What drives gift-giving intention in live streaming? The perspectives of emotional attachment and flow experience}.
\newblock \bibinfo{journal}{\emph{International Journal of Human--Computer Interaction}} \bibinfo{volume}{37}, \bibinfo{number}{14} (\bibinfo{year}{2021}), \bibinfo{pages}{1317--1329}.
\newblock


\bibitem[Litt(2012)]%
        {litt2012knock}
\bibfield{author}{\bibinfo{person}{Eden Litt}.} \bibinfo{year}{2012}\natexlab{}.
\newblock \showarticletitle{Knock, knock. Who's there? The imagined audience}.
\newblock \bibinfo{journal}{\emph{Journal of broadcasting \& electronic media}} \bibinfo{volume}{56}, \bibinfo{number}{3} (\bibinfo{year}{2012}), \bibinfo{pages}{330--345}.
\newblock


\bibitem[Luo et~al\mbox{.}(2020)]%
        {luo2020emotional}
\bibfield{author}{\bibinfo{person}{Mufan Luo}, \bibinfo{person}{Tiffany~W Hsu}, \bibinfo{person}{Joon~Sung Park}, {and} \bibinfo{person}{Jeffrey~T Hancock}.} \bibinfo{year}{2020}\natexlab{}.
\newblock \showarticletitle{Emotional amplification during live-streaming: Evidence from comments during and after news events}.
\newblock \bibinfo{journal}{\emph{Proceedings of the ACM on human-computer interaction}} \bibinfo{volume}{4}, \bibinfo{number}{CSCW1} (\bibinfo{year}{2020}), \bibinfo{pages}{1--19}.
\newblock


\bibitem[Ma et~al\mbox{.}(2023a)]%
        {ma-etal-2023-insightpilot}
\bibfield{author}{\bibinfo{person}{Pingchuan Ma}, \bibinfo{person}{Rui Ding}, \bibinfo{person}{Shuai Wang}, \bibinfo{person}{Shi Han}, {and} \bibinfo{person}{Dongmei Zhang}.} \bibinfo{year}{2023}\natexlab{a}.
\newblock \showarticletitle{{I}nsight{P}ilot: An {LLM}-Empowered Automated Data Exploration System}. In \bibinfo{booktitle}{\emph{Proceedings of the 2023 Conference on Empirical Methods in Natural Language Processing: System Demonstrations}}, \bibfield{editor}{\bibinfo{person}{Yansong Feng} {and} \bibinfo{person}{Els Lefever}} (Eds.). \bibinfo{publisher}{Association for Computational Linguistics}, \bibinfo{address}{Singapore}, \bibinfo{pages}{346--352}.
\newblock
\urldef\tempurl%
\url{https://doi.org/10.18653/v1/2023.emnlp-demo.31}
\showDOI{\tempurl}


\bibitem[Ma et~al\mbox{.}(2023b)]%
        {ma2023multi}
\bibfield{author}{\bibinfo{person}{Renkai Ma}, \bibinfo{person}{Xinning Gui}, {and} \bibinfo{person}{Yubo Kou}.} \bibinfo{year}{2023}\natexlab{b}.
\newblock \showarticletitle{Multi-Platform Content Creation: The Configuration of Creator Ecology through Platform Prioritization, Content Synchronization, and Audience Management}. In \bibinfo{booktitle}{\emph{Proceedings of the 2023 CHI Conference on Human Factors in Computing Systems}}. \bibinfo{pages}{1--19}.
\newblock


\bibitem[Mallari et~al\mbox{.}(2021)]%
        {mallari2021understanding}
\bibfield{author}{\bibinfo{person}{Keri Mallari}, \bibinfo{person}{Spencer Williams}, {and} \bibinfo{person}{Gary Hsieh}.} \bibinfo{year}{2021}\natexlab{}.
\newblock \showarticletitle{Understanding analytics needs of video game streamers}. In \bibinfo{booktitle}{\emph{Proceedings of the 2021 CHI Conference on Human Factors in Computing Systems}}. \bibinfo{pages}{1--12}.
\newblock


\bibitem[Markel et~al\mbox{.}(2023)]%
        {markel2023gpteach}
\bibfield{author}{\bibinfo{person}{Julia~M Markel}, \bibinfo{person}{Steven~G Opferman}, \bibinfo{person}{James~A Landay}, {and} \bibinfo{person}{Chris Piech}.} \bibinfo{year}{2023}\natexlab{}.
\newblock \showarticletitle{GPTeach: Interactive TA Training with GPT Based Students}.
\newblock  (\bibinfo{year}{2023}).
\newblock


\bibitem[Marsden and Haag(2016)]%
        {marsden2016stereotypes}
\bibfield{author}{\bibinfo{person}{Nicola Marsden} {and} \bibinfo{person}{Maren Haag}.} \bibinfo{year}{2016}\natexlab{}.
\newblock \showarticletitle{Stereotypes and politics: reflections on personas}. In \bibinfo{booktitle}{\emph{Proceedings of the 2016 CHI conference on human factors in computing systems}}. \bibinfo{pages}{4017--4031}.
\newblock


\bibitem[McGinn and Kotamraju(2008)]%
        {mcginn2008chi}
\bibfield{author}{\bibinfo{person}{Jennifer~(Jen) McGinn} {and} \bibinfo{person}{Nalini Kotamraju}.} \bibinfo{year}{2008}\natexlab{}.
\newblock \showarticletitle{Data-driven persona development}. In \bibinfo{booktitle}{\emph{Proceedings of the SIGCHI Conference on Human Factors in Computing Systems}} (Florence, Italy) \emph{(\bibinfo{series}{CHI '08})}. \bibinfo{publisher}{Association for Computing Machinery}, \bibinfo{address}{New York, NY, USA}, \bibinfo{pages}{1521–1524}.
\newblock
\showISBNx{9781605580111}
\urldef\tempurl%
\url{https://doi.org/10.1145/1357054.1357292}
\showDOI{\tempurl}


\bibitem[McRoberts et~al\mbox{.}(2016)]%
        {mcroberts2016viewers}
\bibfield{author}{\bibinfo{person}{Sarah McRoberts}, \bibinfo{person}{Elizabeth Bonsignore}, \bibinfo{person}{Tamara Peyton}, {and} \bibinfo{person}{Svetlana Yarosh}.} \bibinfo{year}{2016}\natexlab{}.
\newblock \showarticletitle{Do it for the viewers! Audience engagement behaviors of young YouTubers}. In \bibinfo{booktitle}{\emph{Proceedings of the The 15th International Conference on Interaction Design and Children}}. \bibinfo{pages}{334--343}.
\newblock


\bibitem[Morgenstern(2012)]%
        {Morgenstern_2012}
\bibfield{author}{\bibinfo{person}{Erin Morgenstern}.} \bibinfo{year}{2012}\natexlab{}.
\newblock \bibinfo{booktitle}{\emph{The Night Circus}}.
\newblock \bibinfo{publisher}{Vintage}, \bibinfo{address}{London, England}.
\newblock
\showISBNx{9780099554790}


\bibitem[{\O}rmen and Gregersen(2023)]%
        {ormen2023towards}
\bibfield{author}{\bibinfo{person}{Jacob {\O}rmen} {and} \bibinfo{person}{Andreas Gregersen}.} \bibinfo{year}{2023}\natexlab{}.
\newblock \showarticletitle{Towards the engagement economy: interconnected processes of commodification on YouTube}.
\newblock \bibinfo{journal}{\emph{Media, Culture \& Society}} \bibinfo{volume}{45}, \bibinfo{number}{2} (\bibinfo{year}{2023}), \bibinfo{pages}{225--245}.
\newblock


\bibitem[Park and Choi(2023)]%
        {park2023audilens}
\bibfield{author}{\bibinfo{person}{Jeongeon Park} {and} \bibinfo{person}{DaEun Choi}.} \bibinfo{year}{2023}\natexlab{}.
\newblock \showarticletitle{AudiLens: Configurable LLM-Generated Audiences for Public Speech Practice}. In \bibinfo{booktitle}{\emph{Adjunct Proceedings of the 36th Annual ACM Symposium on User Interface Software and Technology}}. \bibinfo{pages}{1--3}.
\newblock


\bibitem[Park et~al\mbox{.}(2023)]%
        {park2023generative}
\bibfield{author}{\bibinfo{person}{Joon~Sung Park}, \bibinfo{person}{Joseph O'Brien}, \bibinfo{person}{Carrie~Jun Cai}, \bibinfo{person}{Meredith~Ringel Morris}, \bibinfo{person}{Percy Liang}, {and} \bibinfo{person}{Michael~S Bernstein}.} \bibinfo{year}{2023}\natexlab{}.
\newblock \showarticletitle{Generative agents: Interactive simulacra of human behavior}. In \bibinfo{booktitle}{\emph{Proceedings of the 36th Annual ACM Symposium on User Interface Software and Technology}}. \bibinfo{pages}{1--22}.
\newblock


\bibitem[Park et~al\mbox{.}(2022)]%
        {park2022social}
\bibfield{author}{\bibinfo{person}{Joon~Sung Park}, \bibinfo{person}{Lindsay Popowski}, \bibinfo{person}{Carrie Cai}, \bibinfo{person}{Meredith~Ringel Morris}, \bibinfo{person}{Percy Liang}, {and} \bibinfo{person}{Michael~S Bernstein}.} \bibinfo{year}{2022}\natexlab{}.
\newblock \showarticletitle{Social simulacra: Creating populated prototypes for social computing systems}. In \bibinfo{booktitle}{\emph{Proceedings of the 35th Annual ACM Symposium on User Interface Software and Technology}}. \bibinfo{pages}{1--18}.
\newblock


\bibitem[Park et~al\mbox{.}(2024)]%
        {park2024generative}
\bibfield{author}{\bibinfo{person}{Joon~Sung Park}, \bibinfo{person}{Carolyn~Q Zou}, \bibinfo{person}{Aaron Shaw}, \bibinfo{person}{Benjamin~Mako Hill}, \bibinfo{person}{Carrie Cai}, \bibinfo{person}{Meredith~Ringel Morris}, \bibinfo{person}{Robb Willer}, \bibinfo{person}{Percy Liang}, {and} \bibinfo{person}{Michael~S Bernstein}.} \bibinfo{year}{2024}\natexlab{}.
\newblock \showarticletitle{Generative agent simulations of 1,000 people}.
\newblock \bibinfo{journal}{\emph{arXiv preprint arXiv:2411.10109}} (\bibinfo{year}{2024}).
\newblock


\bibitem[Rossi and Rubera(2021)]%
        {rossi2021measuring}
\bibfield{author}{\bibinfo{person}{Federico Rossi} {and} \bibinfo{person}{Gaia Rubera}.} \bibinfo{year}{2021}\natexlab{}.
\newblock \showarticletitle{Measuring competition for attention in social media: National women’s soccer league players on twitter}.
\newblock \bibinfo{journal}{\emph{Marketing Science}} \bibinfo{volume}{40}, \bibinfo{number}{6} (\bibinfo{year}{2021}), \bibinfo{pages}{1147--1168}.
\newblock


\bibitem[Saikia et~al\mbox{.}(2023)]%
        {saikia2023unveiling}
\bibfield{author}{\bibinfo{person}{Krishi~Pallab Saikia}, \bibinfo{person}{Debdeep Mukherjee}, \bibinfo{person}{Saranik Mahapatra}, \bibinfo{person}{Prasun Nandy}, {and} \bibinfo{person}{Rik Das}.} \bibinfo{year}{2023}\natexlab{}.
\newblock \showarticletitle{Unveiling deeper petrochemical insights: navigating contextual question answering with the power of semantic search and LLM fine-tuning}. In \bibinfo{booktitle}{\emph{2023 International Conference on Computing, Communication, and Intelligent Systems (ICCCIS)}}. IEEE, \bibinfo{pages}{881--886}.
\newblock


\bibitem[Salminen et~al\mbox{.}(2020a)]%
        {salminen2020literature}
\bibfield{author}{\bibinfo{person}{Joni Salminen}, \bibinfo{person}{Kathleen Guan}, \bibinfo{person}{Soon-gyo Jung}, \bibinfo{person}{Shammur~A Chowdhury}, {and} \bibinfo{person}{Bernard~J Jansen}.} \bibinfo{year}{2020}\natexlab{a}.
\newblock \showarticletitle{A literature review of quantitative persona creation}. In \bibinfo{booktitle}{\emph{Proceedings of the 2020 CHI Conference on Human Factors in Computing Systems}}. \bibinfo{pages}{1--14}.
\newblock


\bibitem[Salminen et~al\mbox{.}(2020b)]%
        {salminen2020template}
\bibfield{author}{\bibinfo{person}{Joni Salminen}, \bibinfo{person}{Kathleen Guan}, \bibinfo{person}{Lene Nielsen}, \bibinfo{person}{Soon-gyo Jung}, {and} \bibinfo{person}{Bernard~J Jansen}.} \bibinfo{year}{2020}\natexlab{b}.
\newblock \showarticletitle{A template for data-driven personas: analyzing 31 quantitatively oriented persona profiles}. In \bibinfo{booktitle}{\emph{Human Interface and the Management of Information. Designing Information: Thematic Area, HIMI 2020, Held as Part of the 22nd International Conference, HCII 2020, Copenhagen, Denmark, July 19--24, 2020, Proceedings, Part I 22}}. Springer, \bibinfo{pages}{125--144}.
\newblock


\bibitem[Salminen et~al\mbox{.}(2018)]%
        {salminen2018personas}
\bibfield{author}{\bibinfo{person}{Joni Salminen}, \bibinfo{person}{Bernard~J Jansen}, \bibinfo{person}{Jisun An}, \bibinfo{person}{Haewoon Kwak}, {and} \bibinfo{person}{Soon-gyo Jung}.} \bibinfo{year}{2018}\natexlab{}.
\newblock \showarticletitle{Are personas done? Evaluating their usefulness in the age of digital analytics}.
\newblock \bibinfo{journal}{\emph{Persona Studies}} \bibinfo{volume}{4}, \bibinfo{number}{2} (\bibinfo{year}{2018}), \bibinfo{pages}{47--65}.
\newblock


\bibitem[Salminen et~al\mbox{.}(2023)]%
        {salminen2023can}
\bibfield{author}{\bibinfo{person}{Joni Salminen}, \bibinfo{person}{Soon-gyo Jung}, \bibinfo{person}{Hind Almerekhi}, \bibinfo{person}{Erik Cambria}, {and} \bibinfo{person}{Bernard Jansen}.} \bibinfo{year}{2023}\natexlab{}.
\newblock \showarticletitle{How Can Natural Language Processing and Generative AI Address Grand Challenges of Quantitative User Personas?}. In \bibinfo{booktitle}{\emph{International Conference on Human-Computer Interaction}}. Springer, \bibinfo{pages}{211--231}.
\newblock


\bibitem[Salminen et~al\mbox{.}(2019)]%
        {salminen2019detecting}
\bibfield{author}{\bibinfo{person}{Joni Salminen}, \bibinfo{person}{Soon-Gyo Jung}, {and} \bibinfo{person}{Bernard~J Jansen}.} \bibinfo{year}{2019}\natexlab{}.
\newblock \showarticletitle{Detecting demographic bias in automatically generated personas}. In \bibinfo{booktitle}{\emph{Extended Abstracts of the 2019 CHI Conference on Human Factors in Computing Systems}}. \bibinfo{pages}{1--6}.
\newblock


\bibitem[Salminen et~al\mbox{.}(2020c)]%
        {salminen2020persona}
\bibfield{author}{\bibinfo{person}{Joni Salminen}, \bibinfo{person}{Joao~M Santos}, \bibinfo{person}{Haewoon Kwak}, \bibinfo{person}{Jisun An}, \bibinfo{person}{Soon-gyo Jung}, {and} \bibinfo{person}{Bernard~J Jansen}.} \bibinfo{year}{2020}\natexlab{c}.
\newblock \showarticletitle{Persona perception scale: development and exploratory validation of an instrument for evaluating individuals’ perceptions of personas}.
\newblock \bibinfo{journal}{\emph{International Journal of Human-Computer Studies}}  \bibinfo{volume}{141} (\bibinfo{year}{2020}), \bibinfo{pages}{102437}.
\newblock


\bibitem[Shao et~al\mbox{.}(2023)]%
        {shao2023character}
\bibfield{author}{\bibinfo{person}{Yunfan Shao}, \bibinfo{person}{Linyang Li}, \bibinfo{person}{Junqi Dai}, {and} \bibinfo{person}{Xipeng Qiu}.} \bibinfo{year}{2023}\natexlab{}.
\newblock \showarticletitle{Character-llm: A trainable agent for role-playing}.
\newblock \bibinfo{journal}{\emph{arXiv preprint arXiv:2310.10158}} (\bibinfo{year}{2023}).
\newblock


\bibitem[Shin et~al\mbox{.}(2024)]%
        {Shin2024dis}
\bibfield{author}{\bibinfo{person}{Joongi Shin}, \bibinfo{person}{Michael~A. Hedderich}, \bibinfo{person}{Bart\l{}omiej~Jakub Rey}, \bibinfo{person}{Andr\'{e}s Lucero}, {and} \bibinfo{person}{Antti Oulasvirta}.} \bibinfo{year}{2024}\natexlab{}.
\newblock \showarticletitle{Understanding Human-AI Workflows for Generating Personas}. In \bibinfo{booktitle}{\emph{Proceedings of the 2024 ACM Designing Interactive Systems Conference}} (Copenhagen, Denmark) \emph{(\bibinfo{series}{DIS '24})}. \bibinfo{publisher}{Association for Computing Machinery}, \bibinfo{address}{New York, NY, USA}, \bibinfo{pages}{757–781}.
\newblock
\showISBNx{9798400705830}
\urldef\tempurl%
\url{https://doi.org/10.1145/3643834.3660729}
\showDOI{\tempurl}


\bibitem[Suh et~al\mbox{.}(2023)]%
        {suh2023structured}
\bibfield{author}{\bibinfo{person}{Sangho Suh}, \bibinfo{person}{Meng Chen}, \bibinfo{person}{Bryan Min}, \bibinfo{person}{Toby Jia-Jun Li}, {and} \bibinfo{person}{Haijun Xia}.} \bibinfo{year}{2023}\natexlab{}.
\newblock \showarticletitle{Structured Generation and Exploration of Design Space with Large Language Models for Human-AI Co-Creation}.
\newblock \bibinfo{journal}{\emph{arXiv preprint arXiv:2310.12953}} (\bibinfo{year}{2023}).
\newblock


\bibitem[Sun and Li(2024)]%
        {sun2024influence}
\bibfield{author}{\bibinfo{person}{Dan Sun} {and} \bibinfo{person}{Yiping Li}.} \bibinfo{year}{2024}\natexlab{}.
\newblock \showarticletitle{Influence of Strategic Crisis Communication on Public Perceptions during Public Health Crises: Insights from YouTube Chinese Media}.
\newblock \bibinfo{journal}{\emph{Behavioral Sciences}} \bibinfo{volume}{14}, \bibinfo{number}{2} (\bibinfo{year}{2024}), \bibinfo{pages}{91}.
\newblock


\bibitem[Turner and Turner(2011)]%
        {turner2011stereotyping}
\bibfield{author}{\bibinfo{person}{Phil Turner} {and} \bibinfo{person}{Susan Turner}.} \bibinfo{year}{2011}\natexlab{}.
\newblock \showarticletitle{Is stereotyping inevitable when designing with personas?}
\newblock \bibinfo{journal}{\emph{Design studies}} \bibinfo{volume}{32}, \bibinfo{number}{1} (\bibinfo{year}{2011}), \bibinfo{pages}{30--44}.
\newblock


\bibitem[Wei et~al\mbox{.}(2023)]%
        {wei2023leveraging}
\bibfield{author}{\bibinfo{person}{Jing Wei}, \bibinfo{person}{Sungdong Kim}, \bibinfo{person}{Hyunhoon Jung}, {and} \bibinfo{person}{Young-Ho Kim}.} \bibinfo{year}{2023}\natexlab{}.
\newblock \showarticletitle{Leveraging large language models to power chatbots for collecting user self-reported data}.
\newblock \bibinfo{journal}{\emph{arXiv preprint arXiv:2301.05843}} (\bibinfo{year}{2023}).
\newblock


\bibitem[Wei et~al\mbox{.}(2022)]%
        {wei2022chain}
\bibfield{author}{\bibinfo{person}{Jason Wei}, \bibinfo{person}{Xuezhi Wang}, \bibinfo{person}{Dale Schuurmans}, \bibinfo{person}{Maarten Bosma}, \bibinfo{person}{Fei Xia}, \bibinfo{person}{Ed Chi}, \bibinfo{person}{Quoc~V Le}, \bibinfo{person}{Denny Zhou}, {et~al\mbox{.}}} \bibinfo{year}{2022}\natexlab{}.
\newblock \showarticletitle{Chain-of-thought prompting elicits reasoning in large language models}.
\newblock \bibinfo{journal}{\emph{Advances in neural information processing systems}}  \bibinfo{volume}{35} (\bibinfo{year}{2022}), \bibinfo{pages}{24824--24837}.
\newblock


\bibitem[Wilcoxon et~al\mbox{.}(1970)]%
        {wilcoxon1970critical}
\bibfield{author}{\bibinfo{person}{Frank Wilcoxon}, \bibinfo{person}{S Katti}, \bibinfo{person}{Roberta~A Wilcox}, {et~al\mbox{.}}} \bibinfo{year}{1970}\natexlab{}.
\newblock \showarticletitle{Critical values and probability levels for the Wilcoxon rank sum test and the Wilcoxon signed rank test}.
\newblock \bibinfo{journal}{\emph{Selected tables in mathematical statistics}}  \bibinfo{volume}{1} (\bibinfo{year}{1970}), \bibinfo{pages}{171--259}.
\newblock


\bibitem[Wohn et~al\mbox{.}(2018)]%
        {Wohn2018}
\bibfield{author}{\bibinfo{person}{Donghee~Yvette Wohn}, \bibinfo{person}{Guo Freeman}, {and} \bibinfo{person}{Caitlin McLaughlin}.} \bibinfo{year}{2018}\natexlab{}.
\newblock \showarticletitle{Explaining Viewers' Emotional, Instrumental, and Financial Support Provision for Live Streamers}. In \bibinfo{booktitle}{\emph{Proceedings of the 2018 CHI Conference on Human Factors in Computing Systems}} \emph{(\bibinfo{series}{CHI '18})}. \bibinfo{publisher}{Association for Computing Machinery}, \bibinfo{address}{New York, NY, USA}, \bibinfo{pages}{1–13}.
\newblock
\showISBNx{9781450356206}
\urldef\tempurl%
\url{https://doi.org/10.1145/3173574.3174048}
\showDOI{\tempurl}


\bibitem[Wu et~al\mbox{.}(2019)]%
        {wu2019agent}
\bibfield{author}{\bibinfo{person}{Eva~Yiwei Wu}, \bibinfo{person}{Emily Pedersen}, {and} \bibinfo{person}{Niloufar Salehi}.} \bibinfo{year}{2019}\natexlab{}.
\newblock \showarticletitle{Agent, gatekeeper, drug dealer: How content creators craft algorithmic personas}.
\newblock \bibinfo{journal}{\emph{Proceedings of the ACM on Human-Computer Interaction}} \bibinfo{volume}{3}, \bibinfo{number}{CSCW} (\bibinfo{year}{2019}), \bibinfo{pages}{1--27}.
\newblock


\bibitem[Wu et~al\mbox{.}(2022)]%
        {wu2022ai}
\bibfield{author}{\bibinfo{person}{Tongshuang Wu}, \bibinfo{person}{Michael Terry}, {and} \bibinfo{person}{Carrie~Jun Cai}.} \bibinfo{year}{2022}\natexlab{}.
\newblock \showarticletitle{Ai chains: Transparent and controllable human-ai interaction by chaining large language model prompts}. In \bibinfo{booktitle}{\emph{Proceedings of the 2022 CHI conference on human factors in computing systems}}. \bibinfo{pages}{1--22}.
\newblock


\bibitem[Yang et~al\mbox{.}(2022a)]%
        {yang2022doc}
\bibfield{author}{\bibinfo{person}{Kevin Yang}, \bibinfo{person}{Dan Klein}, \bibinfo{person}{Nanyun Peng}, {and} \bibinfo{person}{Yuandong Tian}.} \bibinfo{year}{2022}\natexlab{a}.
\newblock \showarticletitle{Doc: Improving long story coherence with detailed outline control}.
\newblock \bibinfo{journal}{\emph{arXiv preprint arXiv:2212.10077}} (\bibinfo{year}{2022}).
\newblock


\bibitem[Yang et~al\mbox{.}(2022b)]%
        {yang2022re3}
\bibfield{author}{\bibinfo{person}{Kevin Yang}, \bibinfo{person}{Yuandong Tian}, \bibinfo{person}{Nanyun Peng}, {and} \bibinfo{person}{Dan Klein}.} \bibinfo{year}{2022}\natexlab{b}.
\newblock \showarticletitle{Re3: Generating longer stories with recursive reprompting and revision}.
\newblock \bibinfo{journal}{\emph{arXiv preprint arXiv:2210.06774}} (\bibinfo{year}{2022}).
\newblock


\bibitem[Yin and Neubig(2017)]%
        {yin2017syntactic}
\bibfield{author}{\bibinfo{person}{Pengcheng Yin} {and} \bibinfo{person}{Graham Neubig}.} \bibinfo{year}{2017}\natexlab{}.
\newblock \showarticletitle{A syntactic neural model for general-purpose code generation}.
\newblock \bibinfo{journal}{\emph{arXiv preprint arXiv:1704.01696}} (\bibinfo{year}{2017}).
\newblock


\bibitem[Zhang et~al\mbox{.}(2021)]%
        {zhang2021comments}
\bibfield{author}{\bibinfo{person}{Haoxiang Zhang}, \bibinfo{person}{Shaowei Wang}, \bibinfo{person}{Tse-Hsun Chen}, {and} \bibinfo{person}{Ahmed~E Hassan}.} \bibinfo{year}{2021}\natexlab{}.
\newblock \showarticletitle{Are comments on stack overflow well organized for easy retrieval by developers?}
\newblock \bibinfo{journal}{\emph{ACM Transactions on Software Engineering and Methodology (TOSEM)}} \bibinfo{volume}{30}, \bibinfo{number}{2} (\bibinfo{year}{2021}), \bibinfo{pages}{1--31}.
\newblock


\bibitem[Zhang et~al\mbox{.}(2023)]%
        {zhang2023heterogeneous}
\bibfield{author}{\bibinfo{person}{Zhaowei Zhang}, \bibinfo{person}{Ceyao Zhang}, \bibinfo{person}{Nian Liu}, \bibinfo{person}{Siyuan Qi}, \bibinfo{person}{Ziqi Rong}, \bibinfo{person}{Song-Chun Zhu}, \bibinfo{person}{Shuguang Cui}, {and} \bibinfo{person}{Yaodong Yang}.} \bibinfo{year}{2023}\natexlab{}.
\newblock \showarticletitle{Heterogeneous Value Alignment Evaluation for Large Language Models}.
\newblock \bibinfo{journal}{\emph{arXiv preprint arXiv:2305.17147}} (\bibinfo{year}{2023}).
\newblock


\bibitem[Zhu et~al\mbox{.}(2019)]%
        {zhu2019creating}
\bibfield{author}{\bibinfo{person}{Haining Zhu}, \bibinfo{person}{Hongjian Wang}, {and} \bibinfo{person}{John~M Carroll}.} \bibinfo{year}{2019}\natexlab{}.
\newblock \showarticletitle{Creating Persona Skeletons from Imbalanced Datasets-A Case Study using US Older Adults' Health Data}. In \bibinfo{booktitle}{\emph{Proceedings of the 2019 on designing interactive systems conference}}. \bibinfo{pages}{61--70}.
\newblock


\end{thebibliography}

\appendix
\captionsetup{width=.8\textwidth}
\clearpage
\onecolumn

\section{Formative Study}
\label{app:formative-study}

\begin{table}[ht!]
\centering
\scalebox{1}{
\begin{tabular}{ccccc}
\toprule
\textbf{\makecell[c]{Participant ID \\ (Gender, Age)}} & \textbf{\makecell[c]{Channel \\ category}} & \textbf{\makecell[c]{Start \\ date}} & \textbf{\makecell[c]{Total number of\\subscribers}} & \textbf{\makecell[c]{Level of\\commitment}} \\
\midrule
I1 (M, 20s) & Classic music & 2019.03 & 59.5k & Part-time \\
\midrule
I2 (F, 30s) & Nail arts & 2015.12 & 1790k & Part-time \\
\midrule
I3 (M, 30s) & Pop music review & 2018.02 & 16.8k & Part-time \\
\midrule
I4 (M, 30s) & Car review & 2019.05 & 61.5k & Full-time \\
\midrule
I5 (M, 30s) & Game & 2022.01 & 0.3k & Part-time \\
\midrule
I6 (M, 30s) & Couple Vlog & 2020.04 & 4.7k & Part-time \\
\midrule
I7 (M, 20s) & Travel & 2022.08 & 2.3k & Part-time \\
\midrule
I8 (F, 20s) & Beauty \& Fashion & 2016.03 & 67.9k & Full-time \\
\midrule
I9 (F, 30s) & Baking & 2020.10 & 14.1k & Full-time \\
\midrule
I10 (F, 30s) & Family & 2018.01 & 0.3k & Part-time \\
\midrule
I11 (M, 30s) & Lifestyle & 2022.05 & 30k & Part-time \\
\midrule
I12 (M, 20s) & Music Producing Tutorial & 2021.11 & 8.7k & Full-time \\
\midrule
I13 (M, 30s) & Economics & 2021.02 & 53k & Part-time \\
\bottomrule
\end{tabular}
}
\vspace*{2mm}
\caption{
Participants’ demographics, channel information, and their level of commitment at the time of formative studies.
}
\label{tab:formative_study}
\Description{This table summarizes the demographic information, channel categories, channel start dates, total number of subscribers, and commitment levels of 13 participants from the formative studies. Each row provides data for one participant, including their ID, gender, age group, channel type, and whether they are part-time or full-time creators. Notably, Participant I2 runs a nail art channel with the highest subscriber count (1.79M), while Participant I5 has the fewest subscribers (0.3k) on their gaming channel.}
\end{table}

\section{Technical Evaluation Results}
\subsection{Technical Evaluation 1: Quality of Dimensions \& Values}
\label{app:techeval1}

\begin{table}[ht!]
\centering
\scalebox{1}{
\resizebox{\linewidth}{!}{
\begin{tabular}{cccccccc}
\toprule
\textbf{\makecell[c]{Channel}} &\textbf{\makecell[c]{Topic}} &\textbf{\makecell[c]{Number of \\ Dimensions}} & \textbf{\makecell[c]{Number of \\ Values}} & \textbf{\makecell[c]{Relevance of \\ Dimensions}} & \textbf{\makecell[c]{Relevance of\\Values}} & \textbf{\makecell[c]{Number of \\Similar Dimensions}} & \textbf{\makecell[c]{Number of \\Similar Values}} \\
\midrule
Channel A & Baking & 6 & 18 & 3.72 & 3.76 & 2 / 6 & 4 / 18 \\
\midrule
Channel B & Music producing tutorial & 5 & 24 & 3.07 & 3.44 & 0 / 5 & 0 / 24 \\
\midrule
Channel C & Fashion \& Make-up & 5 & 15 & 3.87 & 3.82 & 0 / 5 & 0 / 15 \\
\midrule
Channel D & Electronics & 5 & 15 & 3.93 & 3.82 & 0 / 5 & 0 / 15 \\
\midrule
Channel E & Pop music review & 4 & 17 & 4 & 3.30 & 0 / 4 & 2 / 17 \\
\midrule
Channel F & Interior & 5 & 16 & 3.47 & 3.44 & 0 / 5 & 2 / 16 \\
\bottomrule
\end{tabular}
}
}
\vspace*{2mm}
\caption{
Technical evaluation results for dimension-values generation pipeline. Three human evaluators evaluated (1) the relevance of dimensions/values (5-point likert scale), and (2) the mutual-exclusiveness within dimensions/values (0/1 binary evaluation). For the relevance, we computed the mean of each channel's dimensions and values. For the similarity (mutually-exclusiveness), we ran a majority voting evaluation and counted the overlapping dimensions and values. 
}
\label{tab:techeval1}
\Description{This table presents the technical evaluation results for the dimension-values generation pipeline across six channels, covering topics such as baking, music production tutorials, and fashion. The table includes the number of dimensions and values generated, as well as their evaluated relevance and mutual exclusiveness. Relevance is rated on a 5-point Likert scale, while mutual exclusiveness (similarity) is determined by majority voting. Notably, Channel E has the highest relevance for dimensions (4.00), while Channels B, C, and D have no overlapping dimensions or values, indicating high exclusivity.}
\end{table}

\begin{figure*}[h]
\centering
  \includegraphics[width=\textwidth]{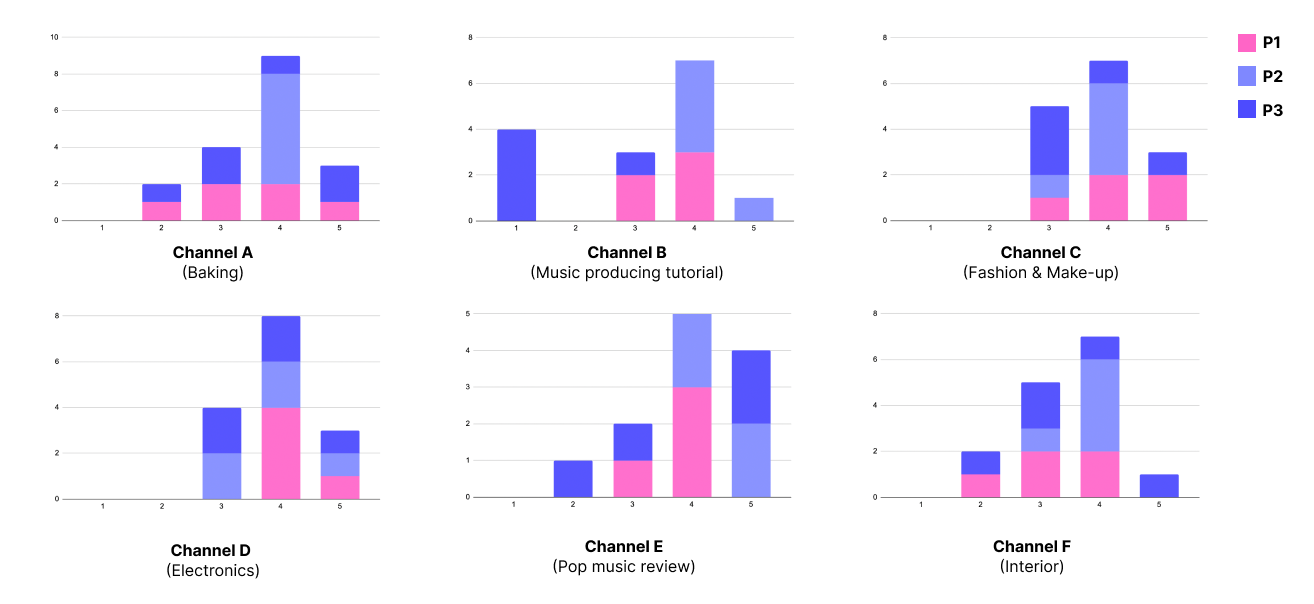}
  \caption{Technical evaluation results for dimension-values generation pipeline. The frequency distribution of relevance score of each dimension. }
  \label{fig:techeval1_dimension}
\Description{Technical evaluation results for dimension-values generation pipeline. The frequency distribution of relevance score of each dimension. Three evaluators' scores are accumulated in each channel's bar graph.}
\end{figure*}

\begin{figure*}[h]
\centering
  \includegraphics[width=\textwidth]{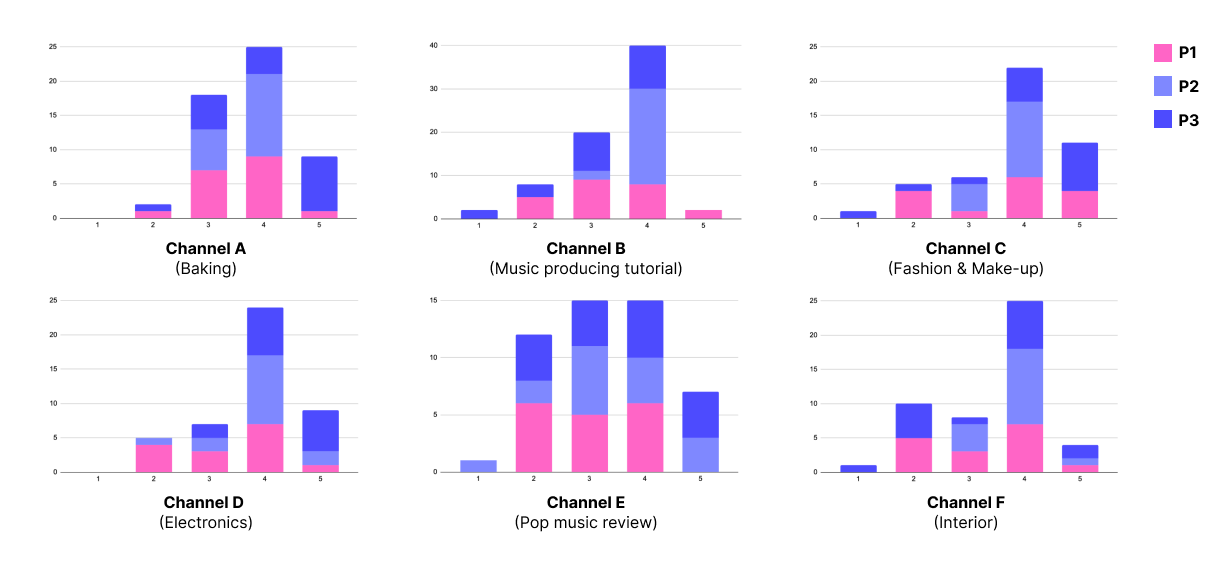}
  \caption{Technical evaluation results for dimension-values generation pipeline. The frequency distribution of relevance score of each value.}
  \label{fig:techeval1_value}
  \Description{Technical evaluation results for dimension-values generation pipeline. The frequency distribution of relevance score of each value. Three evaluators' scores are accumulated in each channel's bar graph.}
\end{figure*}

\clearpage

\subsection{Technical Evaluation 2: Quality of \sysname{} Clustering Method}
\label{app:techeval2}

\begin{figure*}[h]
\centering
\includegraphics[width=0.7\textwidth]{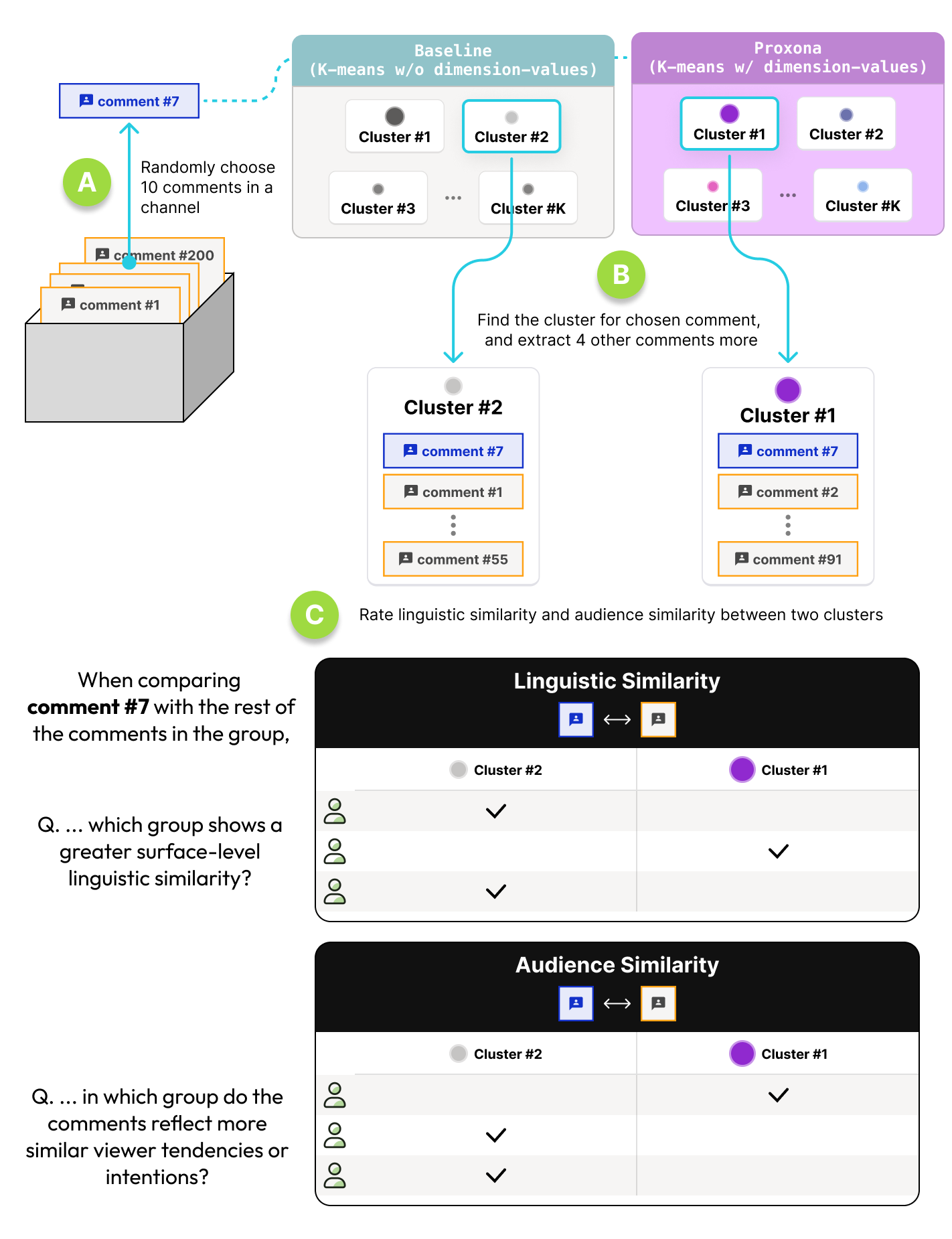}
\caption{The procedure of evaluating audience comment clustering pipeline. We compare the \baseline{} and \sysname{} approach where the data input of embedding is different--- including dimension-values information or not. We first randomly choose 10 comments in a channel. Then, we find the cluster for each chosen comment, and extract four other comments more. We asked three evaluators to rate linguistic similarity and audience similarity, by choosing superiority between two clusters.}
\label{fig:techeval2-procedure}
\Description{The procedure of evaluating audience comment clustering pipeline. We compare the \baseline{} and \sysname{} approach where the data input of embedding is different--- including dimension-values information or not. We first randomly choose 10 comments in a channel. Then, we find the cluster for each chosen comment, and extract four other comments more. We asked three evaluators to rate linguistic similarity and audience similarity, by choosing superiority between two clusters.}
\end{figure*}


\begin{table}[h]
\centering
\scalebox{1}{
\resizebox{\linewidth}{!}{
\begin{tabular}{ccccc}
\toprule
\textbf{\makecell[c]{Channel}} 
&\textbf{\makecell[c]{Topic}} 
&\textbf{\makecell[c]{Number of Superior Clusters \\for Linguistic Similarity}} 
&\textbf{\makecell[c]{Number of Superior Clusters \\for Audience Similarity}} 
&\textbf{\makecell[c]{Number of Superior Clusters \\for Both Similarity}} \\
\midrule
Channel A & Baking & 4 & 6 & 2 \\
\midrule
Channel D & Electronics & 7 & 5 & 5 \\
\midrule
Channel E & Pop music review & 7 & 6 & 6 \\
\midrule
Channel F & Interior & 6 & 7 & 6 \\
\midrule
Channel G & Diary \& Stationery & 5 & 8 & 4 \\
\midrule
Average & - & 5.8 & 6.4 & 4.6 \\
\bottomrule
\end{tabular}
}
}
\vspace*{2mm}
\caption{
Technical evaluation results for audience comment clustering pipeline. We compared \sysname{} pipeline with \baseline{}, and present the number of clusters generated by \sysname{} gaining superiority on each type of similarity.
}
\label{tab:techeval2-result}
\Description{This table displays the technical evaluation results for the audience comment clustering pipeline across five channels with topics ranging from baking to electronics. It shows the number of superior clusters generated by the system for two types of similarity: linguistic and audience similarity, as well as clusters that are superior in both categories. Notably, Channel F (Interior) and Channel E (Pop music review) have the highest number of superior clusters for both similarities, with 6 clusters each. The table also provides an average for each similarity type across all channels.}
\end{table}



\begin{table*}[h]
\resizebox{0.9\textwidth}{!}{%
\begin{tabular}{|p{1.5cm}|p{2cm}|p{6cm}|p{6cm}|}
\hline
\textbf{Group}    & \textbf{Comment} & \textbf{Value set of clustered group of Proxona}                         & \textbf{Value set of clustered group of Baseline}                                           \\ \hline
Comment group \#4 & Comment \#1      & \textbf{Spec Enthusiast}, Feedback Provider, Brand Critic                & Spec Enthusiast, Feedback Provider, Brand Critic                                            \\
                  & Comment \#2      & \textbf{Spec Enthusiast}, Everyday User, Feedback Provider, Brand Critic & \textbf{Brand Analyst}, Quality Investor, Information Seeker, Brand Critic                  \\
                  & Comment \#3      & \textbf{Spec Enthusiast}, Value Seeker, Hobbyist, Information Seeker     & \textbf{Brand Analyst}, Quality Investor, Everyday User, Information Seeker, Brand Agnostic \\
                  & Comment \#4      & \textbf{Spec Enthusiast}, Feedback Provider, Brand Critic                & \textbf{Brand Analyst}, Everyday User, Community Participant, Brand Loyalist                \\
                  & Comment \#5      & \textbf{Spec Enthusiast}, Hobbyist, Information Seeker                   & \textbf{Brand Analyst}, Everyday User, Brand Loyalist                                       \\ \hline
Comment group \#1 & Comment \#1      & \textbf{Spec Enthusiast,} Feedback Provider, Brand Critic                         & Spec Enthusiast, Everyday User, Feedback Provider, Brand Critic                             \\
                  & Comment \#2      & \textbf{Spec Enthusiast}, Everyday User, Feedback Provider, Brand Critic          & Community Participant                                                                       \\
                  & Comment \#3      & \textbf{Spec Enthusiast}, Value Seeker, Hobbyist, Information Seeker              & Information Seeker                                                                          \\
                  & Comment \#4      & \textbf{Spec Enthusiast}, Feedback Provider, Brand Critic                         & Everyday User, Information Seeker, Brand Agnostic                                           \\
                  & Comment \#5      & \textbf{Spec Enthusiast}, Hobbyist, Information Seeker                            & Brand Analyst, Quality Investor, Everyday User, Information Seeker, Brand Loyalist          \\ \hline
\end{tabular}%
}
\caption{Example of dataset for technical evaluation. This data represents two out of ten comment groups of Channel D. It shows the inferred combination of values of each comment within the group. The highlighted values indicate the crucial values that are frequently observed within each group. In comment group 4, a specific value is observed across multiple comments in both conditions (\sysname{}, \baseline{}). However, in comment group 1, the common value is observed only in cases clustered by \sysname{}.} 
\label{tab:techeval2_ex}
\Description{This table presents an example dataset used for the technical evaluation of the Proxona and Baseline clustering methods. It includes two comment groups from Channel D and shows the value sets inferred for each comment by both systems. The highlighted values indicate the most crucial or frequently observed values within each group. For instance, in comment group #4, the value "Spec Enthusiast" is consistently identified across multiple comments by both Proxona and Baseline, while in comment group #1, this consistency is only seen in Proxona's clustering, highlighting its distinct clustering behavior.}
\end{table*}

\clearpage

\subsection{Technical Evaluation 3: Quality of Personas}
\label{app:techeval3}

\begin{table}[ht]
\centering
\resizebox{\textwidth}{!}{
\begin{tabular}{ccccccccccc}
\toprule
\multicolumn{7}{c}{Channel (\# of Evaluators)} & & \multicolumn{2}{c}{Aggregated} \\
\cmidrule{1-7} \cmidrule{9-10}
Channel 1 (5) & Channel 2 (4) & Channel 3 (5) & Channel 4 (5) & Channel 5 (5) & Channel 6 (5) & Channel 7 (5) & & \sysname{} & \baseline{} \\
\midrule
\multicolumn{10}{l}{\textit{1. (PPS - Completeness) Persona profile is detailed enough to make decisions about the audience it describes.}}\\
\cellcolor{yellow!50}\textbf{3/5} & \cellcolor{yellow!50}\textbf{4/5} & \cellcolor{yellow!50}\textbf{4/5} & \cellcolor{yellow!50}\textbf{4/5} & \cellcolor{yellow!50}\textbf{4/5} & \cellcolor{yellow!50}\textbf{4/5} & \cellcolor{yellow!50}\textbf{4/5} & & \cellcolor{yellow!50}\textbf{7} & 0 \\
\midrule
\multicolumn{10}{l}{\textit{2. (PPS - Credibility) Personas seem like real people. }}\\
2/5 & 1/5 & 2/5 & \cellcolor{yellow!50}\textbf{3/5} & 2/5 & 2/5 & \cellcolor{yellow!50}\textbf{3/5} & & 2 & 5 \\
\midrule
\multicolumn{10}{l}{\textit{3. (PPS - Clarity) The information about the personas is well presented. }}\\
\cellcolor{yellow!50}\textbf{3/5} & \cellcolor{yellow!50}\textbf{4/5} & \cellcolor{yellow!50}\textbf{5/5} & \cellcolor{yellow!50}\textbf{4/5} & \cellcolor{yellow!50}\textbf{5/5} & \cellcolor{yellow!50}\textbf{3/5} & \cellcolor{yellow!50}\textbf{3/5} & & \cellcolor{yellow!50}\textbf{7} & 0 \\
\midrule
\multicolumn{10}{l}{\textit{4. (PPS - Empathy) I feel like I understand these personas.}}\\
2/5 & \cellcolor{yellow!50}\textbf{3/5} & \cellcolor{yellow!50}\textbf{3/5} & \cellcolor{yellow!50}\textbf{3/5} & \cellcolor{yellow!50}\textbf{3/5} & 1/5 & \cellcolor{yellow!50}\textbf{3/5} & & \cellcolor{yellow!50}\textbf{5} & 2 \\
\midrule
\multicolumn{10}{l}{\textit{5. (Diversity) Personas are diverse and distinct from each other. }}\\
\cellcolor{yellow!50}\textbf{5/5} & \cellcolor{yellow!50}\textbf{5/5} & \cellcolor{yellow!50}\textbf{3/5} & \cellcolor{yellow!50}\textbf{3/5} & \cellcolor{yellow!50}\textbf{3/5} & 1/5 & \cellcolor{yellow!50}\textbf{4/5} & & \cellcolor{yellow!50}\textbf{6} & 1 \\
\midrule
\multicolumn{10}{l}{\textit{6. (Serendipity) Personas are more surprising and novel compared to just reading comments. }}\\
\cellcolor{yellow!50}\textbf{5/5} & \cellcolor{yellow!50}\textbf{5/5} & \cellcolor{yellow!50}\textbf{5/5} & \cellcolor{yellow!50}\textbf{4/5} & \cellcolor{yellow!50}\textbf{4/5} & 2/5 & \cellcolor{yellow!50}\textbf{4/5} & & \cellcolor{yellow!50}\textbf{6} & 1 \\
\midrule
\multicolumn{10}{l}{\textit{7. (Predictability) Personas were largely predictable or similar to my imagination (R).}}\\
1/5 & 1/5 & 1/5 & \cellcolor{yellow!50}\textbf{3/5} & \cellcolor{yellow!50}\textbf{3/5} & \cellcolor{yellow!50}\textbf{4/5} & 1/5 & & \cellcolor{yellow!50}\textbf{3*} & 4 \\
\midrule
\multicolumn{10}{l}{\textit{8. (Multi-dimensionality) Personas have multidimensional characteristics.}} \\
\cellcolor{yellow!50}\textbf{5/5} & \cellcolor{yellow!50}\textbf{5/5} & \cellcolor{yellow!50}\textbf{5/5} & \cellcolor{yellow!50}\textbf{3/5} & 2/5 & 1/5 & \cellcolor{yellow!50}\textbf{5/5} & & \cellcolor{yellow!50}\textbf{4} & 3 \\
\midrule
\multicolumn{10}{l}{\textit{9. (Comprehension) Personas are easy to compare. }}\\
2/5 & 2/5 & \cellcolor{yellow!50}\textbf{4/5} & 1/5 & 1/5 & 1/5 & \cellcolor{yellow!50}\textbf{5/5} & & 2 & 5 \\
\midrule
\multicolumn{10}{l}{\textit{10. (Generalizability) The personas could be applied universally, regardless of the specific channel (R).}}\\
1/5 & 1/5 & 1/5 & 2/5 & 2/5 & \cellcolor{yellow!50}\textbf{4/5} & 1/5 & & \cellcolor{yellow!50}\textbf{1*} & 6 \\
\bottomrule
\end{tabular}
}
\vspace*{2mm}
\caption{
The majority voting results from participants’ comparative evaluations across different channels and aggregated scores for each question. * stands for the results derived from reverse questions (Q7, Q10).
}
\label{tab:techeval3}
\Description{This table presents the results of majority voting from participants' comparative evaluations across different channels for Proxona and the baseline system. The table covers 10 evaluation criteria, such as completeness, credibility, empathy, diversity, and multidimensionality, with scores given on a scale of 1 to 5 by multiple evaluators for each channel. The highlighted cells indicate higher scores achieved by Proxona across different channels. Additionally, aggregated scores for Proxona and Baseline are shown for each evaluation criterion, with Proxona generally performing better in most areas. Reverse questions (Q7 and Q10) are marked with an asterisk, where lower scores indicate better results for Proxona.}
\end{table}

\begin{table}[h]
\centering
\scalebox{1}{
\resizebox{0.7\linewidth}{!}{
\begin{tabular}{ccccc}
\toprule
\textbf{\makecell[c]{Question \#}} 
&\textbf{\makecell[c]{\sysname{} Avg. (SD)}} 
&\textbf{\makecell[c]{\baseline{} Avg. (SD)}} 
&\textbf{\makecell[c]{Significance}} \\
\midrule
Q1: PPS-Completeness & 4.06 (0.639) &  3.49 (0.818) & ** \\
\midrule
Q2: PPS-Credibility & 3.51 (0.818) & 3.89 (0.718) & {} \\
\midrule
Q3: PPS-Clarity & 4.09 (0.612) & 3.83 (0.664) & {} \\
\midrule
Q4: PPS-Empathy & 3.83 (0.707) & 3.86 (0.55) & {} \\
\midrule
Q5: Diversity & 4.06 (0.968) & 3.4 (0.881) & ** \\
\midrule
Q6: Serendipity & 3.94 (0.802) & 3.06 (0.938) & ** \\
\midrule
Q7: Predictability (R) & 3.17 (0.891) & 3.77 (0.69) & ** \\
\midrule
Q8: Multi-dimensionality & 3.74 (0.78) & 3.26 (0.78) & * \\
\midrule
Q9: Comprehension & 3.43 (0.815) & 3.4 (0.812) & {} \\
\midrule
Q10: Generalizability (R) & 3.29 (0.75) & 3.51 (0.818) & {} \\
\bottomrule
\end{tabular}
}
}
\vspace*{2mm}
\caption{
Technical evaluation results for standalone survey evaluations where participants assessed with 5-point Likert scale for each condition (\sysname{} and \baseline{}). The total mean and standard deviation score based on 35 data points (5 evaluation * 7 channels) are shown in the table. The statistical significances between conditions are shown on the last column (Wilcoxon-signed rank test).  
}
\label{tab:techeval3-descriptive}
\Description{This table summarizes the results of a standalone survey evaluation where participants rated \sysname{} and the baseline system across 10 questions on a 5-point Likert scale. The table presents the average score (Avg.) and standard deviation (SD) for each question, comparing the two systems. Significant differences between \sysname{} and the baseline are highlighted in the last column, with asterisks indicating statistical significance based on the Wilcoxon-signed rank test. Notably, \sysname{} outperforms the baseline in areas like completeness, diversity, serendipity, and predictability.}
\end{table}

\clearpage

\section{User Study}
\label{app:userstudy}

\begin{table*}[h]
\centering
\scalebox{1}{
\begin{tabular}{ccc}
\toprule
\textbf{\makecell[c]{Channel}} 
&\textbf{\makecell[c]{Task 1 topic}} 
&\textbf{\makecell[c]{Task 2 topic}} \\
\midrule
P1 
& Nespresso coffee machine 
& Nike Run Club app \\
\midrule
P2 
& Nike Run Club app 
& Massage chair \\
\midrule
P3 
& Vitamin supplements 
& Running shoes \\
\midrule
P4 
& Grocery shopping app 
& Nespresso coffee machine \\
\midrule
P5 
& Vitamin supplements 
& Nespresso coffee machine \\
\midrule
P6 
& Running shoes 
& Grocery shopping app \\
\midrule
P7 
& Nike Run Club app 
& Modern art exhibition \\
\midrule
P8 
& Nespresso coffee machine 
& Language learning app \\
\midrule
P9 
& Massage chair 
& Vitamin supplements \\
\midrule
P10 
& Language learning app 
& Running shoes \\
\midrule
P11 
& Nike Run Club app 
& Language learning app \\
\bottomrule
\end{tabular}
}
\vspace*{2mm}
\caption{Topics used for Task 1 and Task 2 in the user study across 11 participants (P1 \- P11). Each participant was assigned two different topics, including consumer products (e.g., Nespresso coffee machine, running shoes) and digital tools (e.g., Nike Run Club app, language learning app).}
\label{tab:user_study_topics}
\Description{This table lists the topics used for Task 1 and Task 2 in the user study across 11 participants (P1 to P11). Each participant was assigned two different topics, including consumer products (e.g., Nespresso coffee machine, running shoes) and digital tools (e.g., Nike Run Club app, language learning app). The table provides an overview of the task topics participants interacted with during the study, showcasing the diversity of items evaluated. For instance, P1 evaluated a Nespresso coffee machine for Task 1 and the Nike Run Club app for Task 2.}
\end{table*}

\begin{figure*}[h]
\centering
  \includegraphics[width=0.8\textwidth]{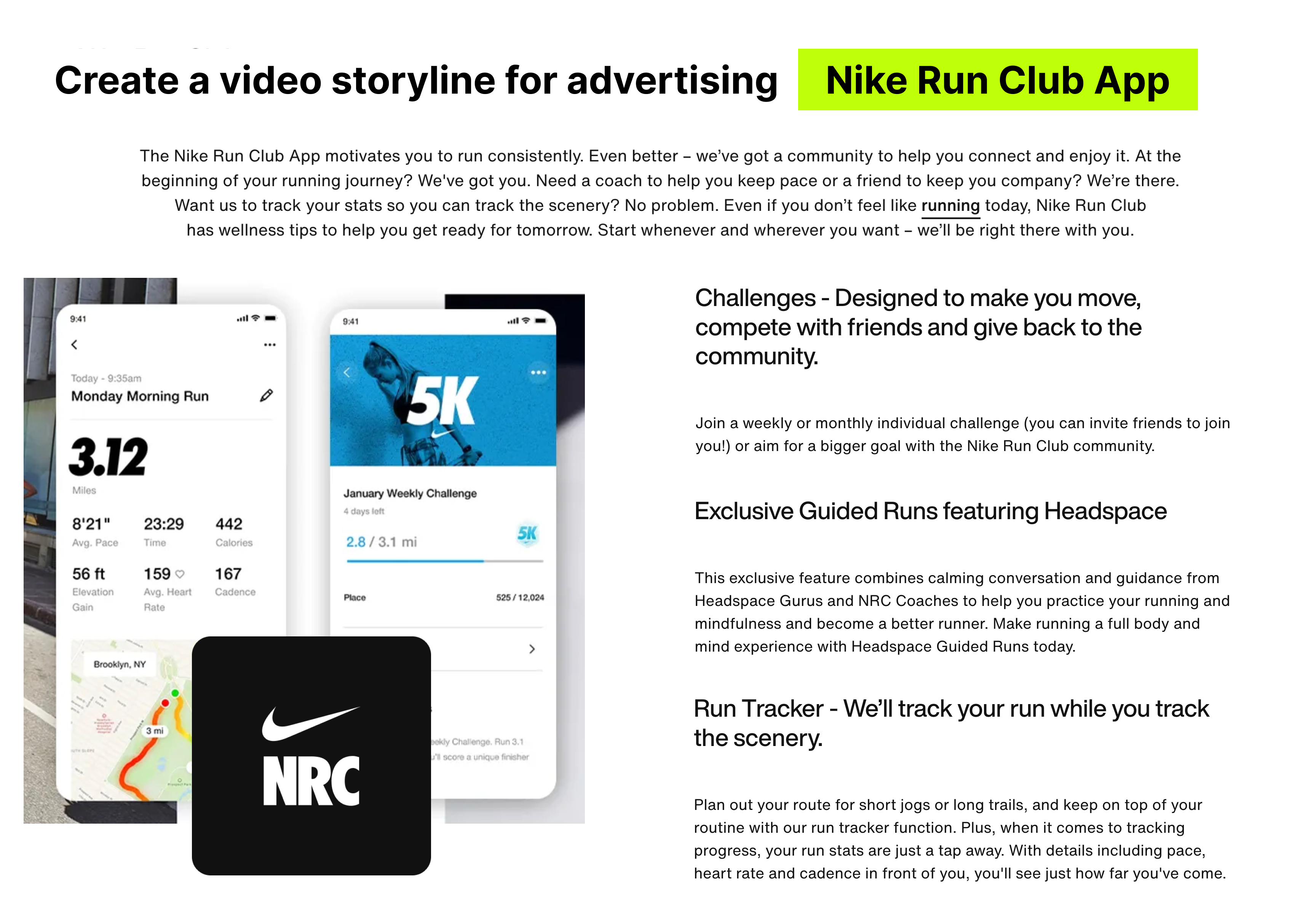}
  \caption{Example task for user study: Create a video storyline for advertising Nike Run Club application.}
  \label{fig:userstudy-task}
  \Description{This figure illustrates an example task from the user study where participants were asked to create a video storyline for advertising the Nike Run Club application. The task required participants to use system features to explore audience personas and feedback tools to inform their video content creation process.}
\end{figure*}

\begin{table*}[h]
\centering
\scalebox{1}{
\resizebox{\linewidth}{!}{
\begin{tabular}{c|ccc|ccc}
\toprule
\textbf{\makecell[c]{Channel}} 
& \textbf{\makecell[c]{Number of chat\\(Phase 1: Exploration)}} 
& \textbf{\makecell[c]{Number of chat\\(Phase 2: Creation)}} 
& \textbf{\makecell[c]{Number of chat\\(Total)}} 
& \textbf{\makecell[c]{Number of feedback\\(Suggestion)}} 
& \textbf{\makecell[c]{Number of feedback\\(Evaluation)}} 
& \textbf{\makecell[c]{Number of feedback\\(Total)}} \\
\midrule
P1 
& 5 & 0 & 5 
& 0
& 1
& 1 \\
\midrule
P2 
& 3 & 0 & 3 
& 0
& 5
& 5 \\
\midrule
P3 
& 8 & 4 & 12 
& 1
& 0
& 1 \\
\midrule
P4 
& 11 & 2 & 13 
& 2
& 1
& 3 \\
\midrule
P5 
& 3 & 0 & 3 
& 1
& 5
& 6 \\
\midrule
P6 
& 11 & 1 & 12 
& 1
& 1
& 2 \\
\midrule
P7 
& 5 & 1 & 6 
& 1
& 1
& 2 \\
\midrule
P8 
& 7 & 0 & 7 
& 0
& 0
& 0 \\
\midrule
P9 
& 9 & 1 & 10 
& 0
& 0
& 0 \\
\midrule
P10 
& 6 & 1 & 7 
& 2
& 2
& 4 \\
\midrule
P11 
& 6 & 0 & 6 
& 3
& 1
& 4 \\
\midrule
Mean (SD)
& 6.73 (2.8)	& 0.91 (1.2)	& 7.64 (3.6)	& 1.00 (1.0)	& 1.55 (1.8)	& 2.55 (2.0) \\
Min / Median / Max
& 3/6/11	&0/1/4	&3/7/13	&0/1/3	&0/1/5	&0/2/6 \\
\midrule
Total
& 74	&10	&84	&11	&17	&28 \\
\bottomrule
\end{tabular}
}
}
\vspace*{2mm}
\caption{Descriptive analysis of interaction feature usages including chat and feedback. The number of chats [left] indicates the frequency of using the conversation space in each phase, while the number of feedback instances [right] reflects the frequency of using the feedback feature in the text editor during phase 2.}
\label{tab:user_study}
\Description{This table presents a descriptive analysis of interaction feature usage, broken down into chat and feedback instances during two phases of the user study. The table records the number of chats during Phase 1 (understanding) and Phase 2 (creation), as well as the number of feedback instances categorized into suggestions and evaluations during Phase 2. Each row represents data for one participant (P1 to P11). For example, P4 had the highest number of interactions, with 13 chats in total and 3 feedback instances. The table also includes summary statistics, such as the mean, standard deviation, minimum, median, and maximum values across all participants.}
\end{table*}

\clearpage

\section{User Study Results}
\subsection{Overall Usability (7-point Likert Scale)}
\label{app:question_usability}


\begin{table}[ht!]
    \centering
    \begin{tabular}{p{0.75\columnwidth}|c |c}
    \hline
    \toprule
    Criteria & \makecell{\baseline{} \\ Avg. (SD)} &  \makecell{\sysname{} \\ Avg. (SD)} \\
    \midrule
    Q1. I was able to understand viewers sufficiently.*  & 4.73 (1.42)  & 6.09 (0.70)\\ \hline
    Q2. I was able to plan the videos based on a understanding of viewers.*  & 5.00 (1.41)  & 6.00 (0.77)\\ \hline
    Q3. I was able to clearly understand what type of content viewers enjoy.  & 5.36 (1.03)  & 5.82 (1.47)\\ \hline
    Q4. I was able to plan videos with sufficient evidence to satisfy viewers.*  & 4.55 (1.51)  & 5.73 (1.35)\\ \hline
    Q5. I could make decisions with confidence about the viewers.*  & 4.55 (1.75)  & 5.82 (0.98)\\ \hline
    Q6. I was able to apply viewers' perspectives to video planning.  & 5.18 (1.40)  & 6.00 (1.26)\\ \hline
    Q7. I was satisfied with the experience of video planning (creating storylines).*  & 5.18 (1.08)  & 6.36 (0.81)\\ \hline
    Q8. Please evaluate the completeness of the video storyline you created (0-100).**  & 73.0 (12.21)  & 86.82 (10.78)\\
    \bottomrule
    \end{tabular}
    \caption{Overall usability survey results. *, ** indicates statistically significant (\emph{p} < 0.05, \emph{p} < 0.01.)}
    \label{tab:question_usability}
    \Description{This table presents the results of an overall usability survey comparing \sysname{} and the baseline system across eight questions related to video planning and understanding of viewers. The table shows average scores and standard deviations (SD) for each system. Notable differences include statistically significant improvements with \sysname{} in questions about understanding viewers (Q1), planning videos based on viewer insights (Q2), and satisfaction with the video planning experience (Q7). Additionally, participants rated the completeness of their video storylines higher when using \sysname{} (86.82) compared to the baseline (73.0). Statistically significant results are marked with * (\emph{p} < 0.05) and ** (\emph{p} < 0.01).}

\end{table}

\subsection{Nasa-TLX (7-point Likert Scale)}

\begin{table}[h]
    \centering
    \begin{tabular}{p{0.75\columnwidth}|c |c}
    \hline
    \toprule
    Criteria & \makecell{\baseline{} \\ Avg. (SD)} &  \makecell{\sysname{} \\ Avg. (SD)} \\
    \midrule
    NasaTLX-Q1. Mental Demand: How mentally demanding was the task?  & 3.09 (1.81)  & 2.73 (1.85)\\ \hline
    NasaTLX-Q2. Physical Demand: How physically demanding was the task?  & 2.09 (1.76)  & 1.82 (1.08)\\ \hline
    NasaTLX-Q3. Temporal Demand: How hurried or rushed was the pace of the task?  & 2.09 (0.83)  & 2.73 (2.00)\\ \hline
    NasaTLX-Q4. Performance: How successful were you in accomplishing what you were asked to do?*  & 4.91 (1.30)  & 5.73 (1.01)\\ \hline
    NasaTLX-Q5. Effort: How hard did you have to work to accomplish your level of performance?  & 3.55 (1.37)  & 3.73 (1.27)\\ \hline
    NasaTLX-Q6. Frustration: How insecure, discouraged, irritated, stressed and annoyed were you?  & 2.09 (1.51)  & 1.91 (1.45)\\
    \bottomrule
    \end{tabular}
    \caption{Nasa-TLX survey  questionnaires and results. * indicates statistically significant (\emph{p} < 0.05)}
    \label{tab:question_nasatlx}
    \Description{This table presents the results of the NASA-TLX survey, which measures six dimensions of task load for both \sysname{} and the baseline system. The criteria include mental demand, physical demand, temporal demand, performance, effort, and frustration, with average scores and standard deviations (SD) reported for each system. A lower score indicates a lower perceived task load, except for performance, where a higher score reflects greater perceived success. The results show that participants found the task less mentally and physically demanding with \sysname{}. Notably, \sysname{} yielded significantly higher performance scores (5.73 vs. 4.91), marked with an asterisk (*) to indicate statistical significance (\emph{p} < 0.05).}
\end{table}

\clearpage

\subsection{Quality of Dimensions and Values (7-point Likert Scale, adjusted from ~\cite{suh2023structured})}
\label{app:question_dimension_values}

\begin{table}[ht!]
    \centering
    \begin{tabular}{p{0.65\columnwidth}|c}
    \toprule
    Criteria ``The generated dimensions (top categories) ...'' & Avg. (SD) \\ \midrule
    
    Q9. help to identify what elements are important.  & 4.27 (1.01)\\ \hline
    Q10. are useful in understanding my audience.  & 4.36 (0.67)\\ \hline
    Q11. are relevant to my audience.  & 4.55 (0.93)\\ \hline
    Q12. are mutually exclusive.  & 3.36 (1.21)\\ \hline
    Q13. are clear.  & 3.91 (0.83)\\ \hline
    Q14. are novel, providing new perspectives.  & 3.91 (0.94)\\ \hline
    Q15. are diversely composed.  & 4.36 (0.67)\\ \bottomrule

    \end{tabular}
    \caption{Quality of dimensions survey questionnaires and results.}
    \label{tab:question_dimension}
    \Description{This table shows the results of a survey evaluating the quality of the dimensions (or top categories) generated by the system. Participants rated each criterion on a scale, and the table presents the average score (Avg.) and standard deviation (SD) for each question. The criteria assess the usefulness, relevance, clarity, novelty, and diversity of the generated dimensions. Notably, the dimensions were rated highest for their relevance to the audience (4.55) and usefulness in understanding the audience (4.36), while mutual exclusiveness received the lowest score (3.36), indicating room for improvement in distinguishing between categories.}

\end{table}


\begin{table}[ht!]
    \centering
    \begin{tabular}{p{0.65\columnwidth}|c}
    \toprule
    Criteria ``The generated values (subcategories) ...'' & Avg. (SD) \\ \midrule
    
    Q16. are useful in understanding my audience.  & 4.18 (0.98)\\ \hline
    Q17. are relevant to my audience.  & 4.45 (0.69)\\ \hline
    Q18. are mutually exclusive.  & 3.64 (1.12)\\ \hline
    Q19. are specific.  & 4.64 (0.50)\\ \hline
    Q20. are novel, providing new perspectives.  & 4.36 (0.81)\\ \hline
    Q21. are diversely composed.  & 4.73 (0.47)\\ \bottomrule

    \end{tabular}
    \caption{Quality of Values survey questionnaires and results.}
    \label{tab:question_value}
    \Description{This table presents the results of a survey evaluating the quality of the values (or subcategories) generated by the system. Participants rated six criteria, such as usefulness, relevance, mutual exclusiveness, specificity, novelty, and diversity, on a scale. The average score (Avg.) and standard deviation (SD) are provided for each question. The values were rated highest for their diversity (4.73) and specificity (4.64), while mutual exclusiveness received the lowest score (3.64), indicating potential overlap between some subcategories. The values were also seen as highly relevant (4.45) and providing new perspectives (4.36).}

\end{table}

\clearpage

\subsection{Quality of Audience Persona Chat and Feedback (7-point Likert Scale)}
\label{app:question_persona_chat}

\begin{table}[ht!]
    \centering
    \begin{tabular}{p{0.65\columnwidth}|c}
    \toprule
    Criteria ``The conversation with the audience persona ...'' & Avg. (SD) \\ \midrule
    
    Q22. was consistent.  & 6.55 (0.69)\\ \hline
    Q23. clearly reflected the perspective of specific personas.  & 6.36 (0.81)\\ \hline
    Q24. was natural.  & 5.45 (1.37)\\ \hline
    Q25. was sufficiently reliable.  & 5.27 (1.35)\\ \hline
    Q26. provided evidence (e.g., video content) in the conversation, when necessary.  & 5.91 (1.38)\\ \bottomrule

    \end{tabular}
    \caption{Quality of audience persona chat survey  questionnaires and results.}
    \label{tab:question_persona_chat}
    \Description{This table presents the results of a survey assessing the quality of conversations with audience personas. Participants rated five criteria on a 7-point Likert scale, with the average score (Avg.) and standard deviation (SD) shown for each. The conversations were rated highly for consistency (6.55) and for reflecting the perspective of specific personas (6.36). The scores for naturalness (5.45) and reliability (5.27) were slightly lower but still positive. Participants also found that the personas provided relevant evidence, such as video content, when necessary (5.91).}

\end{table}


\begin{table}[ht!]
    \centering
    \begin{tabular}{p{0.65\columnwidth}|c}
    \toprule
    Criteria ``The feedback from the audience persona ...'' & Avg. (SD) \\ \midrule
    
    Q27. was clear.  & 5.82 (1.08)\\ \hline
    Q28. clearly reflected the perspective of specific personas.  & 6.36 (0.50)\\ \hline
    Q29. was diverse.  & 5.00 (1.55)\\ \hline
    Q30. was sufficiently reliable.  & 5.18 (1.60)\\ \hline
    Q31. was given in a sufficiently applicable form.  & 5.82 (1.94)\\ \bottomrule

    \end{tabular}
    \caption{Quality of feedback survey questionnaires and results.}
    \label{tab:question_persona_feedback}
    \Description{This table presents the results of a survey evaluating the quality of feedback provided by audience personas. Participants rated five criteria on a scale, with the average score (Avg.) and standard deviation (SD) displayed for each. The feedback was rated highly for reflecting the perspective of specific personas (6.36) and for clarity (5.82). Lower scores were given for the diversity of feedback (5.00) and reliability (5.18), though the feedback was considered sufficiently applicable (5.82).}

\end{table}

\subsection{AI Chains (7-point Likert Scale, adjusted from ~\cite{wu2022ai})}
\label{app:ai_chains}

\begin{table}[ht!]
    \centering
    \begin{tabular}{p{0.65\columnwidth}|c}
    \toprule
    Criteria & Avg. (SD) \\ \midrule
    
    Q32. Match goal: I am satisfied with the final result obtained using the system and was able to achieve my work goal.  & 6.00 (0.77)\\ \hline
    Q33. Think through: The system helped me think about the desired outcome for achieving the given work goal, allowing me to contemplate how to complete the task.  & 6.27 (0.79)\\ \hline
    Q34. Transparent: I felt that the process leading to the final result was clearly shown by the system, and I could generally track the progress.  & 5.91 (1.14)\\ \hline
    Q35. Controllable: I felt I had enough control while using the system. In other words, I could steer the system in the direction I wanted to achieve my work goals.  & 5.18 (1.66)\\ \hline
    Q36. Collaborative: I felt like I was collaborating with the system to create the outcome, as if working together with the system.  & 6.18 (1.17)\\ \bottomrule

    \end{tabular}
    \caption{AI Chain survey  questionnaires and results.}
    \label{tab:ai_chains}

    \Description{This table presents the results of a survey evaluating user satisfaction with the system in terms of its ability to help achieve work goals, provide transparency, and offer control during the task. Participants rated five criteria, with average scores (Avg.) and standard deviations (SD) provided for each. The highest scores were given for the system helping users think through their tasks (6.27) and for its collaborative nature (6.18). Users were also satisfied with the final result (6.00) and felt that the system provided clear progress tracking (5.91), though they rated the level of control slightly lower (5.18).
}

\end{table}



\appendix

\end{document}